\documentclass{aa}
\newcommand{\kms}{km~s$^{-1}$}
\usepackage{natbib}
\usepackage{graphicx}
\usepackage{amssymb}
\usepackage{txfonts}
\begin{document}

\title{PROSAC: A Submillimeter Array survey of low-mass protostars} 
\subtitle{II. The mass evolution of envelopes, disks, and stars from the Class
  0 through I stages}
\titlerunning{The mass evolution of embedded protostars}
\authorrunning{J.~K. J{\o}rgensen et al.}
\author{Jes K. J{\o}rgensen\inst{1} \and
  Ewine F. van Dishoeck\inst{2,3} \and Ruud Visser\inst{2}  \and Tyler L. Bourke\inst{4} \and
  David J. Wilner\inst{4} \and Dave Lommen\inst{2} \and Michiel
  R. Hogerheijde\inst{2} \and Philip C. Myers\inst{4}}

\institute{Argelander-Institut f\"{u}r Astronomie, University of Bonn,
  Auf dem H\"{u}gel 71, 53121 Bonn, Germany \and Leiden Observatory,
  Leiden University, PO Box 9513, NL-2300 RA Leiden, The Netherlands
  \and Max-Planck Institut f\"{u}r Extraterrestrische Physik,
  Giessenbachstrasse 1, 85748 Garching, Germany \and
  Harvard-Smithsonian Center for Astrophysics, 60 Garden Street MS-42,
  Cambridge, MA 02138, USA}

\abstract{The key question about early protostellar evolution is how
  matter is accreted from the large-scale molecular cloud, through the
  circumstellar disk onto the central star.}  {We constrain the masses
  of the envelopes, disks, and central stars of a sample of low-mass
  protostars and compare the results to theoretical models for the
  evolution of young stellar objects through the early protostellar
  stages.}  {A sample of 20 Class 0 and I protostars has been observed
  in continuum at (sub)millimeter wavelengths at high angular
  resolution (typically 2$''$) with the Submillimeter Array. Using
  detailed dust radiative transfer models of the interferometric data,
  as well as single-dish continuum observations, we have developed a
  framework for disentangling the continuum emission from the
  envelopes and disks, and from that estimated their masses. For the
  Class I sources in the sample HCO$^+$ 3--2 line emission has
  furthermore been observed with the Submillimeter Array. Four of
  these sources show signs of Keplerian rotation, making it possible
  to determine the masses of the central stars. In the other sources
  the disks are masked by optically thick envelope and outflow
  emission.}  {Both Class 0 and I protostars are surrounded by disks
  with typical masses of about 0.05~$M_\odot$, although significant
  scatter is seen in the derived disk masses for objects within both
  evolutionary stages. No evidence is found for a correlation between
  the disk mass and evolutionary stage of the young stellar
  objects. This contrasts the envelope mass, which decreases sharply
  from $\sim 1$~$M_\odot$ in the Class 0 stage to $\lesssim
  0.1$~$M_\odot$ in the Class I stage. Typically, the disks have
  masses that are 1--10\% of the corresponding envelope masses in the
  Class~0 stage and 20--60\% in the Class~I stage. For the Class~I
  sources for which Keplerian rotation is seen, the central stars
  contain 70--98\% of the total mass in the star-disk-envelope system,
  confirming that these objects are late in their evolution through
  the embedded protostellar stages, with most of the material from the
  ambient envelope accreted onto the central star. Theoretical models
  tend to overestimate the disk masses relative to the stellar masses
  in the late Class~I stage.}  {The results argue in favor of a
  picture in which circumstellar disks are formed early during the
  protostellar evolution (although these disks are not necessarily
  rotationally supported) and rapidly process material accreted from
  the larger scale envelope onto the central star.}

\keywords{star: formation -- circumstellar matter -- planetary systems: protoplanetary disks -- radiative transfer}

\offprints{Jes K.\,J{\o}rgensen} \mail{jes@astro.uni-bonn.de}
\maketitle

\section{Introduction}
The fundamental problem in studies of the evolution of low-mass young
stellar objects is how material is accreted from the larger scale
envelope through the protoplanetary disk onto the central star -- and,
in particular, what evolutionary time-scales determine the formation
and growth of the disk and star and the dissipation of envelopes. Our
theoretical understanding of the evolution of YSOs has typically
relied on the coupling between theoretical studies and empirical
schemes, based, for example, on the spectral energy distributions or
colors or on indirect measures, e.g., of the accretion rates. The most
commonly used distinction between the embedded Class 0 and I
protostars is that the former emit more than 0.5\% of their luminosity
at wavelengths longer than 350~$\mu$m. This in turn is thought to
reflect that they have accreted less than half of their final mass
\citep[e.g.,][]{andre93,andreppiv}, but few hard constraints on the
actual evolution of the matter through the embedded protostellar
stages exist. Direct measurements of the stellar, disk, and envelope
masses during these pivotal stages can place individual protostars in
their proper evolutionary context by addressing questions such as what
fraction of the initial core mass has been accreted onto the central
star, what fraction is being carried away by the action of the
protostellar outflows and what fraction has (so-far) ended up in a
disk. In this paper, we analyze single-dish and interferometric
observations of emission from dust around a sample of 20 Class 0 and I
low-mass protostars, as well as observations of a dense gas tracer,
HCO$^+$ 3--2, for the Class I sources. By interpreting continuum data
using dust radiative transfer models, we disentangle the contributions
from the envelopes and disks to the emission and estimate their
masses. Together with stellar masses inferred from Keplerian rotation
in resolved line maps \citep[see also][]{brinch07b,lommen08}, we
directly constrain the evolution of matter during the earliest
embedded stages of young stellar objects.

The formation of a circumstellar disk is clearly a key event in the
evolution of young stellar objects, providing the route through which
material is accreted from the large-scale envelope onto the central
star. Theoretically a rotationally supported circumstellar disk is
expected to form during the collapse of a dense core due to its
initial angular momentum: material falling in from the outer regions
of the core will be deflected away from the central star into a flat
circumstellar disk structure at the ``centrifugal radius'' where
gravity is balanced by rotation
\citep[e.g.,][]{ulrich76,cassen81,terebey84,basu98}. The centrifugal
radius will grow in time, with the exact pace depending on the
distribution of angular momentum in the infalling core, e.g., whether
it can be described as solid-body rotation \citep[e.g.,][]{terebey84}
or as differential rotation as expected in magnetized infalling cores
\citep[e.g.,][]{basu98}. Even earlier in the evolution of low-mass
YSOs, an unstable pseudo-disk may develop as magnetic fields in the
core deflect infalling material away from the radial direction toward
the central disk plane \citep[][]{galli93a,galli93b}.

The existence of circumstellar disks has been used to solve the
so-called ``luminosity problem'' \citep[e.g.][]{kenyon90} in which
embedded young stellar objects are observed to be under-luminous
compared to the expected luminosity due to release of gravitational
energy under constant mass accretion onto the central star. The
solution proposed by \cite{kenyon90} is that material is accreted onto
the circumstellar disk (and thus at larger radii with a smaller
release of gravitational energy) and from there episodically onto the
central star, e.g., related to the observed FU Orionis outbursts. The
recent compilation of YSO luminosities from the \emph{Spitzer Space
  Telescope} ``Cores to Disks (c2d)'' legacy program \citep{evans09}
shows that even the most deeply embedded YSOs are under-luminous
compared to the predictions from models -- perhaps suggesting that
already in these early phases material is not accreted directly on to the
central protostar, but rather accumulated either in a circumstellar
disk or accreted in another episodic manner.

The main issue about inferring the properties of circumstellar disks
around embedded YSOs remains to disentangle the contribution from the
larger scale envelopes and circumstellar disks. The continuum emission
from dust on the other hand provides an optically thin tracer, whose
strength mainly depends on the column density and temperature of the
emitting material -- and thus potentially provides a direct probe of
the dust mass of envelope and/or disks if these can be separated.
High angular resolution line observations provide a means to complete
the picture of the structure of young stellar objects and their
evolution by constraining the dynamical structure of the inner
envelope and disk and through that infer the central stellar masses
\citep[e.g.,][]{brown99chandler,brinch07b,lommen08}. In particular,
aperture synthesis observations at submillimeter wavelengths are
important for this goal: the higher excitation transitions of common
molecules such as HCO$^+$, HCN and CS observed at 1~mm and shorter
wavelengths, probe densities $\ge$10$^5$~cm$^{-3}$ and temperatures
$\ge 25$~K \citep[e.g.,][]{evans99} and are thus much less sensitive
to the structure of the outer envelope/ambient cloud compared to
ground-state and low-excitation transitions observable at longer
wavelengths.

For more evolved YSOs in fairly isolated regions, single-dish
continuum observations have been used for comparative studies of
statistically significant samples of sources
\citep[e.g.,][]{beckwith90,andre94,andrews05,andrews07}. Subarcsecond
resolution imaging using (sub)millimeter wave interferometers has been
done for smaller samples of objects
\citep[e.g.,][]{lay94,mundy96,hogerheijde97ttau,looney00}. For
embedded protostars, the issue arises that both the disks and
envelopes may contribute significantly to the emission within the
typical single-dish beam and for such sources, interferometric imaging
is needed to address the dust distribution on the size scales where
disks may be forming. \cite{keene90} demonstrated the potential of
such high angular resolution millimeter observations using data from
the OVRO millimeter array: by comparing the interferometric data to
models for the collapsing envelope around the L1551-IRS5 Class I
protostar, inferred from modeling its SED, \citeauthor{keene90} found
that a central unresolved component was required to explain the
observed brightness profiles in the interferometric data, such as
expected from a central circumstellar disk. In the case of L1551-IRS5,
this compact component accounts for about 50\% of the total dust
continuum emission -- or if interpreted in the context of a
circumstellar disk and envelope -- a circumstellar disk containing
about 20--25\% of the total mass in the
system. \cite{hogerheijde97,hogerheijde98} used the OVRO Millimeter
Array to survey the continuum and line emission at 3~mm from a sample
of 9 Class~I protostellar sources. They showed that most of the
continuum emission was associated with a compact, unresolved
($<$3\arcsec) component. More recently these kind of studies have seen
significant progress with the advent of new data and more detailed
dust radiative transfer models
\citep[e.g.,][]{hogerheijde00sandell,looney03,n1333i2art,iras2sma,l483art,hotcorepaper,eisner05,brinch07b,lommen08}. In
many cases, evidence is found for similarly compact components -
although the exact interpretation of their nature still depends on
assumptions about the properties of the larger scale envelopes. Still,
despite these efforts the studied samples have been limited to at most
a handful of sources.

In this paper, we combine continuum and line observations for a large
sample of embedded protostars to constrain the masses of their three
main components: the envelopes, disks and central stars. Specifically,
we discuss single-dish and interferometric observations of a sample of
20 Class 0 and I young stellar objects covering size scales from a few
thousand down to a few hundred AU. We use continuum data together with
detailed dust radiative transfer to constrain the masses of the larger
scale envelopes and circumstellar disks, thereby providing constraints
on the evolution of the material through these embedded stages. In
addition, we report observations of the HCO$^+$ 3--2 line at 267~GHz
tracing dense gas in the envelopes around the Class I protostars - in
a few cases showing evidence for Keplerian rotation in the inner
envelopes/circumstellar disks. This paper is laid out as follows:
\S\ref{data} introduces the selected samples of Class 0 and I
protostars. The details of the new SMA observations of the Class~I
sources in this sample are presented in \S\ref{classIobs}; for the
Class~0 sources we refer to
\cite{prosacpaper}. \S\ref{continuumanalysis} discusses the
constraints on the envelope and disk properties from the continuum
data compared to generic dust radiative transfer models of
protostellar envelope. \S\ref{lineanalysis} discusses the results of
the line observations of the Class I sources - in particular, the use
of HCO$^+$ 3--2 as a tracer of Keplerian motions in the inner
envelopes and disks around four of these sources. A discussion of
outflow cavities observed in the HCO$^+$ 3--2 emission for another
three of the Class I sources is deferred to the Appendix where also
details of the individual sources are given. Finally,
\S\ref{discussion} compares the inferred masses to models for the
evolution of the envelope, disk and stellar masses and discusses the
assumptions in, e.g., the assumed dust opacities, before
\S\ref{summary} summarizes the main conclusions of the paper.

\section{Sample and data}\label{data}
\subsection{Sample}
Our data consist of line and continuum observations of 10
Class~0 sources and 9 Class~I sources observed as part of the
Submillimeter Array (SMA; \citealt{ho04})\footnote{The Submillimeter
  Array is a joint project between the Smithsonian Astrophysical
  Observatory and the Academia Sinica Institute of Astronomy and
  Astrophysics, and is funded by the Smithsonian Institution and the
  Academia Sinica.} key project, Protostellar Submillimeter Array
Campaign (PROSAC; \citealt{prosacpaper}).

The Class~0 sources were observed in 3 different spectral setups and
continuum at 1.3~mm (225~GHz) and 0.85~mm (345~GHz) between November
2004 and January 2006, and the data for those were discussed in
\citep{prosacpaper}. 

The Class I YSOs were selected amongst known Class I sources in the
Ophiuchus and Taurus star forming regions also surveyed by
\cite{hogerheijde97} and \cite{vankempen09}. The sample was
constructed to span the expected evolutionary range of Class I sources
based on their bolometric temperatures, $T_{\rm bol}$
\citep{myers93}. In total eight fields were selected encompassing 9
YSOs previously classified as embedded Class I protostars. In
addition, we include the Class~I protostar L1489-IRS, which was
observed with the SMA in compact and extended configuration in the
same lines as the remaining Class~I sources and analyzed using
detailed 2D radiative transfer models by \cite{brinch07b}. These
surveys provide (sub)millimeter continuum data of 1--3\arcsec\
resolution with typical sensitivities of a few mJy~beam$^{-1}$. The
primary beam field of view of the SMA is 43--51$''$ at 1.1--1.3~mm.

An overview of the total sample is given in Table~\ref{sample}. In
this table we give the positions of the sources from fits to the
(sub)millimeter continuum position as well as bolometric temperatures
and luminosities for the sources -- predominantly based on the
compilation from the Spitzer/c2d legacy program \citep{evans09}.
\begin{table*}
\caption{Sample of embedded protostars.}\label{sample}
\begin{center}
\begin{tabular}{llllll}\hline\hline
Source              & RA [J2000] & DEC [J2000] & $T_{\rm bol}$ [K] & $L_{\rm bol}$ [$L_\odot$]& $d$ [pc] \\ \hline
\multicolumn{6}{c}{\emph{Class 0}}\\[1.0ex]
L1448-mm            &  03 25 38.9  & $+$30 44 05.4  &   77  & 5.3    & 220 \\[1.0ex]
\multicolumn{5}{l}{NGC~1333:} \\
\phantom{xx}-IRAS2A             &  03 28 55.6  & $+$31 14 37.1  &   57  & 20$^{a}$    & 220 \\
\phantom{xx}-IRAS4A-SE          &  03 29 10.5  & $+$31 13 31.0  &   43  & 5.8     & 220 \\
\phantom{xx}-IRAS4A-NW          &  03 29 10.4  & $+$31 13 32.3  &   --  & --      & 220 \\
\phantom{xx}-IRAS4B             &  03 29 12.0  & $+$31 13 08.1  &   55  & 3.8$^{b}$ & 220\\
\phantom{xx}-IRAS4B$'$          &  03 29 12.8  & $+$31 13 07.0  &   --  & --      & 220 \\[1.0ex]
L1527               &  04 39 53.9  & $+$26 03 09.8  &   59  & 2.0     & 140 \\
L483                &  18 17 29.9  & $-$04 39 39.6  &   60  & 9.0     & 200 \\
B335                &  19 37 00.9  & $+$07 34 09.7  &   60  & 3.0     & 250 \\
L1157               &  20 39 06.3  & $+$68 02 16.0  &   42  & 6.0     & 325 \\[1.0ex]
\multicolumn{6}{c}{\emph{Class I}}\\[1.0ex]
L1489-IRS           &  04 04 42.9  & $+$26 18 56.3  &  238  & 3.7     & 140 \\
TMR1                &  04 39 13.9  & $+$25 53 20.6  &  144  & 3.7     & 140 \\
TMC1A               &  04 39 35.2  & $+$25 41 44.4  &  172  & 2.2     & 140 \\
GSS30-IRS3          &  16 26 21.7  & $-$24 22 50.6  &  300  & 14$^{c}$ & 125 \\
GSS30-IRS1          &  16 26 22.2  & $-$24 23 01.9  &  --   & --      & 125 \\
WL~12                &  16 26 44.2  & $-$24 34 48.7  &  440  & 2.6     & 125 \\
Elias 29            &  16 27 09.4  & $-$24 37 20.0  &  350  & 41    & 125 \\
IRS~43               &  16 27 26.9  & $-$24 40 50.6  &  310  & 6.0     & 125 \\
IRS~54               &  16 27 51.8  & $-$24 31 45.4  &  740  & 2.5     & 125 \\
IRS~63               &  16 31 35.7  & $-$24 01 29.6  &  530  & 3.3     & 125 \\ \hline
\end{tabular}				            
\end{center}
      
$^{a}$For NGC~1333-IRAS2A \cite{evans09} quote a luminosity of 76~$L_\odot$. Estimating the luminosity from scaling the 70~$\mu$m flux using the approach of \cite{dunham08} gives a luminosity of 19.6~$L_\odot$, which agrees well with previous estimates from modeling the SED in the range 16--20~$L_\odot$. The higher luminosity in the c2d compilation is a result of the SED of IRAS2A being unconstrained at 24~$\mu$m (M.~Dunham, priv. comm) -- and we therefore adopt a luminosity of 20~$L_\odot$ for this source. $^{b}$The Spitzer maps of NGC~1333-IRAS4B shows significant emission from a nearby outflow shock within the typical single-dish beam. As total luminosity for this source, we adopt the sum of the two components, SSTc2d J032912.06+311305.4 and SSTc2d J032912.06+311301.7, from \cite{evans09}. $^{c}$The emission from GSS30-IRS3 (LFAM1) and GSS30-IRS1 are not separated in the Spitzer catalog (the latter dominates the emission and its luminosity is quoted here).
\end{table*}

In addition to the SMA data we collected additional near-infrared data
for each Class~I source: Hubble Space Telescope (HST) 1.6~$\mu$m
images from the Near Infrared Camera and Multi-Object Spectrometer
(NICMOS) camera were obtained from the HST archive at ESO for TMR1,
TMC1A, IRS~43 and GSS30. The images were registered using existing
near-infrared data \citep[e.g.,][]{barsony97}. We also include single
pointing HCO$^+$ 3--2 spectra from the James Clerk Maxwell Telescope
(JCMT) archive that were originally presented by \cite{hogerheijde97}
and unpublished observations that were obtained in connection with the
HCO$^+$ 4--3 study by \cite{vankempen09}. For both Class~0 and I
sources, recalibrated single-dish submillimeter continuum maps
(Fig.~\ref{scubamaps}) were obtained through the JCMT/SCUBA archive
legacy project \citep{difrancesco08}.

\section{SMA observations of Class I sources}\label{classIobs}
\subsection{Observations}
The observations of the Class I sources were performed between May
2006 and July 2007 with the Submillimeter Array (SMA) in two
variations of its compact configuration. For all observations, all 8
antennas were available in the array, providing projected baselines of
9 to 64~k$\lambda$ (May 2006 and January 2007 observations) and 9 to
104~k$\lambda$ (June and July 2007 observations). Typically, two sources
were observed per track with the exception of IRS~63 and Elias 29 that
were observed in separate tracks as discussed by \cite{lommen08}.

The complex gains were calibrated by observations of 1.5--3~Jy quasars
typically located within 15$^\circ$ of the targeted sources once every
20 minutes. The bandpass was calibrated through observations of strong
quasars and planets at the beginning and end of each track. The quasar
fluxes were bootstrapped through observations of Uranus with a
  resulting $\approx$~20\% flux calibration uncertainty.

The SMA receivers and correlator was configured to observe the lines
of HCN $J=3\to 2$ (265.886444~GHz) and HCO$^+$ $J=3\to 2$
(267.557648~GHz) together with the continuum at 1.1~mm: 1 chunk of 512
channels was centered on each of the lines, resulting in a spectral
resolution of 0.2~MHz or 0.23~km~s$^{-1}$. The remainder of the 2~GHz
bandwidth in each sideband was used to record the continuum.

The initial flagging of the data as well as bandpass, flux and
phase/amplitude calibrations were performed in the Mir package
\citep{qimir} and subsequent imaging and cleaning in Miriad
\citep{sault95}. Table~\ref{obsdetails} summarizes the details of the
observations.
\begin{table*}\centering
\caption{Log of observations of Class I sources.\label{obsdetails}}
\begin{tabular}{lllll}\hline\hline
Sources        & Date        & Gain calibrator (flux)                 & Beam size              & Continuum RMS \\ \hline
IRS~63         & 2006 May 15 & J1626-298 (1.3~Jy), J1517-243 (2.4~Jy) & 4.0$''$$\times$2.3$''$ & 5.5 mJy beam$^{-1}$ \\
Elias 29       & 2006 May 17 & J1626-298 (1.7~Jy), J1517-243 (3.2~Jy) & 4.0$''$$\times$2.3$''$ & 3.6 mJy beam$^{-1}$ \\
TMR1, TMC1A    & 2007 Jan 02 & J0530+135 (2.8~Jy), 3c111 (2.3~Jy)     & 2.7$''$$\times$2.4$''$ & 2.6 mJy beam$^{-1}$ \\
IRS~43, IRS~54 & 2007 Jun 27 & J1626-298 (1.5~Jy), J1517-243 (1.6~Jy) & 3.9$''$$\times$2.0$''$ & 2.8 mJy beam$^{-1}$ \\
WL~12, GSS30   & 2007 Jul 04 & J1626-298 (1.3~Jy), J1517-243 (1.4~Jy) & 2.8$''$$\times$2.7$''$ & 2.6 mJy beam$^{-1}$ \\ \hline
\end{tabular}
\end{table*}

\begin{figure*}
\resizebox{0.25\hsize}{!}{\includegraphics{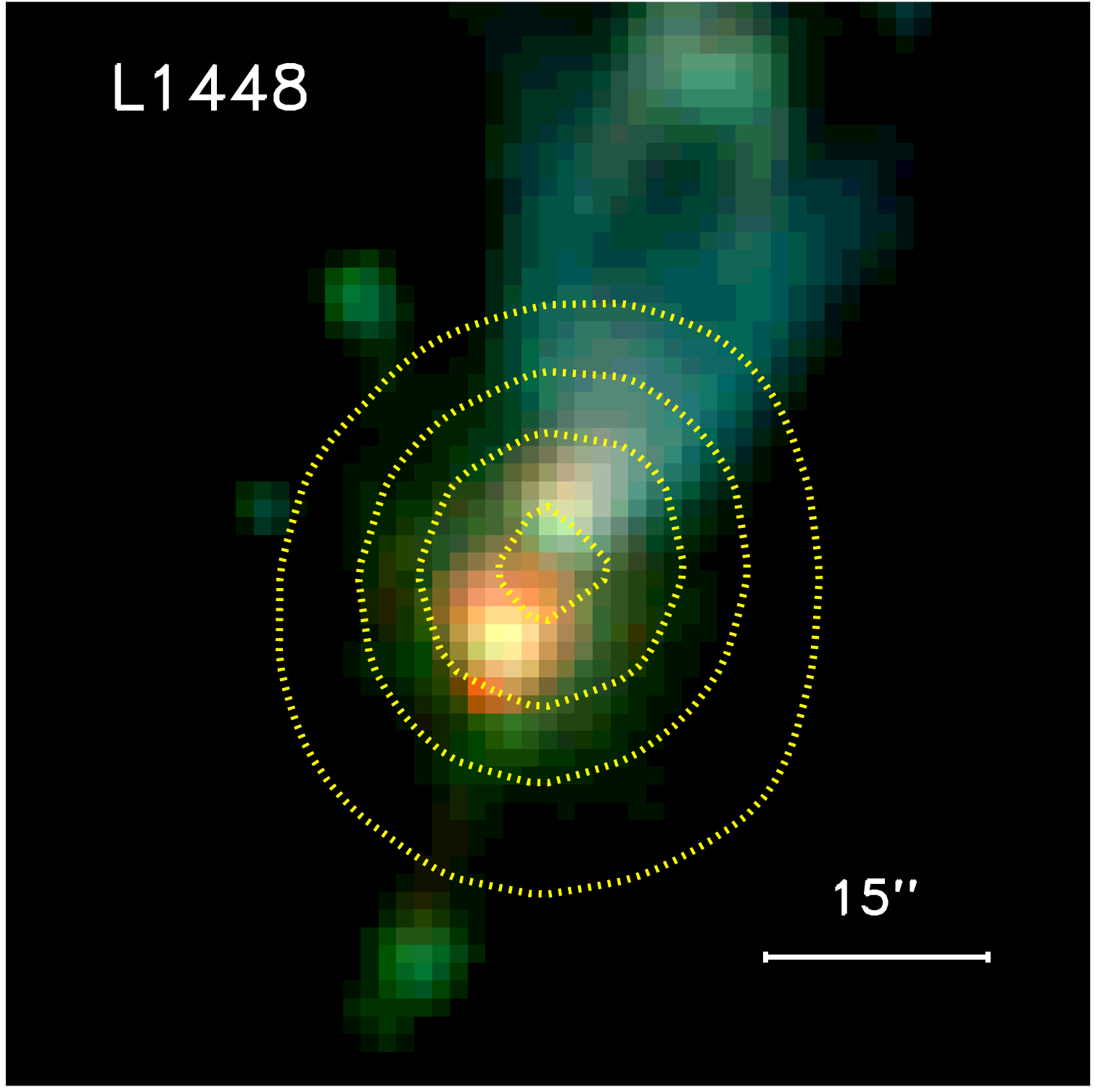}}\resizebox{0.25\hsize}{!}{\includegraphics{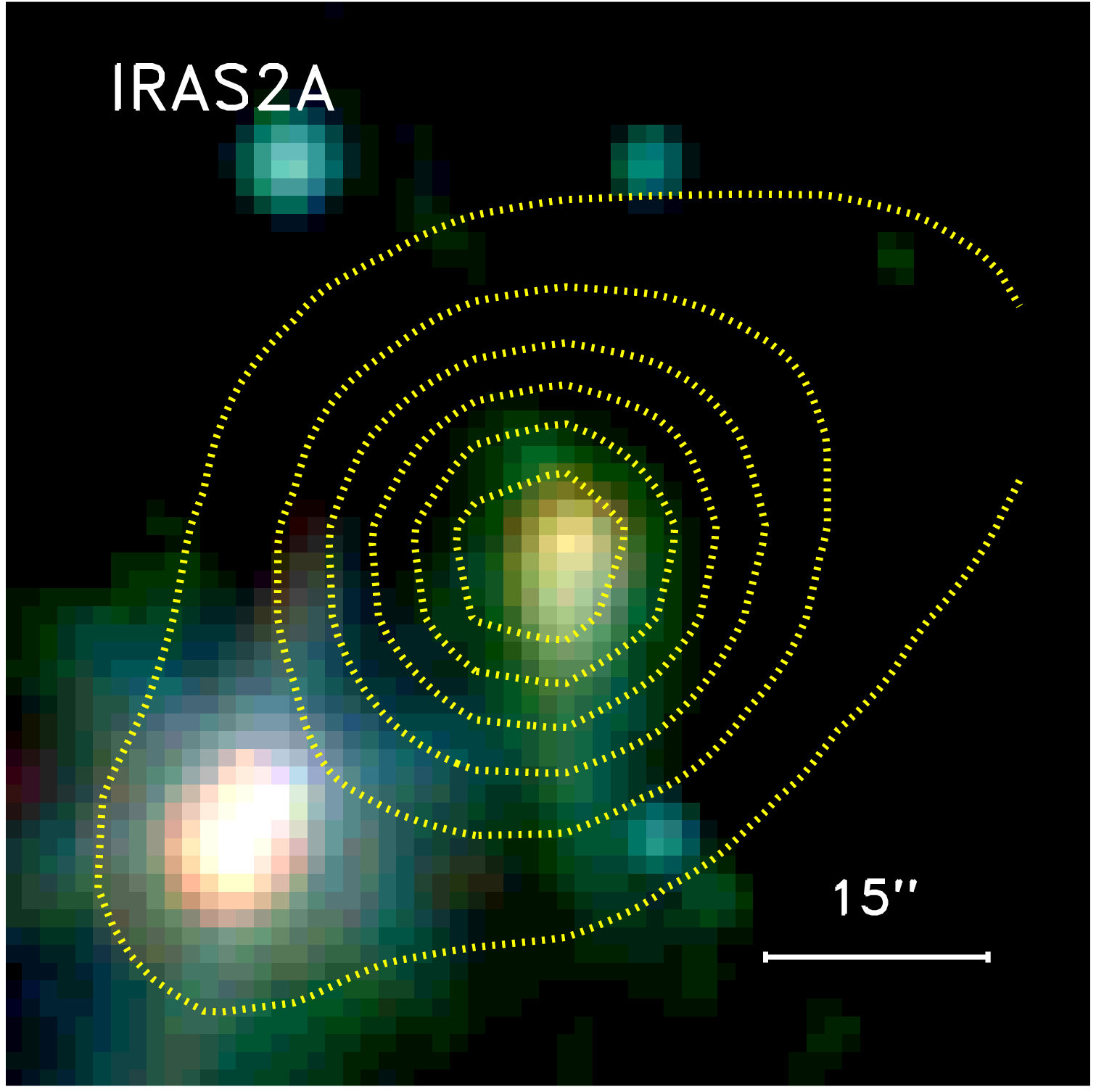}}\resizebox{0.25\hsize}{!}{\includegraphics{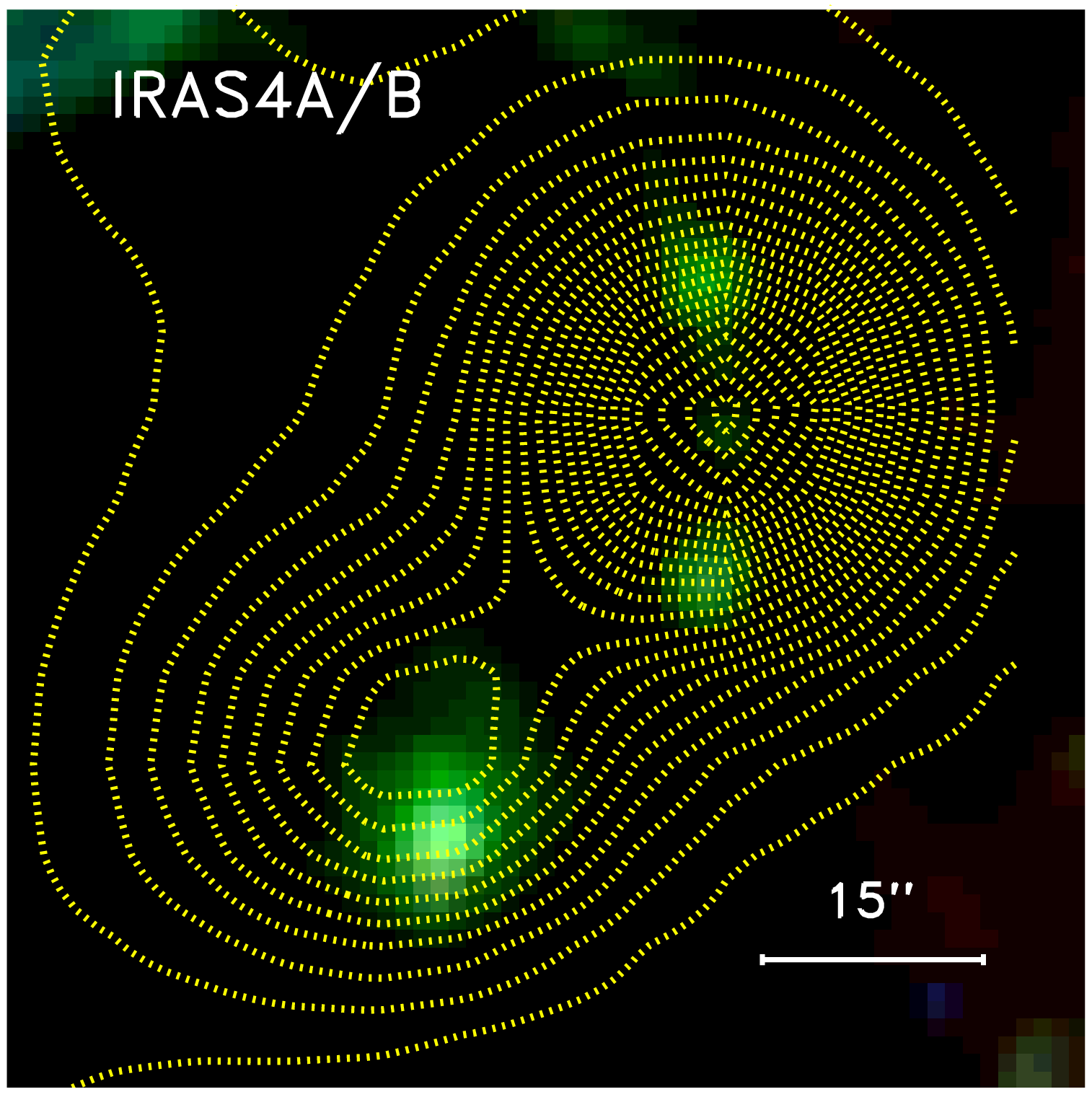}}\resizebox{0.25\hsize}{!}{\includegraphics{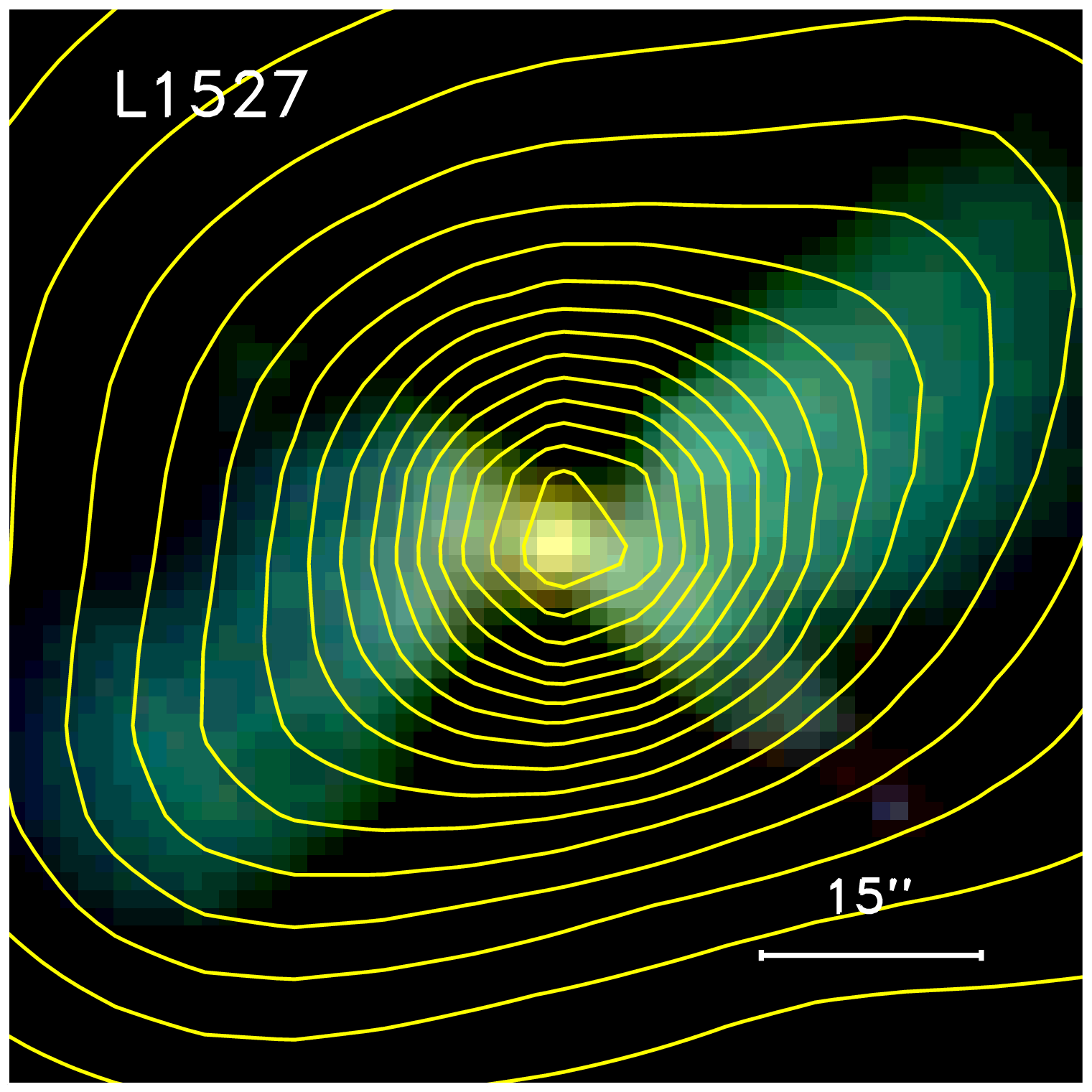}}
\resizebox{0.25\hsize}{!}{\includegraphics{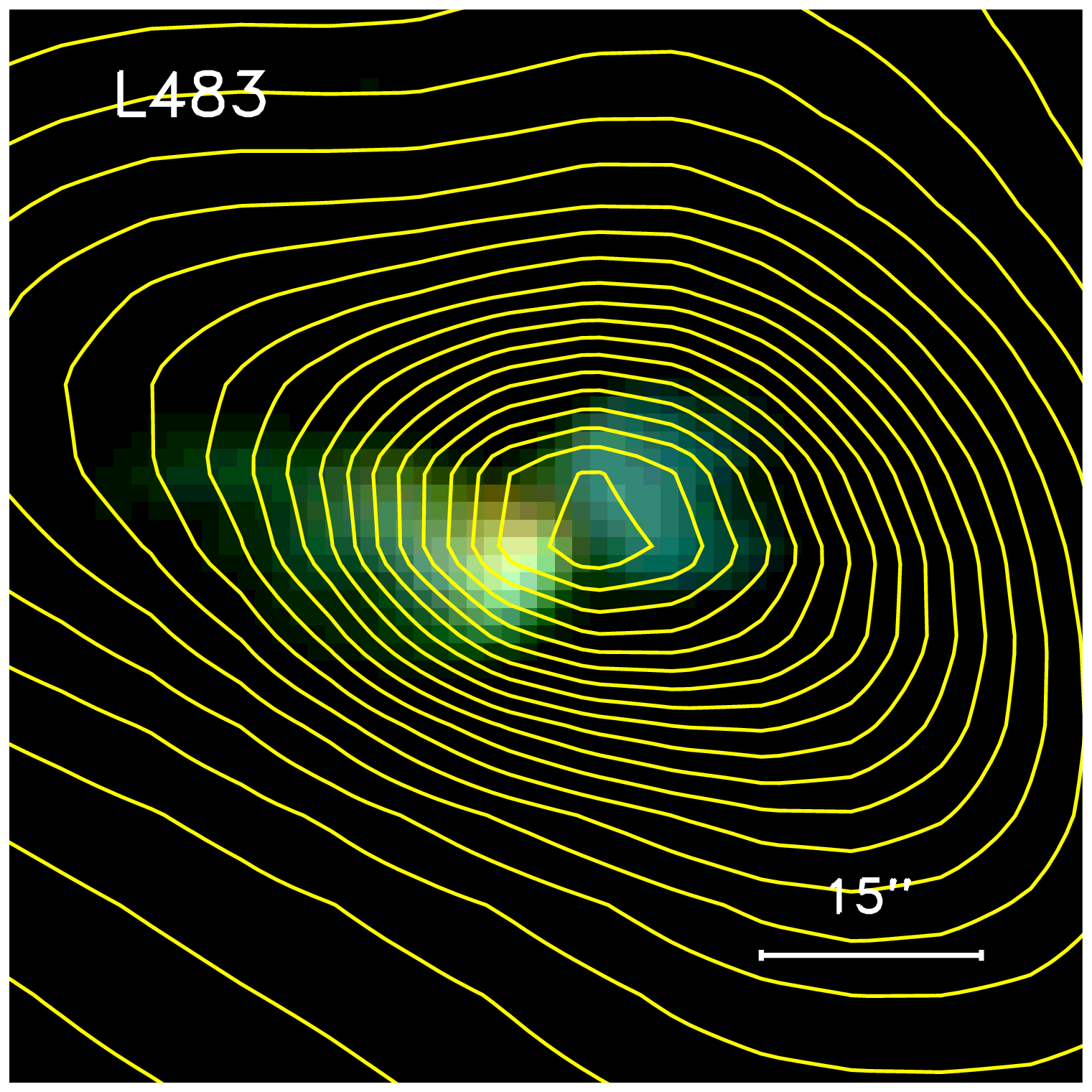}}\resizebox{0.25\hsize}{!}{\includegraphics{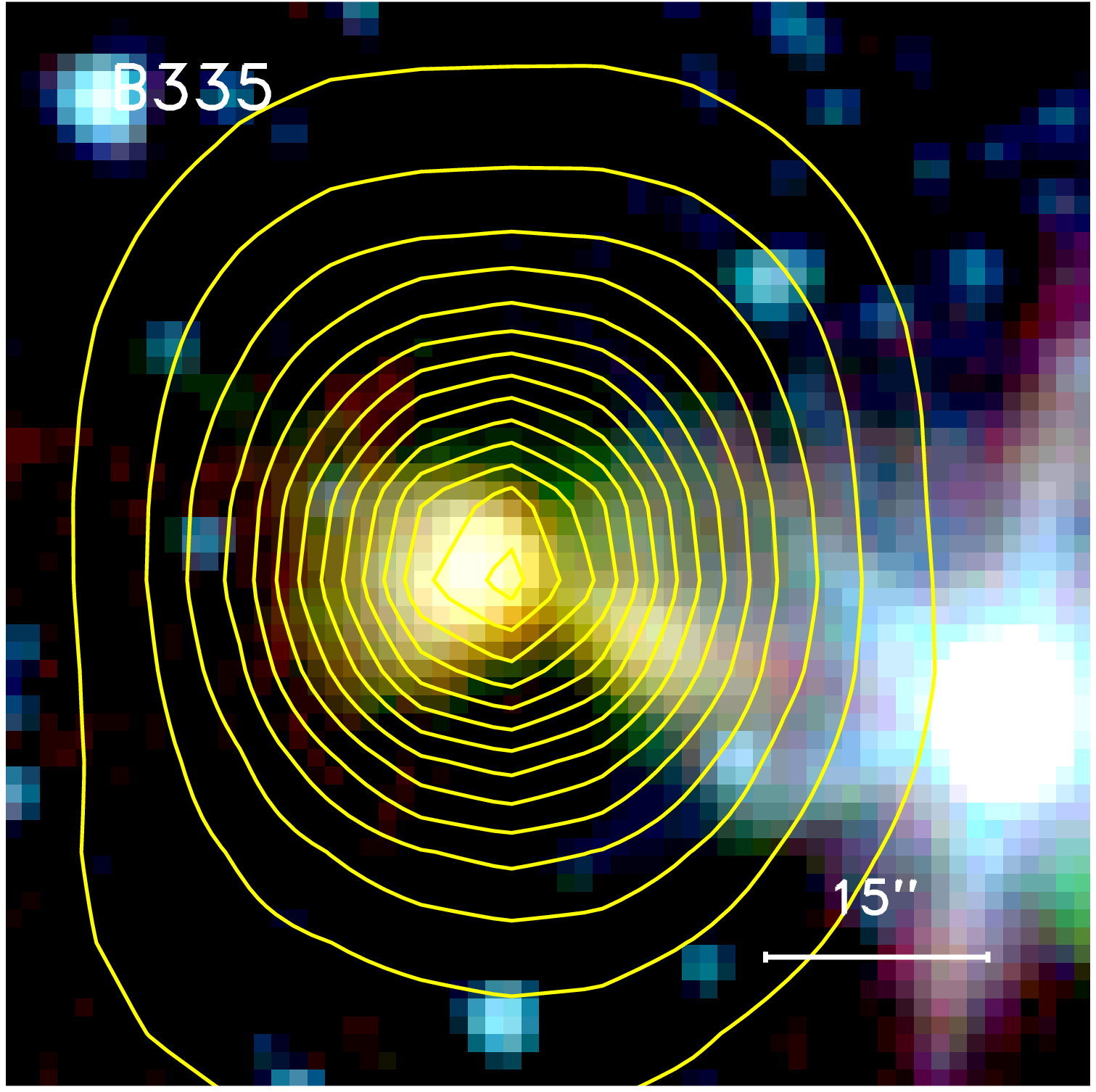}}\resizebox{0.25\hsize}{!}{\includegraphics{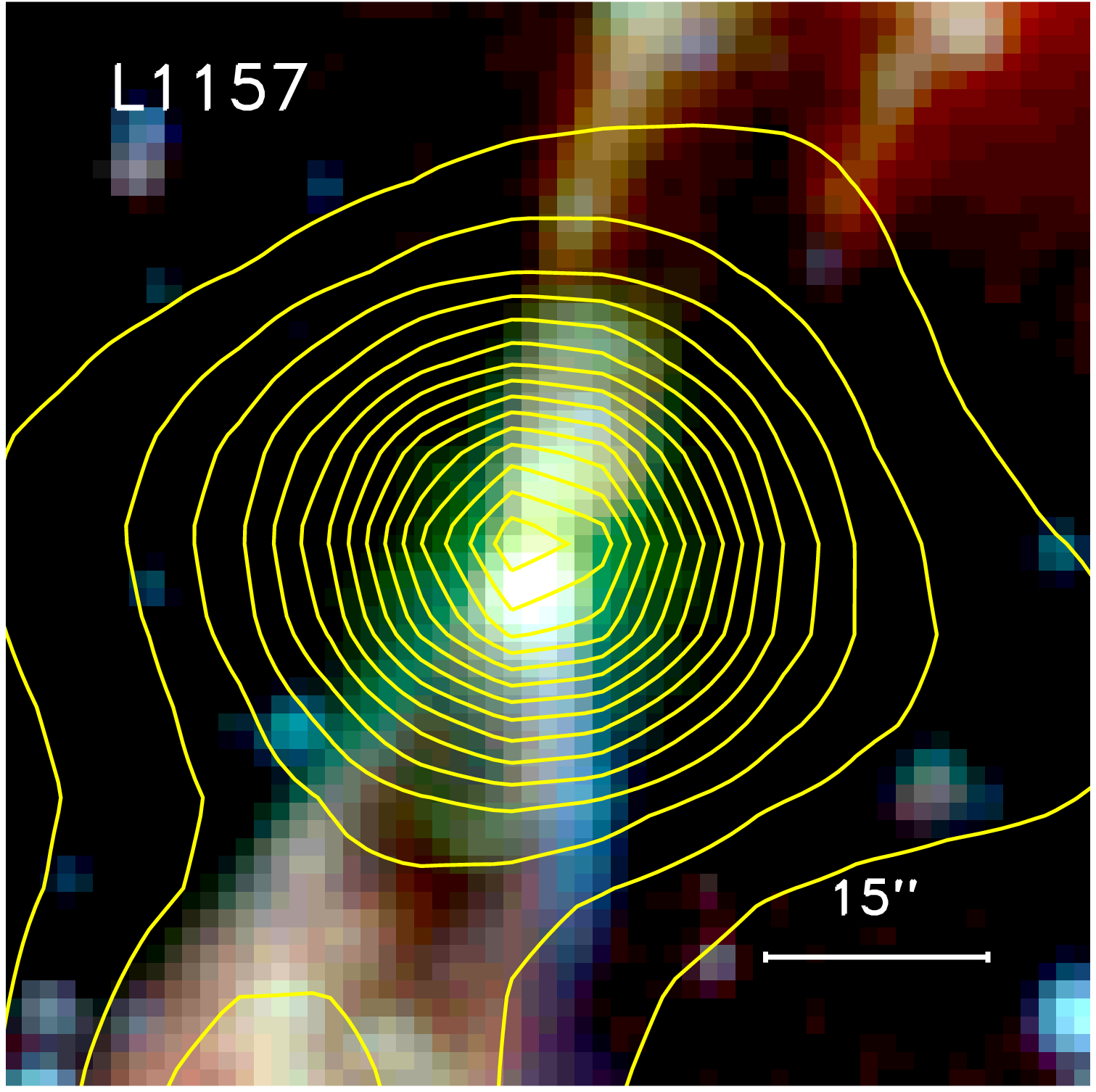}}\resizebox{0.25\hsize}{!}{\includegraphics{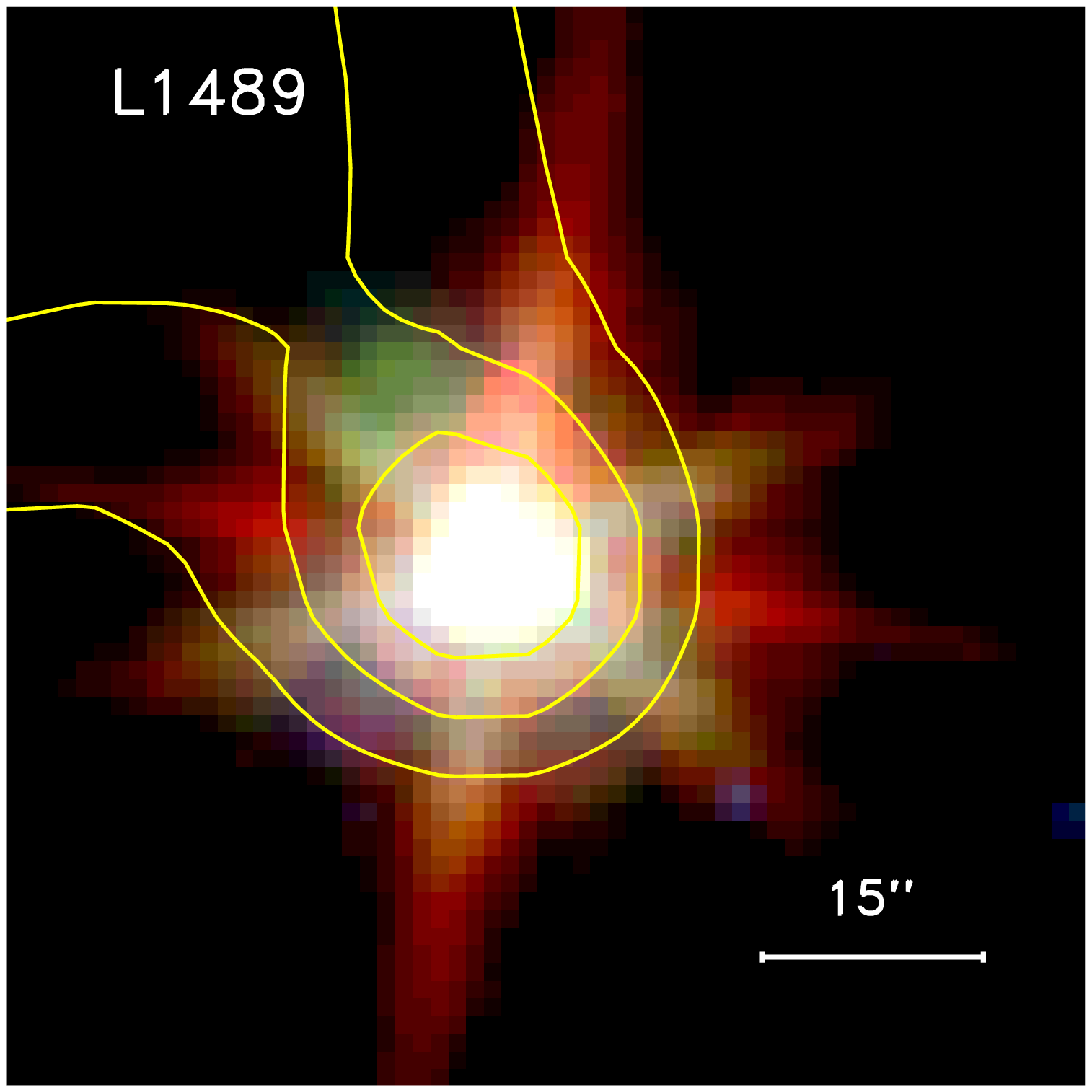}}
\resizebox{0.25\hsize}{!}{\includegraphics{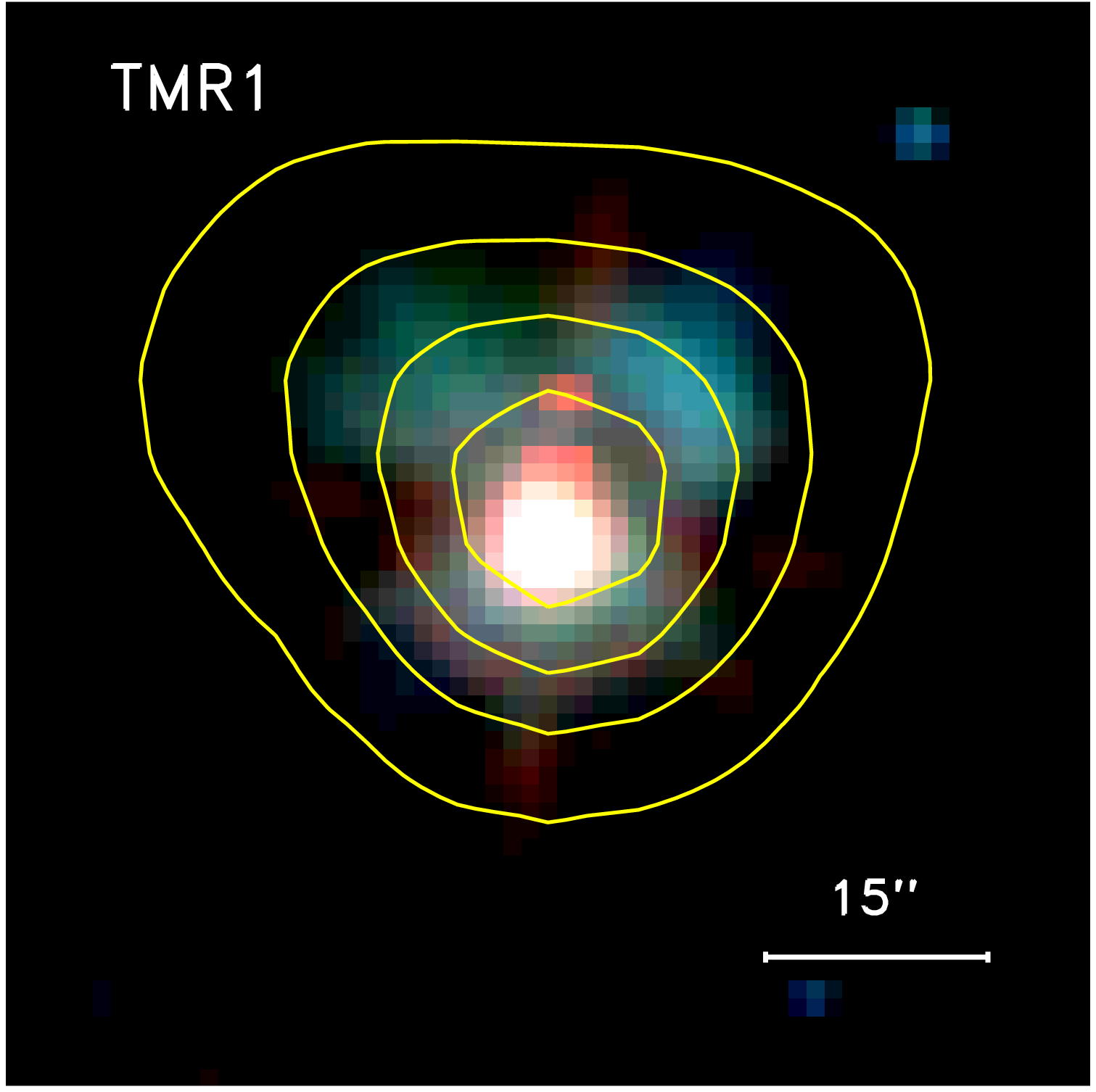}}\resizebox{0.25\hsize}{!}{\includegraphics{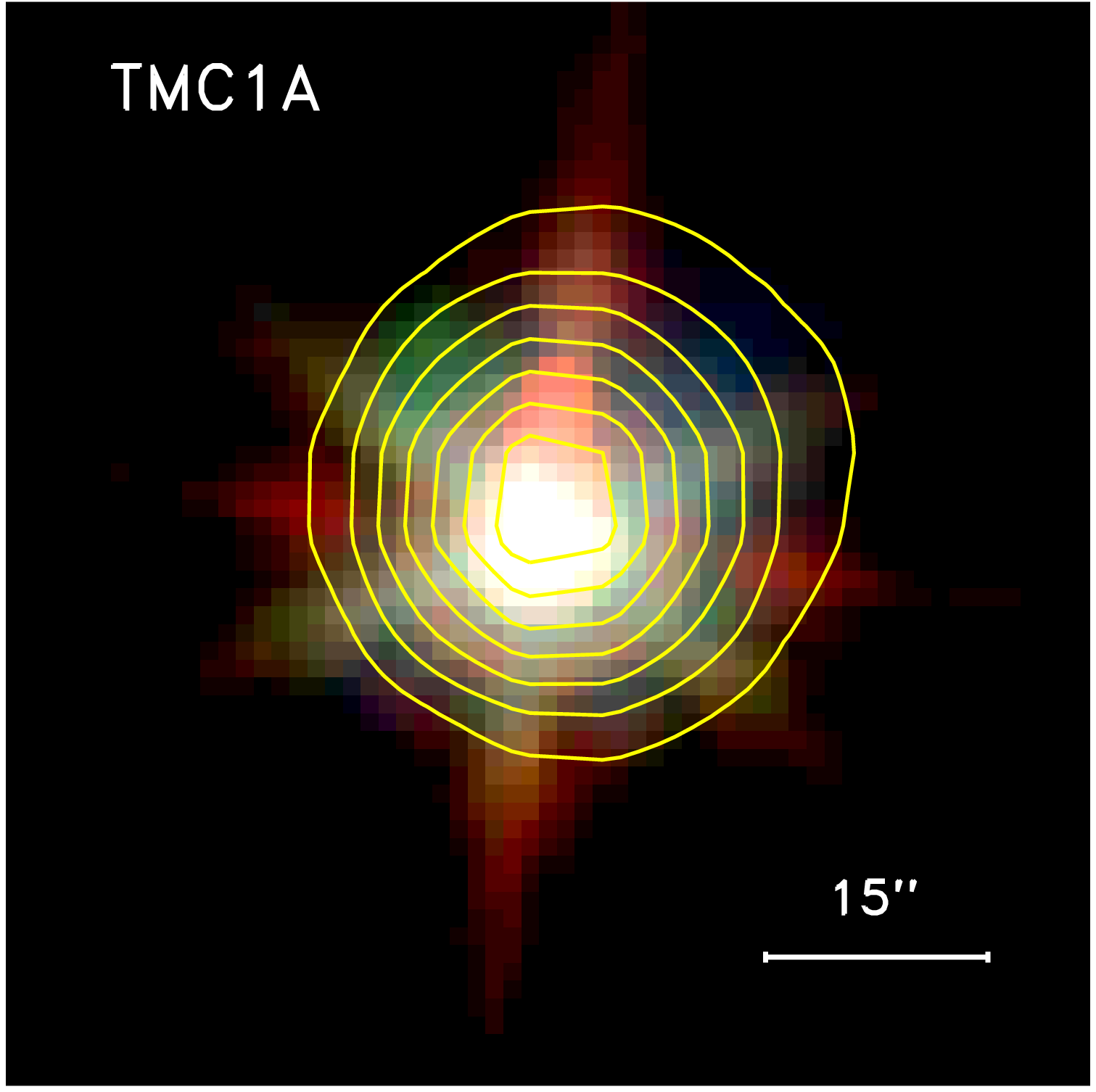}}\resizebox{0.25\hsize}{!}{\includegraphics{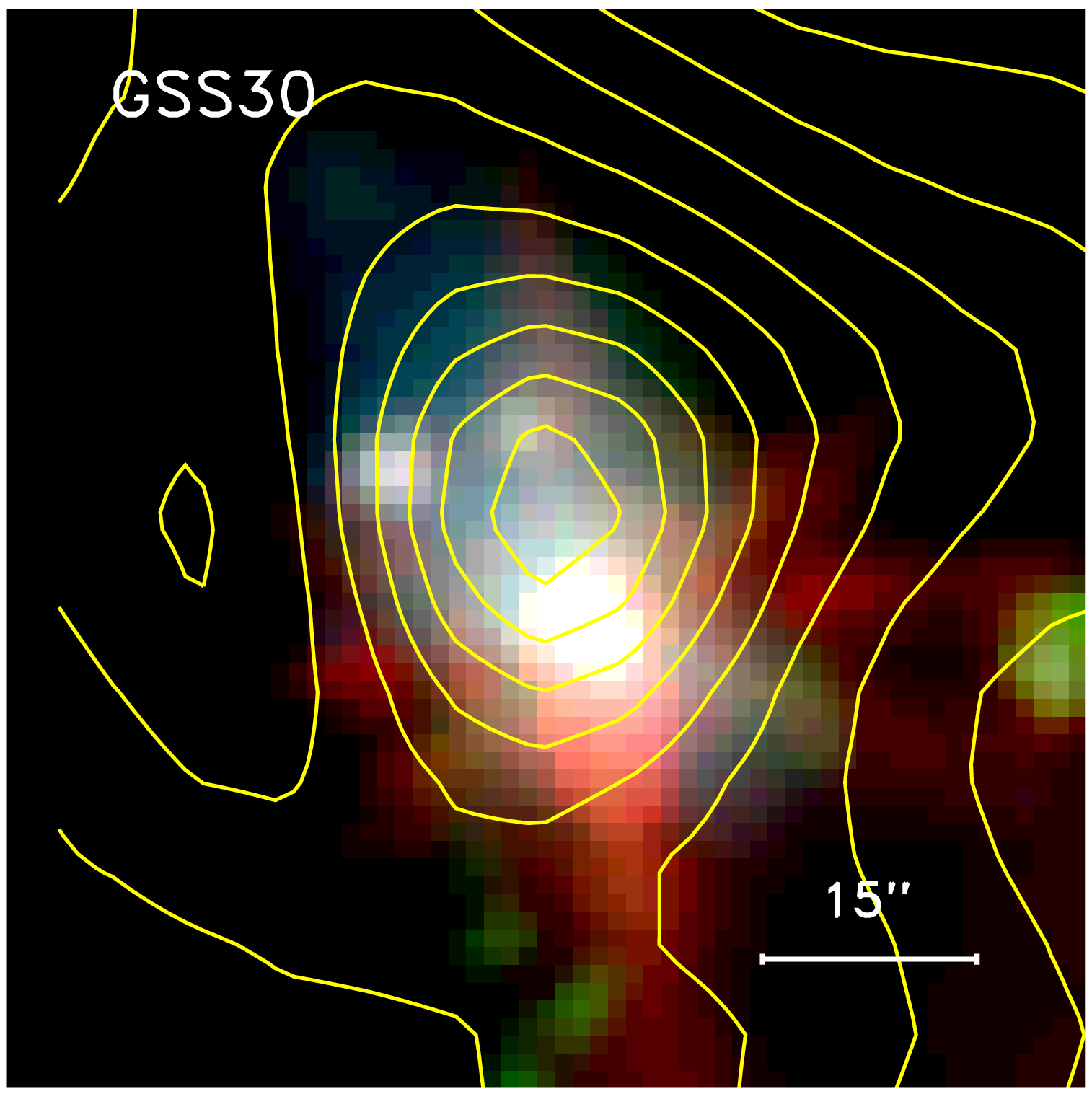}}\resizebox{0.25\hsize}{!}{\includegraphics{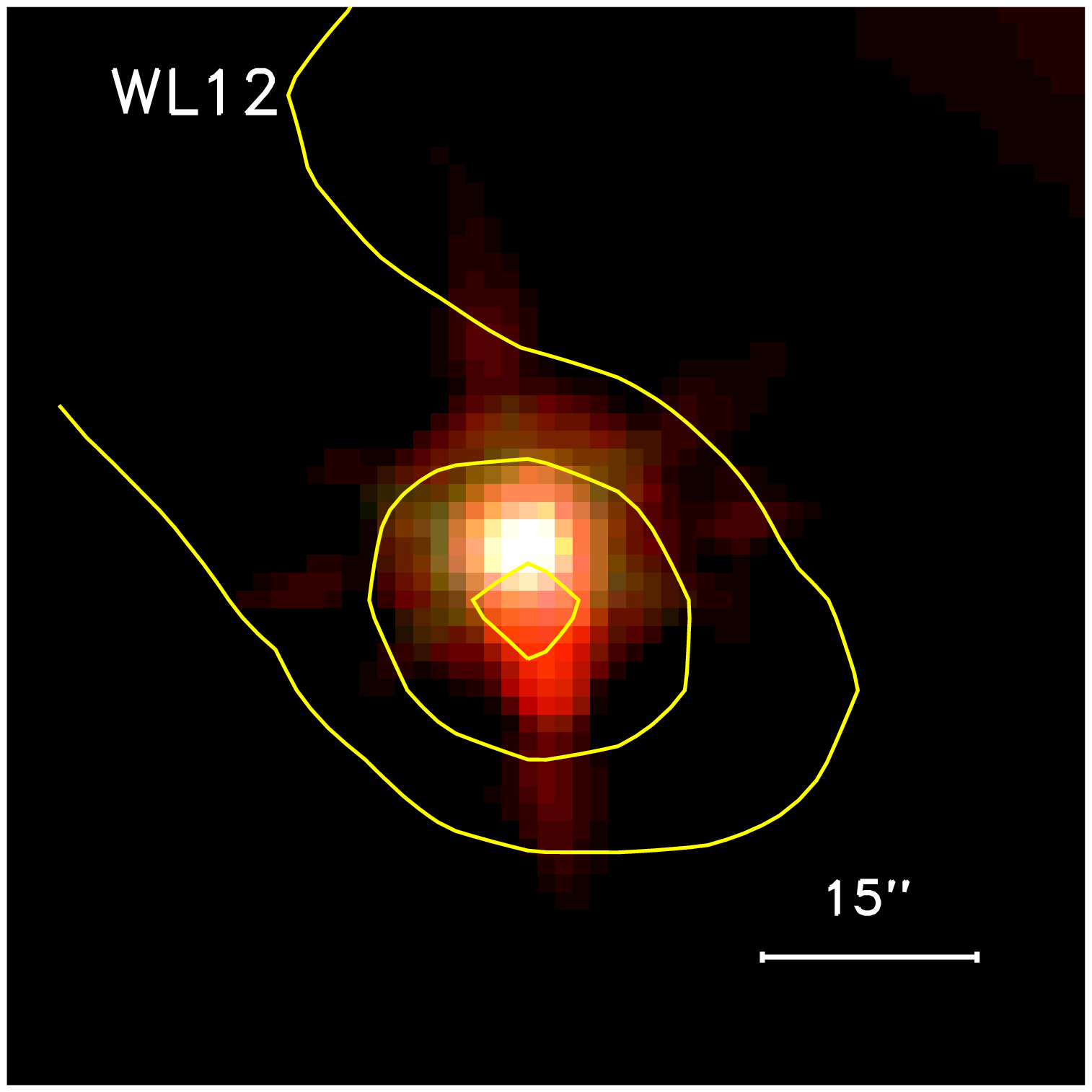}}
\resizebox{0.25\hsize}{!}{\includegraphics{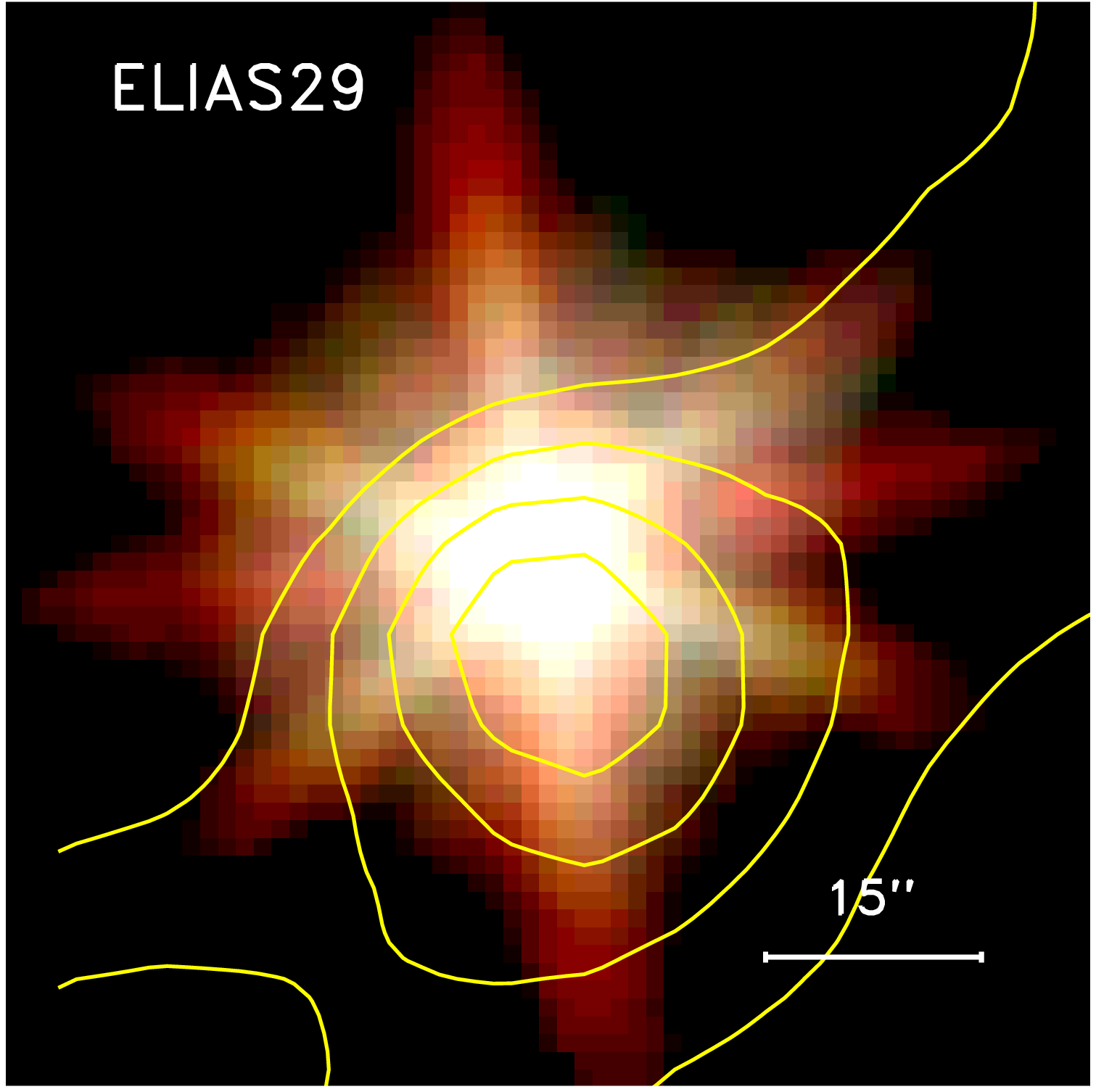}}\resizebox{0.25\hsize}{!}{\includegraphics{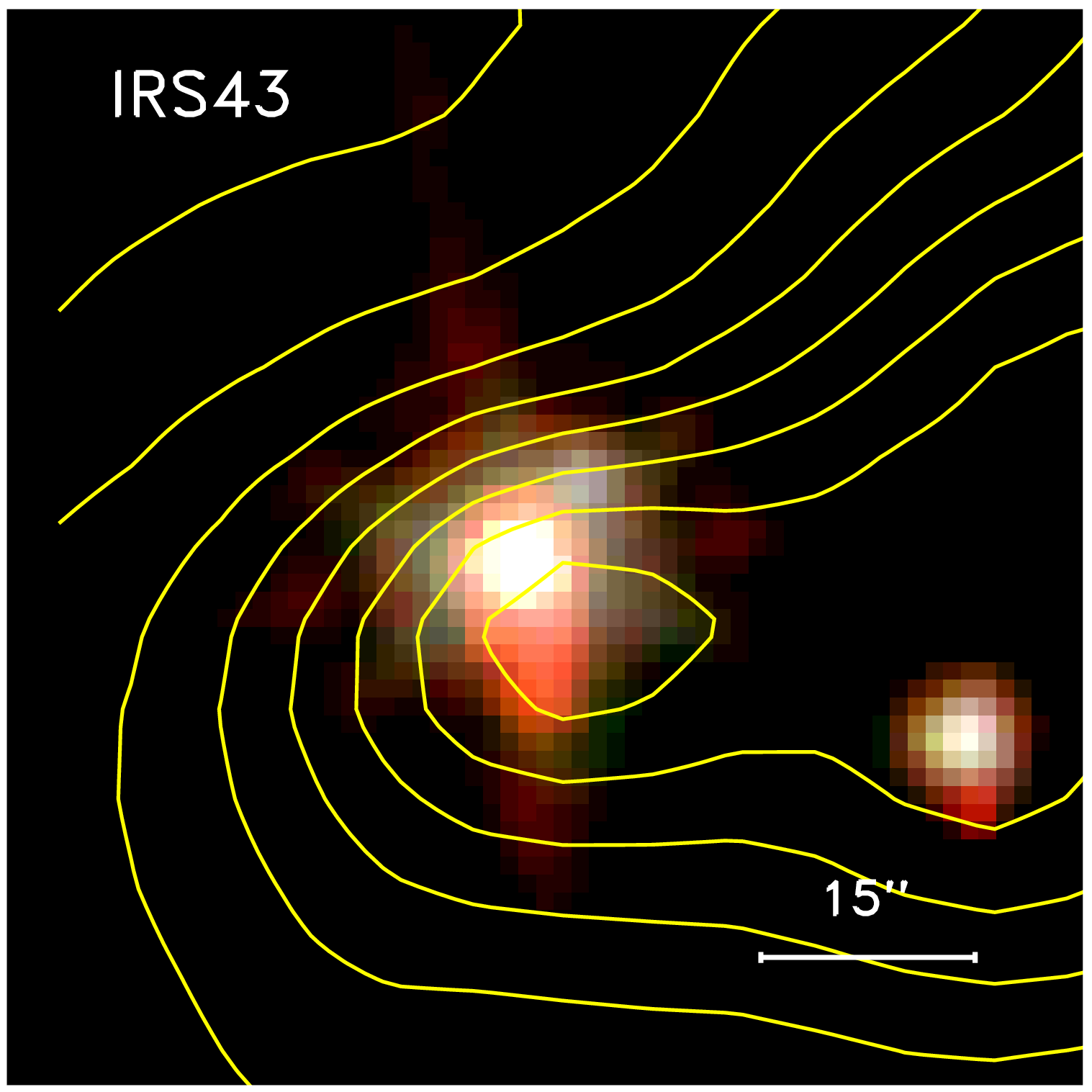}}\resizebox{0.25\hsize}{!}{\includegraphics{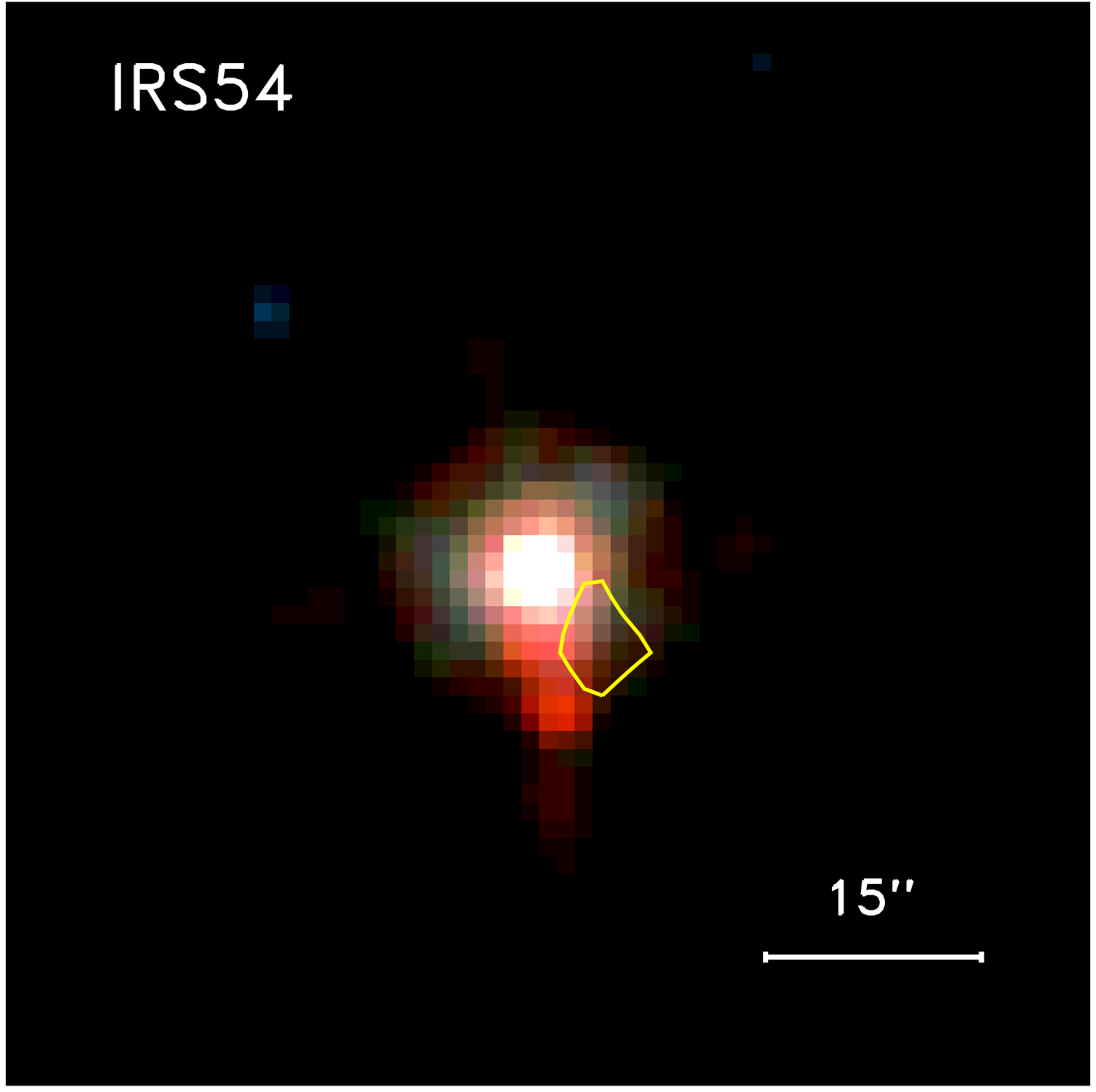}}\resizebox{0.25\hsize}{!}{\includegraphics{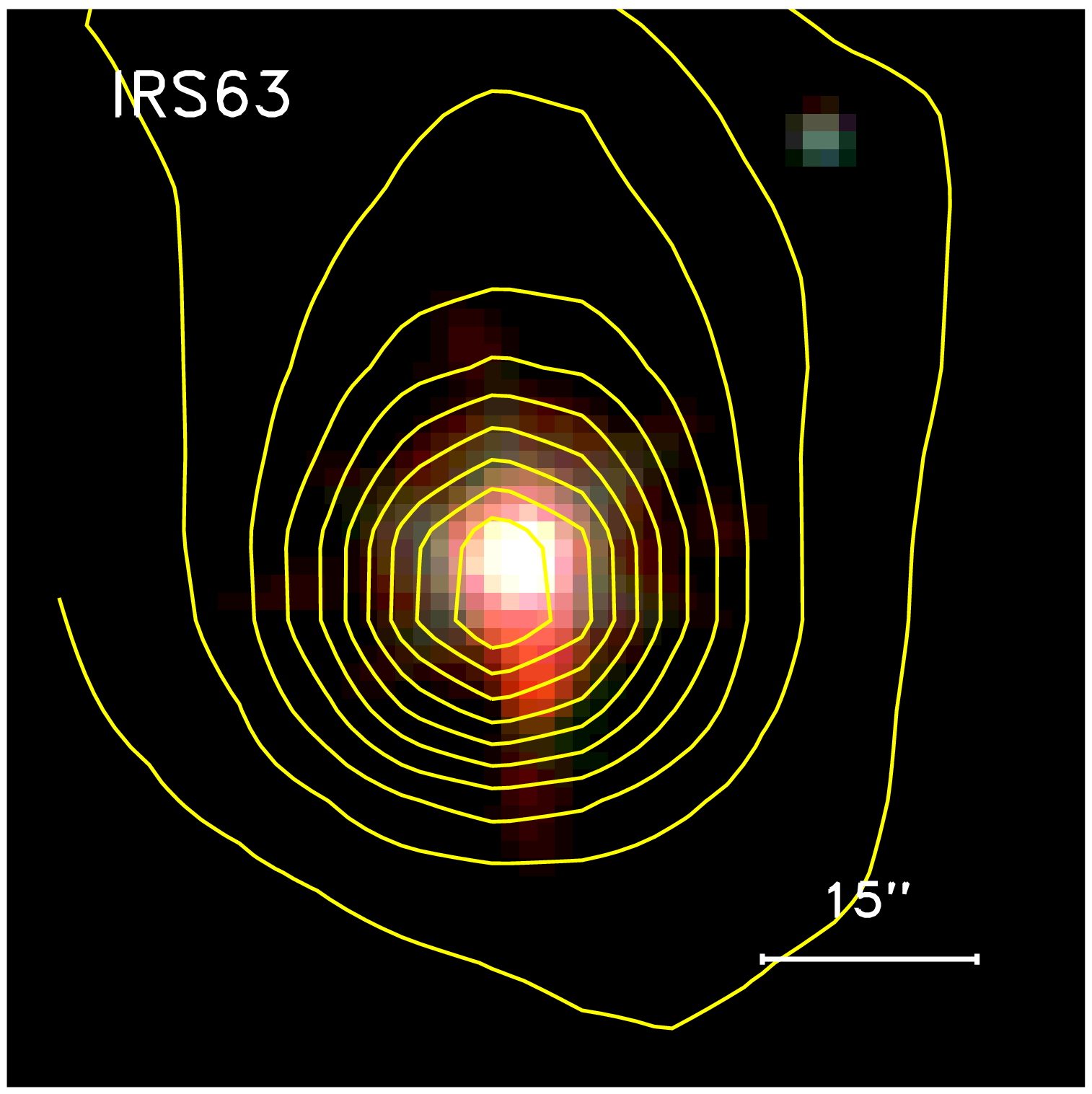}}
\caption{Spitzer Space Telescope 3.6, 4.5 and 8.0~$\mu$m images (blue, green and red) of each of our
  sources with SCUBA 850~$\mu$m emission \citep{difrancesco08} shown as
  contours. The contours are given in steps of 0.1~Jy~beam$^{-1}$ -- except for L1448, IRAS2A, and IRAS4A/B where the (dotted) contours are given in steps of 0.3~Jy~beam$^{-1}$. For further details about Spitzer observations of individual sources see: \emph{Perseus:} \cite{perspitz,scubaspitz} and \cite{gutermuth08ngc1333}, \emph{Taurus:} 
  \cite{padgett08}, \emph{Ophiuchus:} \cite{evans09} and \cite{scubaspitz2}, \emph{L1527:} \cite{tobin08}, \emph{B335:} \cite{stutz08}, and \emph{L1157:} \cite{looney07}.}\label{scubamaps}
\end{figure*}

\subsection{Continuum data overview}\label{classIcont}
\begin{figure*}
\resizebox{0.5\hsize}{!}{\includegraphics{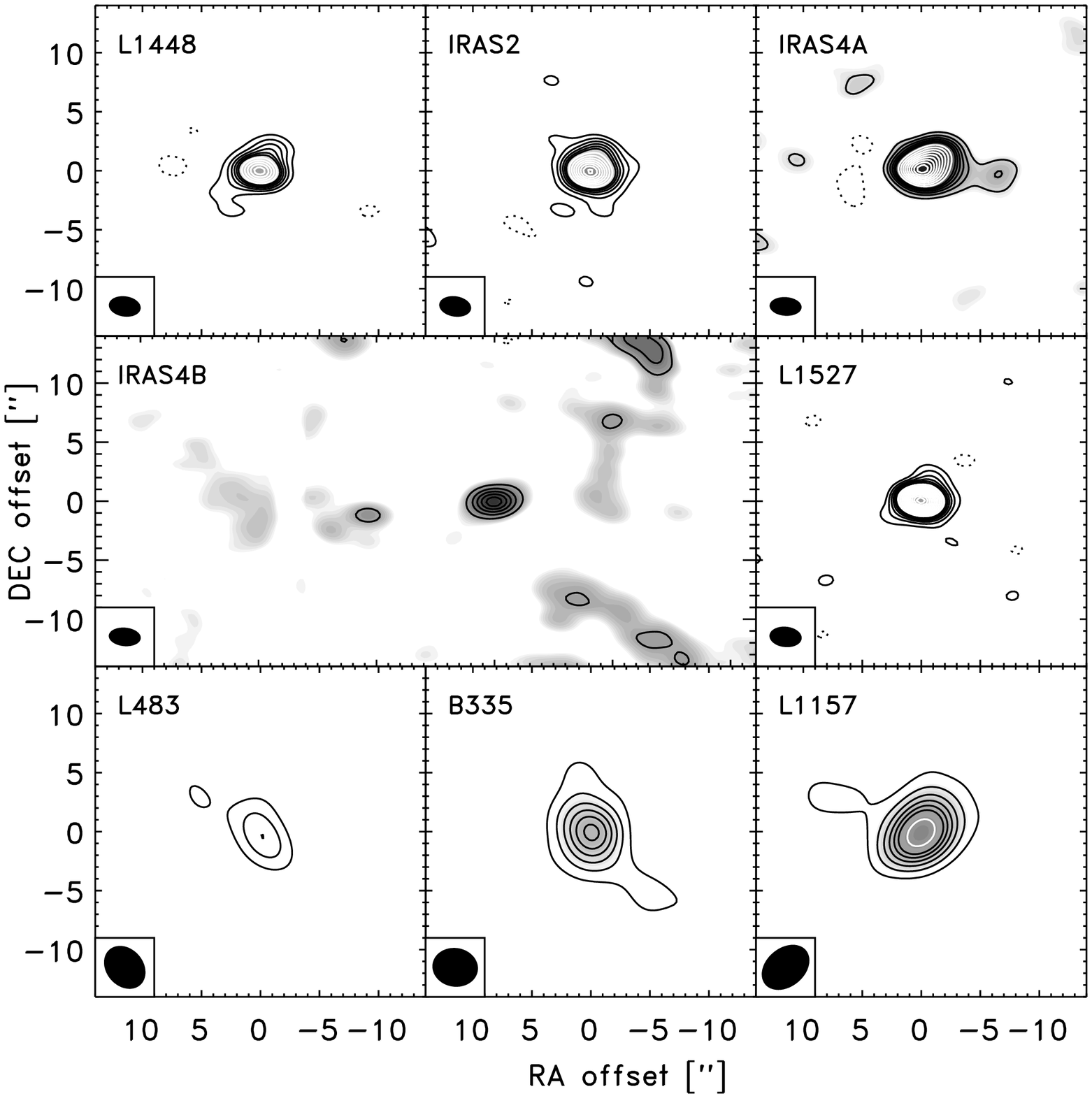}}\resizebox{0.5\hsize}{!}{\includegraphics{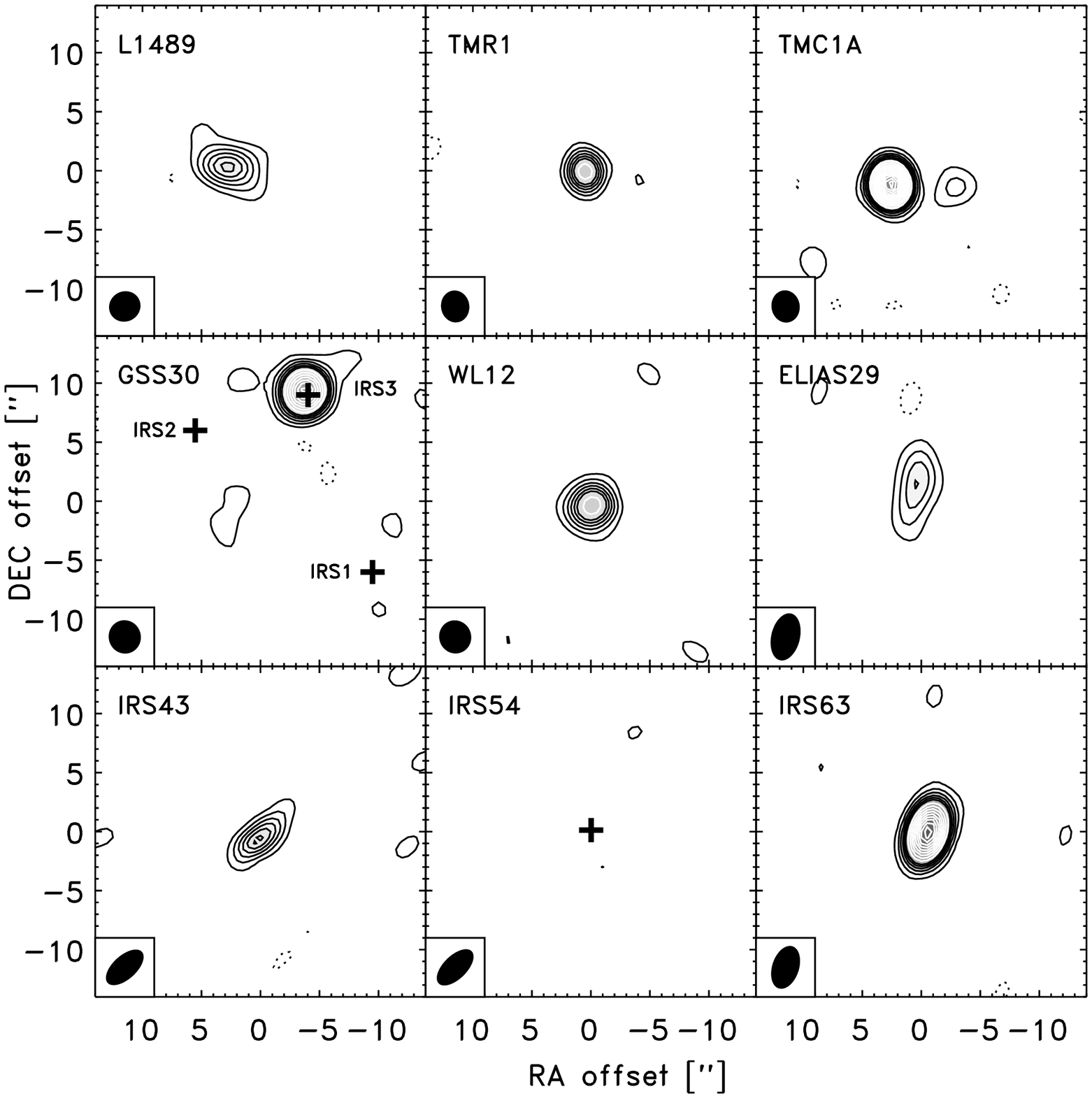}}
\caption{Overview of the 1.3~mm continuum emission from the Class~0
  sources (left group of panels; see also \citealt{prosacpaper}) and
  1.1~mm continuum emission from the sample of Class I sources (right
  group of panels; this paper). Contours are given from 3$\sigma$ to
  18$\sigma$ in steps of 3$\sigma$ (black contours) and onwards from
  there in steps of 6$\sigma$ (white contours). In the panels for
  IRS~54 and GSS30 the Spitzer positions for the known YSOs are
  indicated with the plus-signs.}\label{cont_overview}
\end{figure*}

Fig.~\ref{cont_overview} shows the SMA continuum maps of the embedded
protostars in the sample at 1.1~mm for the Class I sources as well as
the 1.3~mm continuum maps for the Class~0 sources previously presented
by \cite{prosacpaper}. All sources, except IRS~54, are detected in
continuum: one source, TMC1A, shows a fainter secondary component,
TMC1A-2 at a separation of 5.5$''$ ($\approx 770$~AU) whereas the
GSS~30-IRS1, -IRS2, -IRS3 system only shows a continuum detection for
GSS30-IRS3.

\begin{figure}
\resizebox{\hsize}{!}{\includegraphics{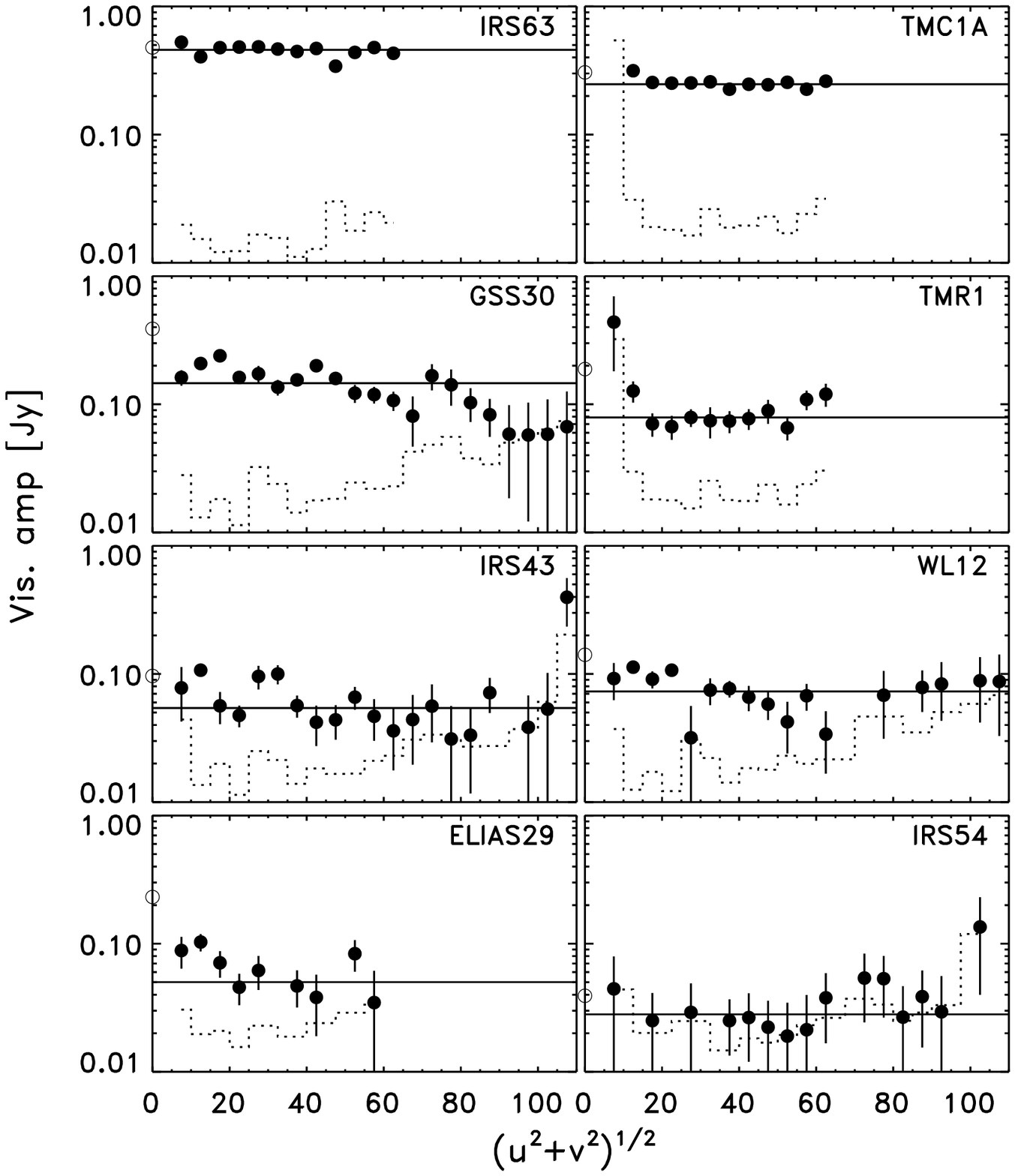}}
\caption{Plot of the continuum flux as function of projected baseline
  length. The dotted histograms indicate the zero-amplitude signal,
  i.e., the anticipated signal in the absence of source emission. The
  open circles indicate the single-dish peak flux [in Jy~beam$^{-1}$]
  from the JCMT/SCUBA maps toward each source extrapolated to
  1.1~mm.}\label{uv_continuum}
\end{figure}

Most of the observed continuum emission has its origin in the inner
few hundred AU of the embedded protostars. This is clearly
demonstrated in plots of the observed visibilities as function of
projected baseline length (Fig.~\ref{uv_continuum}). Most sources are
roughly consistent with a point, or marginally resolved, source at the
resolution of the SMA. These plots are made using circular averages in
the $(u,v)$-plane, i.e., implicitly assume that the emission is
spherically symmetric.

To estimate the structure of each source an elliptical Gaussian was
fitted to the continuum flux in the $(u,v)$ plane for baselines longer
than 20~k$\lambda$, where the contribution from the larger scale
collapsing envelope becomes negligible and the plots of visibility
amplitude vs. projected baseline length flattens
(Fig.~\ref{uv_continuum}). The results are given in
Table~\ref{continuum_gauss}. The derived fluxes are slightly lower
than the single-dish measurements by \cite{andrews05,andrews07oph}
extrapolated to 1.1~mm assuming optical thin dust with $\kappa
  \propto \nu^\beta$ and $\beta=1$ -- but still more than 50\% of
the total flux for each source is recovered. This suggests that
although the envelope itself contributes on scales smaller than the
approximately 15$''$ of the JCMT beam, the flux from the central
compact component dominates. The sources that show the biggest
discrepancies between the single-dish fluxes and interferometric
fluxes (IRS~63, WL~12, GSS30 and Elias 29) are those that show clear
evidence of extended emission in Fig.~\ref{scubamaps}. The deconvolved
sizes of the continuum sources from the Gaussian fits suggest that
they represent structures with sizes of up to 250-300~AU diameter.

\begin{table*}
  \caption{Results of elliptical Gaussian and point source fits to the continuum visibilities for the Class~I sources.}\label{continuum_gauss}
\begin{tabular}{lllllllll} \hline\hline\small
Source     & \multicolumn{2}{c}{Position} & Size$^{a}$, FWHM & \multicolumn{2}{c}{Flux (density)} \\
           & RA     & DEC      &  $\theta_{1.1{\rm~mm}}$ & $F_{\rm 1.1~mm}$$^{b}$ & $S_{\rm 850~\mu{\rm m}}$(15$''$)$^c$ \\
           & (J2000.0)    & (J2000.0)     & [$''$]   & [mJy]            & [mJy beam$^{-1}$]         \\ \hline
TMR1       &   04 39 13.91           & $+$25 53 20.63           & $\ldots$$^d$ &   79$^{d}$      & 480 \\
TMC1A      &   04 39 35.20           & $+$25 41 44.35           & 0.69$\times$0.40   & 256 (246)    & 780 \\
TMC1A-2    &   04 39 34.80           & $+$25 41 43.87           & 3.7$\times$1.1   & 48 (26)      & $\ldots$ \\
IRS~43      &   16 27 26.91           & $-$24 40 50.62           & 1.9$\times$0.13   & 75 (56)      & 990 \\
IRS~54      &   \emph{16 27 51.80}$^d$& \emph{$-$24 31 45.4}$^d$ & $\ldots$           & $\ldots$         & $\ldots$ \\
IRS~63      &   16 31 35.65           & $-$24 01 29.55           & 0.63$\times$0.0096 & 473 (459)    & 1220 \\
GSS30-IRS3 &   16 26 21.71           & $-$24 22 50.63           & 1.4$\times$1.3   & 204 (144)    & 990 \\
WL~12       &   16 26 44.19           & $-$24 34 48.74           & 1.9$\times$1.5   & 104 (64)     & 360 \\
Elias 29   &   16 27 09.44           & $-$24 37 19.99           & 5.8$\times$1.8   & 109 (48)     & 590 \\ \hline
\end{tabular}

$^{a}$Deconvolved size from Gaussian fit. $^{b}$Flux density from Gaussian fit to SMA data in the $(u,v)$-plane. Number in parentheses from
point source fit. $^c$Peak flux at 850~$\mu$m in JCMT 15$''$
single-dish beam. $^d$TMR1 unresolved in SMA observations. Only result from point source fit listed. $^e$IRS~54 undetected at the SMA; position from Spitzer Space
Telescope observations. 
\end{table*}

\subsection{Line data overview}\label{lineresults}
\begin{figure*}
\resizebox{\hsize}{!}{\includegraphics{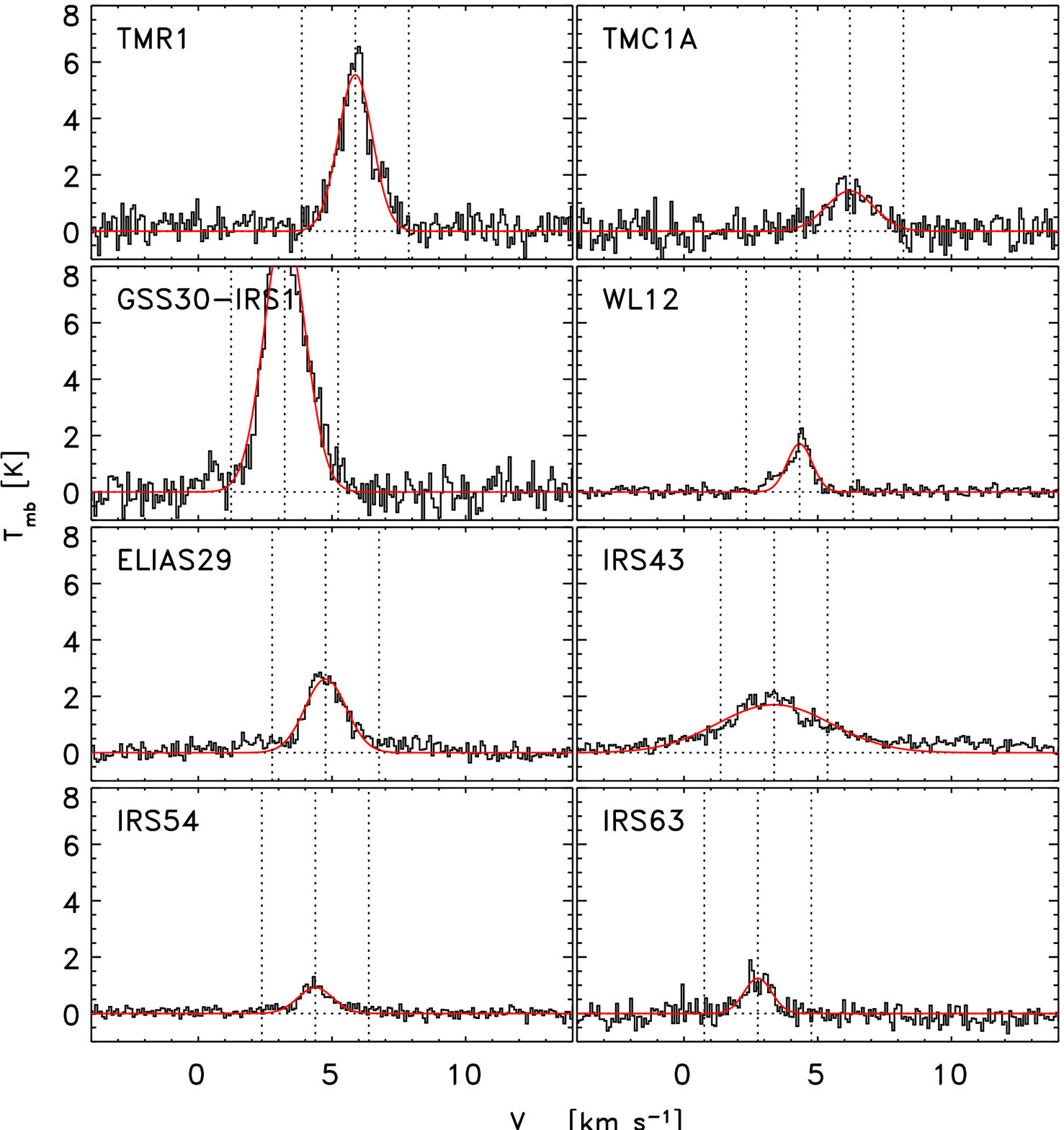}\includegraphics{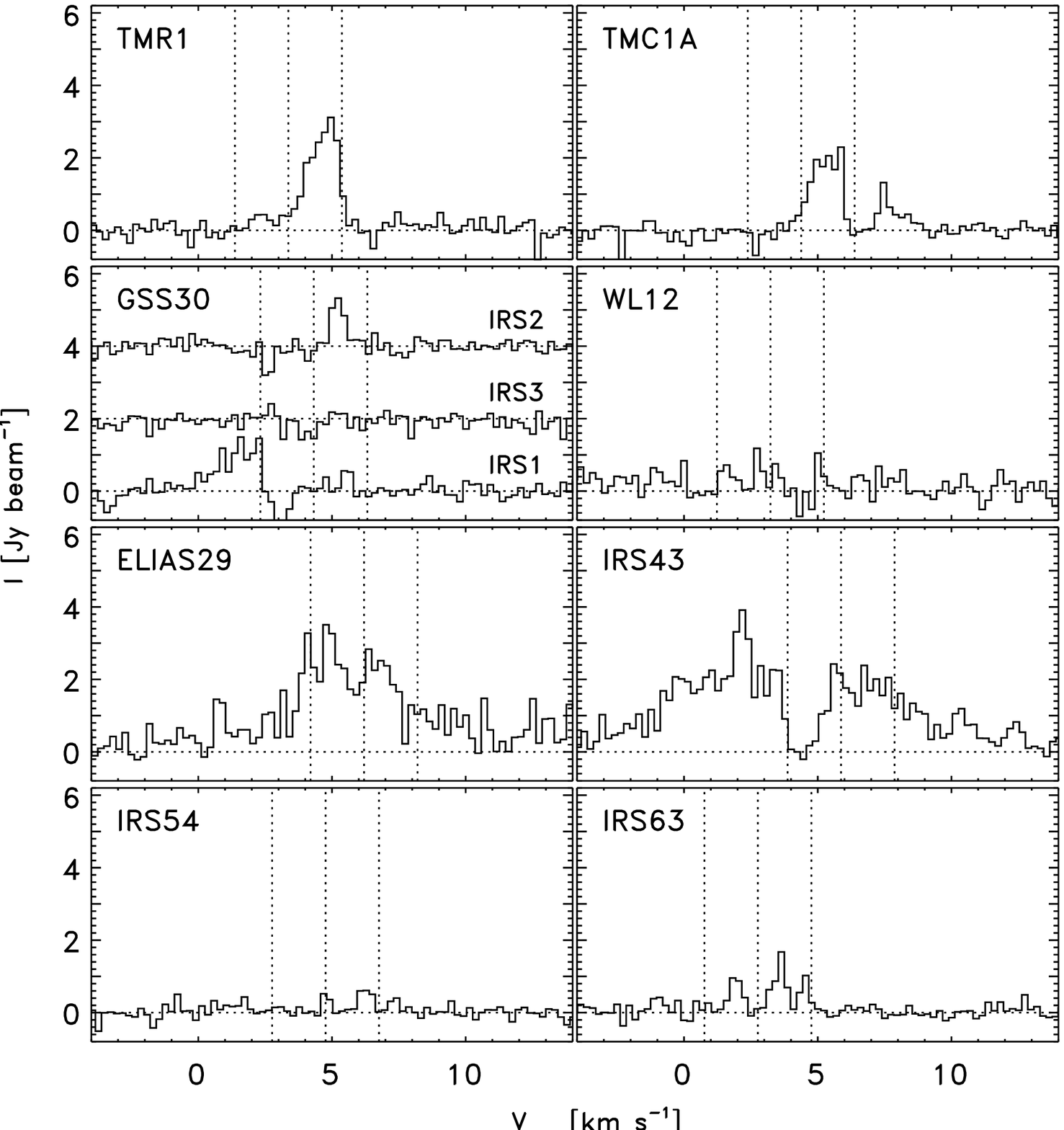}}
\caption{HCO$^+$ 3--2 single-dish spectra from the JCMT archive \emph{(left)}
  and toward the continuum positions of each source from the SMA
  observations \emph{(right)}. A Gaussian was fitted to each single-dish
  spectrum for each source and the centroid velocity as well as
  velocities $\pm 2$~\kms\ from this are shown with dotted lines in
  each panel and used to calculate moment maps in
  Fig.~\ref{hcoP_moment_overview}.}\label{hcoP_overview}\label{hcoP_sd}
\end{figure*}

\begin{figure*}
\resizebox{\hsize}{!}{\includegraphics{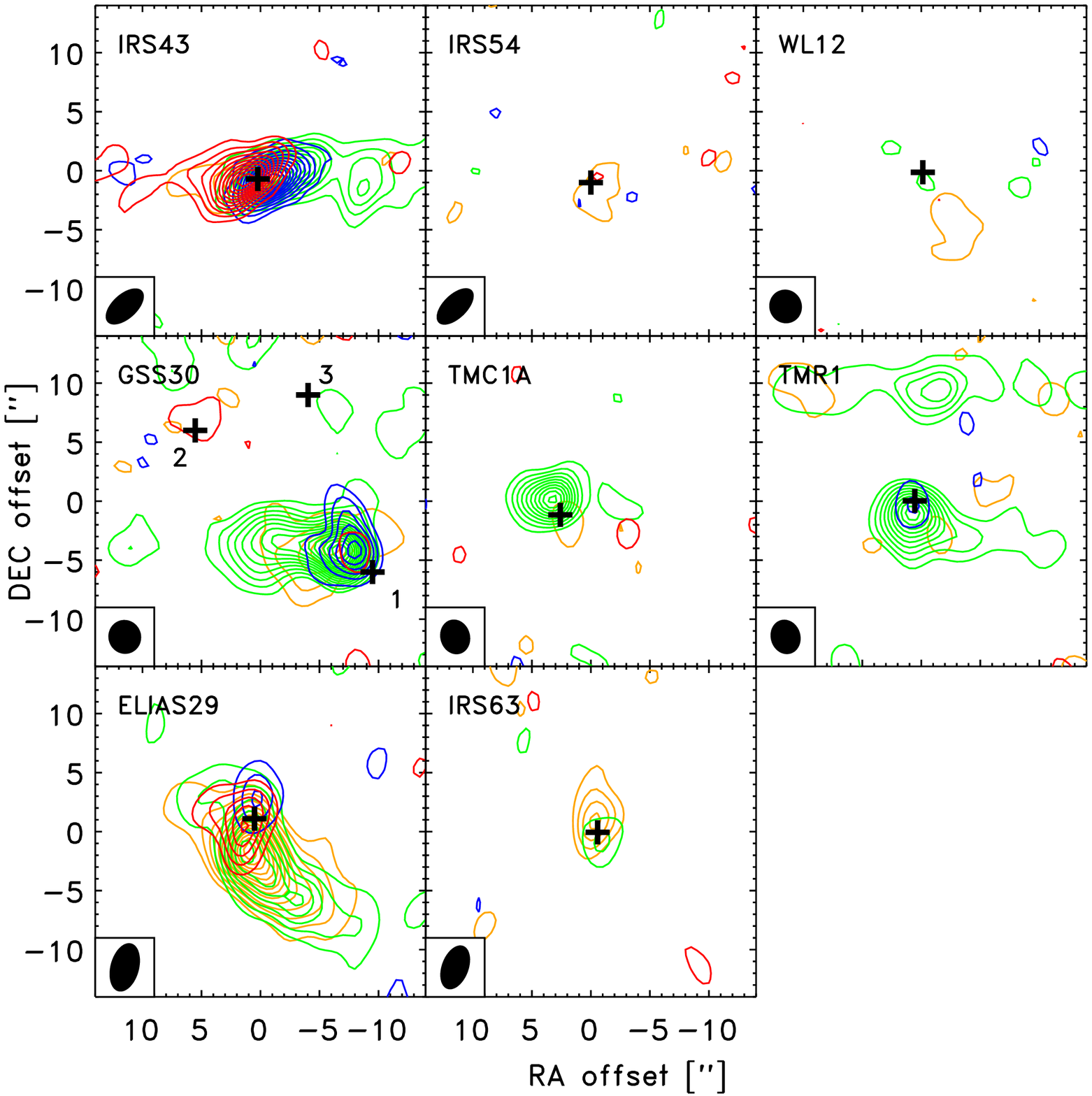}\includegraphics{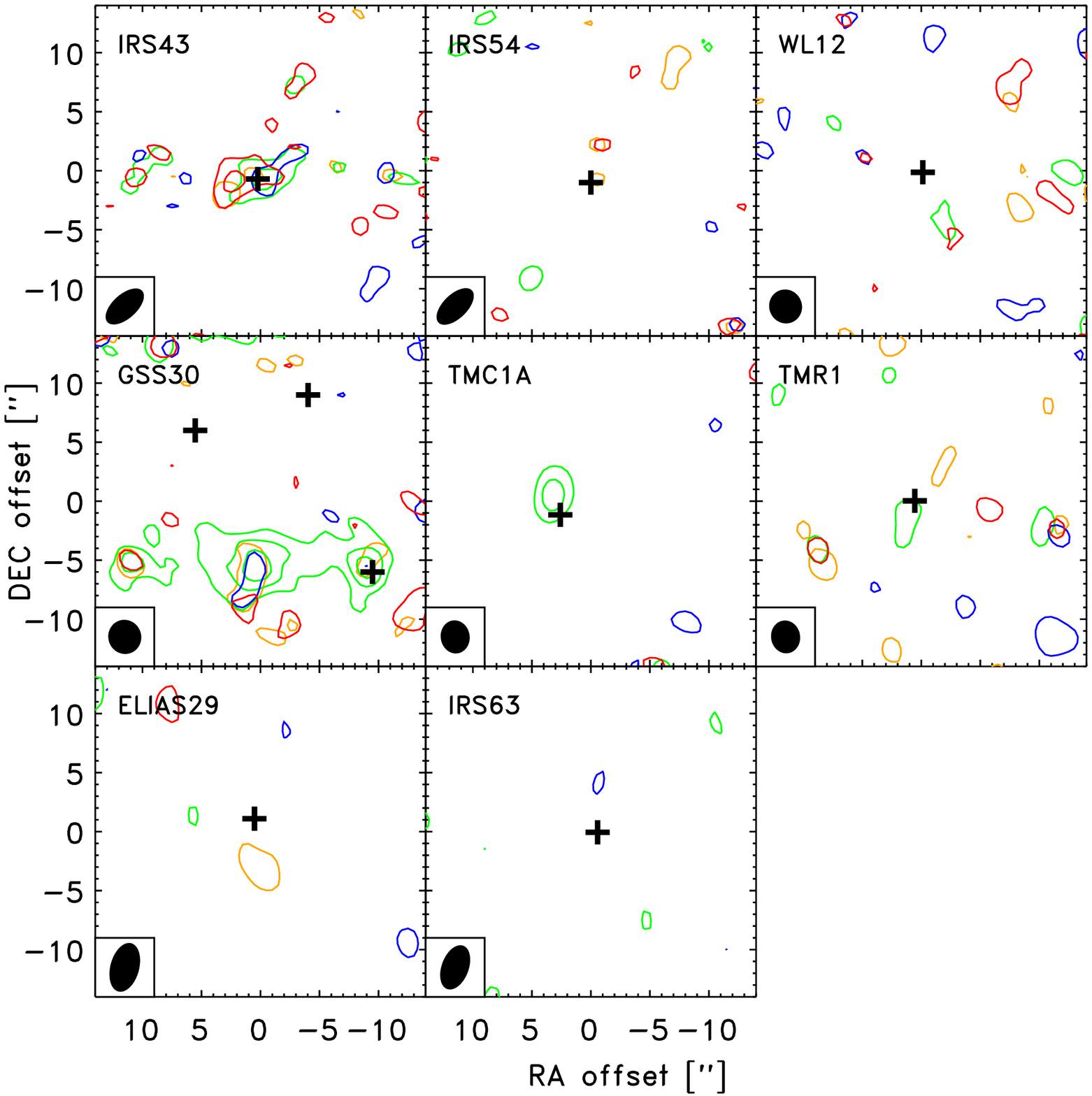}}
\caption{Overview of the HCO$^+$ 3--2 \emph{(left)} and HCN 3--2 \emph{(right)} emission from the sample of Class I
sources. The contours (shown in 3$\sigma$ intervals) indicate emission
integrated over intervals from -4 to -2~\kms\ (blue), -2 to 0~\kms\ (green), 0
to +2~\kms\ (orange) and +2 to +4~\kms\ (red) relative to the systemic
velocities from the Gaussian fits to the single-dish
spectra.}\label{hcoP_moment_overview}\label{hcn_moment_overview}
\end{figure*}

Fig.~\ref{hcoP_overview} shows an overview of the HCO$^+$ 3--2 spectra
toward the continuum positions of each source from the interferometric
observations, as well as single-dish spectra from the JCMT
archive. The HCO$^+$ 3--2 line profiles vary significantly in strength
between different sources, but also relatively between the single-dish
and interferometric data. In the single-dish data all sources show
HCO$^+$ 3--2 lines with intensities of 1.5--10~K~km~s$^{-1}$, but in
the interferometer data only 6 of the 9 Class~I sources have clearly
detected lines. IRS~43 shows the highest intensity and broadest line
profile in the HCO$^+$ SMA observations. Interestingly, the
interferometric spectra toward TMR1 and TMC1A only probe the
blue-shifted emission relative to the single-dish systemic
velocity. Fig.~\ref{hcoP_moment_overview} shows SMA maps of the
HCO$^+$ and HCN emission integrated over 4 velocity intervals around
the systemic velocity. Again, clear differences are found between the
sources. IRS~63 and IRS~43 show red- and blue-shifted emission around
the continuum position whereas TMR1, TMC1A and GSS30-IRS1 display
emission extended from one side relative to the source
position. Toward Elias 29 a complex structure with large-scale
extended HCO$^+$ emission from the nearby ridge is seen. For that
source material with the most extreme red/blue-shifted velocities
around the continuum position trace the Keplerian rotation in the
inner envelope/disk \citep[see][]{lommen08}.

\section{Analysis of continuum data: disk vs. envelope
  masses}\label{continuumanalysis}
With large continuum surveys at hand the first step is to disentangle
the envelope and disk contributions to the dust continuum emission and
from this derive the disk and envelope masses. Ideally fully
self-consistent models can be derived for all sources taking into
account the full spectral energy distributions as well the
interferometric data \citep[e.g.,][]{iras2sma,lommen08}. However, much
information can be derived from just simple comparisons of the
single-dish and interferometric fluxes.

As discussed by, e.g., \cite{terebey93} the continuum emission from a
cloud core or circumstellar disk is well approximated by low optical
depth in which case the intensity profile is simply found by
integrating the temperature and density profile of the core or disk
along the line of sight (e.g., Eq.~1 of \citealt{terebey93}). The
complication naturally arises that both the density and temperature
vary throughout protostellar disks and envelopes. Dust radiative
transfer models calculate the dust temperature distribution
self-consistently and furthermore produce images at different
wavelengths, which can be directly compared to the observations.

In this section we use such dust radiative transfer models to derive
the masses of the envelopes and disks around the sample of Class~0 and
I protostars. We first explore one-dimensional dust radiative transfer
models of protostellar envelopes: we compare the emission from a
spherically symmetric envelope around a central source of heating
(i.e., protostar) at the scales observed with single-dish telescopes
compared to our interferometric observations (\S\ref{models:envelope})
and furthermore use the models to establish a direct relationship
between the single-dish submillimeter flux and mass of an envelope
around a given protostar as function of its luminosity and distance
(\S\ref{models:empiricalrelation}). Finally, we use these results to
separate the envelope and disk emission from the envelopes and disks
and estimate their masses for each of the objects in our sample
(\S\ref{models:separation} and \S\ref{models:results}).

\subsection{Dust radiative transfer models; how much does an envelope contribute to the compact emission?}\label{models:envelope}
First, to test the hypothesis that most of the flux in the
interferometric data at baselines of 50~k$\lambda$ is in fact from the
central compact component we calculated a set of self-consistent
models for the dust in protostellar envelopes using the
\emph{Dusty} radiative transfer code \citep{dusty}. \emph{Dusty}
solves the temperature profile for a spherically symmetric envelope
  around a central source of heating given the power-law envelope density
  profile, input spectrum of the central heating source and dust
  properties. The code furthermore calculates images at
user-specified wavelengths, which can be compared to observed
  images to iteratively place constraints on, e.g., the density
  profile. It has previously been used to constrain the physical
structures of the envelopes around a large sample of
deeply embedded protostars \citep[][]{jorgensen02}.

We fixed the effective temperature of the central black-body to 1500~K
mimicking the combined contribution from a low effective temperature
young star and a disk. Since most of this emission is reprocessed
  in the larger-scale cold envelope, the exact shape and temperature
  of the input spectrum does not affect the distribution and strength
  of the emerging submillimeter emission (see also
  \citealt{jorgensen02,iras16293letter,schoeier02,shirley02}).  We
furthermore fixed the outer envelope radius to 8000~AU (typical of
envelopes in clustered regions such as Ophiuchus and Perseus) and
adopted envelope density profiles with $n\propto r^{-p}$ with
$p=1.5$. Such density profiles are expected for free-falling
envelopes, e.g., within the collapse expansion wave in the inside-out
collapse model of \cite{shu77}.  Detailed models of envelopes of Class
0 and I sources constrained by SCUBA mapping observations find density
profiles with $p=1.5-2.0$
\citep[e.g.,][]{jorgensen02,shirley02,young03}. Those models do not
include the contributions from a central disk, which can steepen the
derived density profile index artificially by 0.2--0.5. A power-law
index of 1.5--1.8 for the envelope density profile therefore appears
to be consistent with the envelope structure at these scales. As in
previous work, we adopt the dust opacity law of \cite{ossenkopf94} for
grains with thin ice mantles, coagulated at a density of $n_{\rm H_2}
\sim 10^6\, {\rm cm}^{-3}$. This dust opacity law has $\kappa_{850
  \mu{\rm m}} = 0.0182$~cm$^{2}$~g$^{-1}$ (dust+gas) and scales
approximately with frequency as $\kappa_\nu \propto \nu^{1.7}$ at
1~mm. Since we for now just deal with the fluxes at submillimeter
wavelengths, the derived masses can be directly rescaled using a
different dust opacity law.

With these fixed parameters we then calculated a set of dust radiative
transfer models varying the inner radius, luminosity and mass of the
envelope. We tested models with inner radii, $R_i=10, 25, 50,
  100, 200$~AU, envelope masses between 0.01 and 0.5~$M_\odot$ and
  luminosities of 1 and 5~$L_\odot$. Together these parameter ranges
  cover models representative for the majority of our sources.

For each model, the radiative transfer code calculates the
  temperature profile self-consistently as well as images at the
wavelengths of our single-dish and interferometric observations
(850~$\mu$m and 1.1~mm, respectively). The images at the wavelength of
our interferometric data were subsequently multiplied by the SMA
primary beam and Fourier transformed, mimicking the interferometric
observations. From these observations we extracted the flux at
baselines of 50~k$\lambda$ (3.5$''$ diameter) as an estimate of the
contribution to the SMA 1.1~mm flux from the envelope. Likewise,
images at 850~$\mu$m were convolved with the SCUBA beam and the peak
flux in the central beam was extracted as an estimate of its
single-dish 850~$\mu$m peak flux.

\begin{figure}\centering
\resizebox{\hsize}{!}{\includegraphics{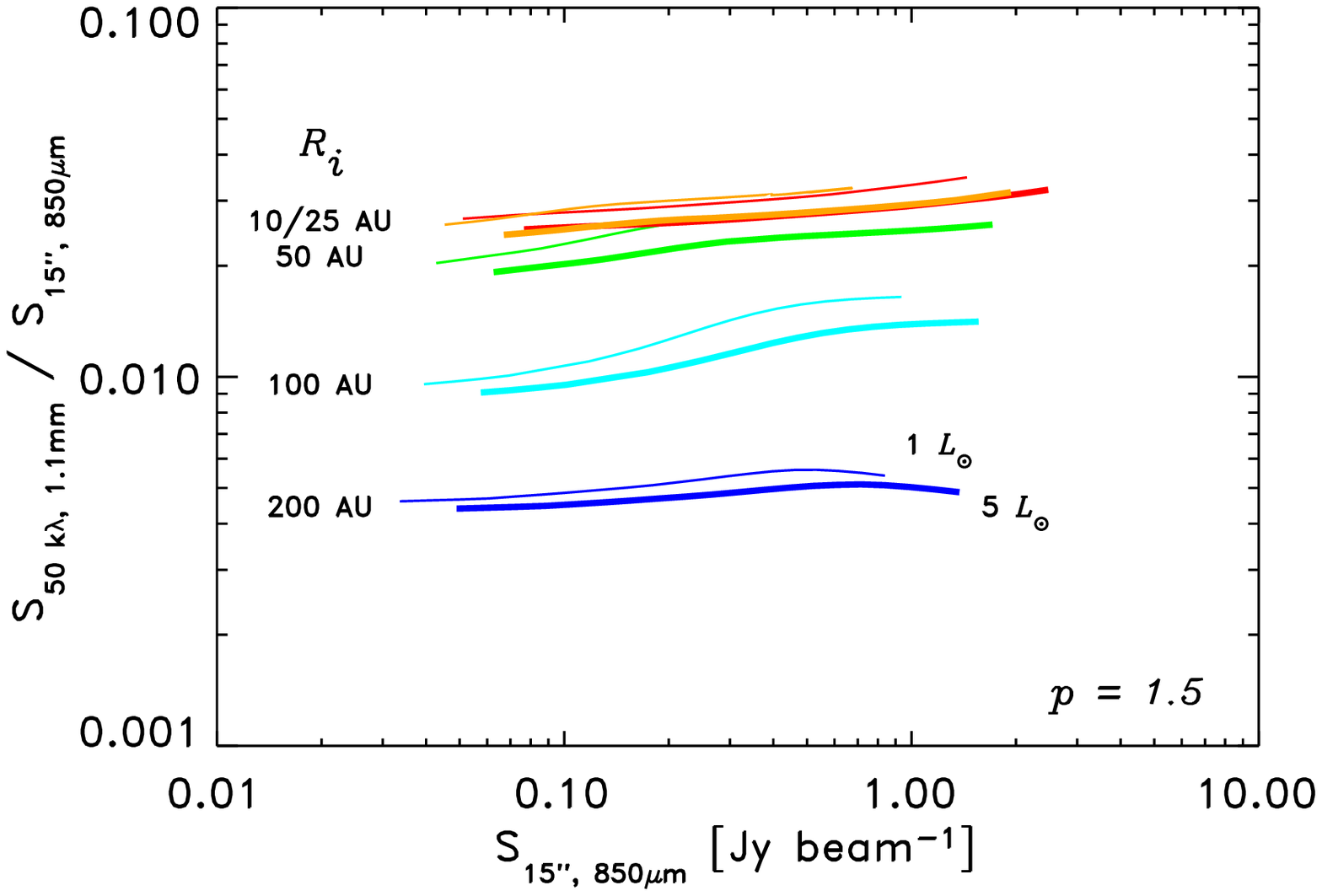}}
\resizebox{\hsize}{!}{\includegraphics{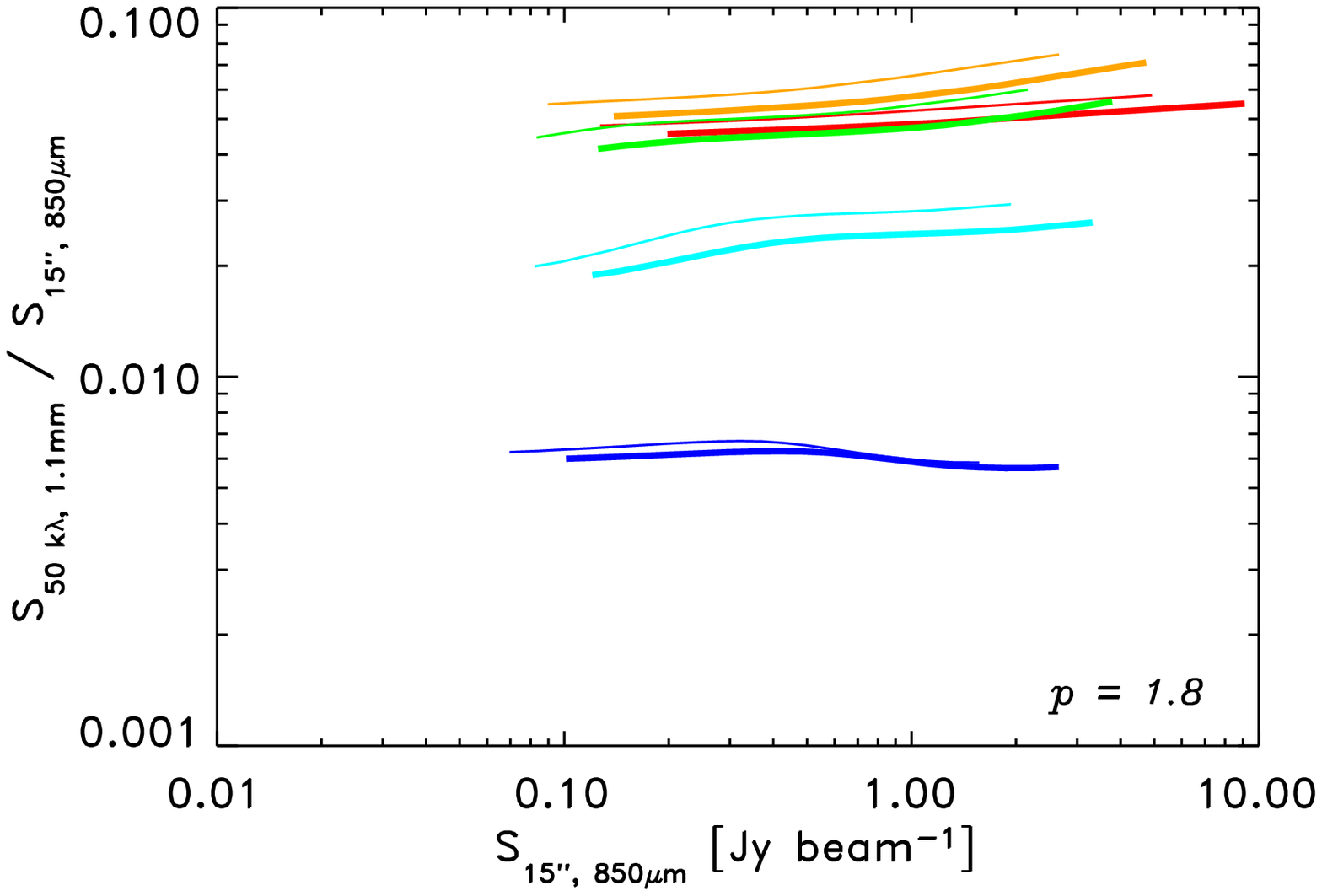}}
\caption{Ratio of interferometric fluxes (1.1~mm flux at
  50~k$\lambda$; $S_{50 k\lambda,{\rm 1.1~mm}}$) and single-dish fluxes
  (850~$\mu$m peak flux in 15$''$ SCUBA beam; $S_{15\arcsec,
    {\rm 850 \mu m}}$) as function of single-dish flux for a set of models
  varying the inner radius (10~AU, 25~AU, 50~AU, 100~AU and
    200~AU shown with red, orange, green, light blue and dark blue
    lines, respectively; see also upper panel), source luminosity (1
    and 5~$L_\odot$ shown with thin and thick lines, respectively) and
    envelope masses (along each model line). Envelopes have density
  profiles, $n\propto r^{-p}$ with $p=1.5$ \emph{(upper)} and $p=1.8$
  \emph{(lower)} and masses ranging from 0.01~$M_\odot$ at low
  $S_{15''}$ to 0.5~$M_\odot$ at high $S_{15''}$.}\label{comp_ext}
\end{figure}

Fig.~\ref{comp_ext} shows the ratio of the 1.1~mm interferometric
fluxes at baselines of 50~k$\lambda$ and single-dish peak fluxes at
850~$\mu$m as function of single-dish peak flux. As shown, the
contribution of the envelope to 1.1~mm emission at baselines of
50~k$\lambda$ is at most only 4\% of the modeled 850~$\mu$m peak
flux. If the envelope is steeper, $n \propto r^{-1.8}$, which seems to
be the steepest envelope structure supported by the modeling of the
single-dish data, the envelope contribution to the 1.1~mm emission at
baselines of 50~k$\lambda$ increases to 8\% of the peak single-dish
flux at 850~$\mu$m.

As expected models with large inner radii resolved by the
interferometer show less compact flux lowering the ratio of the
interferometer-to-single-dish flux. This increase in flux with
decreasing inner radius reaches its maximum of 4\% for an inner radius
of about 25~AU. Material at smaller inner radii is not probed by the
existing interferometric observations: this reflects the distribution
of mass in the power-law envelope ensuring that most mass is at large
scales, or that most of the dust continuum emission is due to mass
located at the typical size scale probed by the
interferometer.

\begin{figure}
\resizebox{\hsize}{!}{\includegraphics{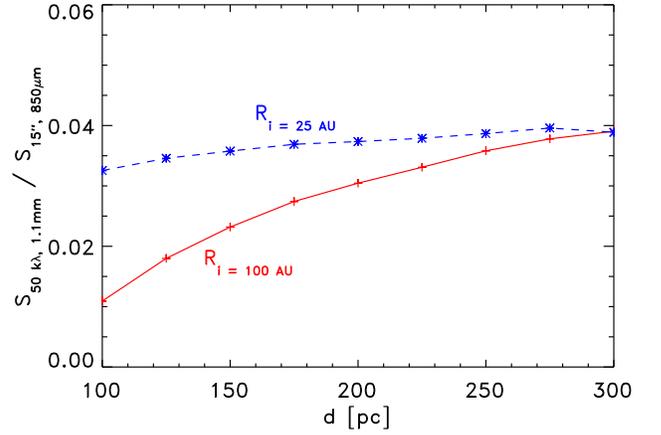}}
\caption{Ratio between the interferometer flux at 1.1~mm and
  single-dish beam peak flux at 850~$\mu$m as function of distance to
  the source for an envelope with $p=1.5$, $R_{\rm out} = 8000$~AU,
  $R_{\rm i}=25$~AU (dashed line) and 100~AU (solid line) and an
  envelope mass $\approx 0.3~M_\odot$.}\label{fluxratio_distance}
\end{figure}

Using these results we can estimate an upper limit to the envelope
contribution at 50~k$\lambda$ for the observed sources by taking 4\%
of the 850~$\mu$m JCMT peak flux and comparing that directly to the
observed 1.1~mm interferometric flux at 50~k$\lambda$: the ratio
between these two numbers then gives the relative envelope
contribution to the 1.1~mm interferometric data. This comparison
confirms that the 1.1~mm SMA measurements of the Class~I sources
(Table~\ref{continuum_gauss}) predominantly contain emission from the
central disk: in median, 87\% of the observed interferometer flux is
from the disk, as also suggested by their flat visibility amplitudes
as function of projected baseline length in
Fig.~\ref{uv_continuum}. For comparison, the Class~0 sources have a
median disk contribution of 68\% of their observed interferometer
flux. The results in Fig.~\ref{comp_ext} are shown for a distance
  of 125~pc. The ratio between the interferometer and single-dish peak
  flux measures the relative surface brightnesses on the spatial
  scales probed by each and does not change significantly when the
  distance increases -- as long as the inner envelope cavity is not
  large enough to be resolved (Fig.~\ref{fluxratio_distance}).

\subsection{Calibration of envelope mass - continuum flux relation}\label{models:empiricalrelation}
With a separation of the envelope and disk contribution to the
single-dish and interferometric observations, the next task is to
derive masses from each component using the observed fluxes. For
prestellar cores it is usually a good approximation to assume that
their dust is optically thin and isothermal with a temperature of
10--15~K and from that estimate the mass of the core from single-dish
observations of their submillimeter continuum fluxes. However, an
envelope with an internal heating source, such as a central protostar,
will have a distribution of temperatures -- which will strongly
depend on the source luminosity. 

\begin{figure}
\resizebox{\hsize}{!}{\includegraphics{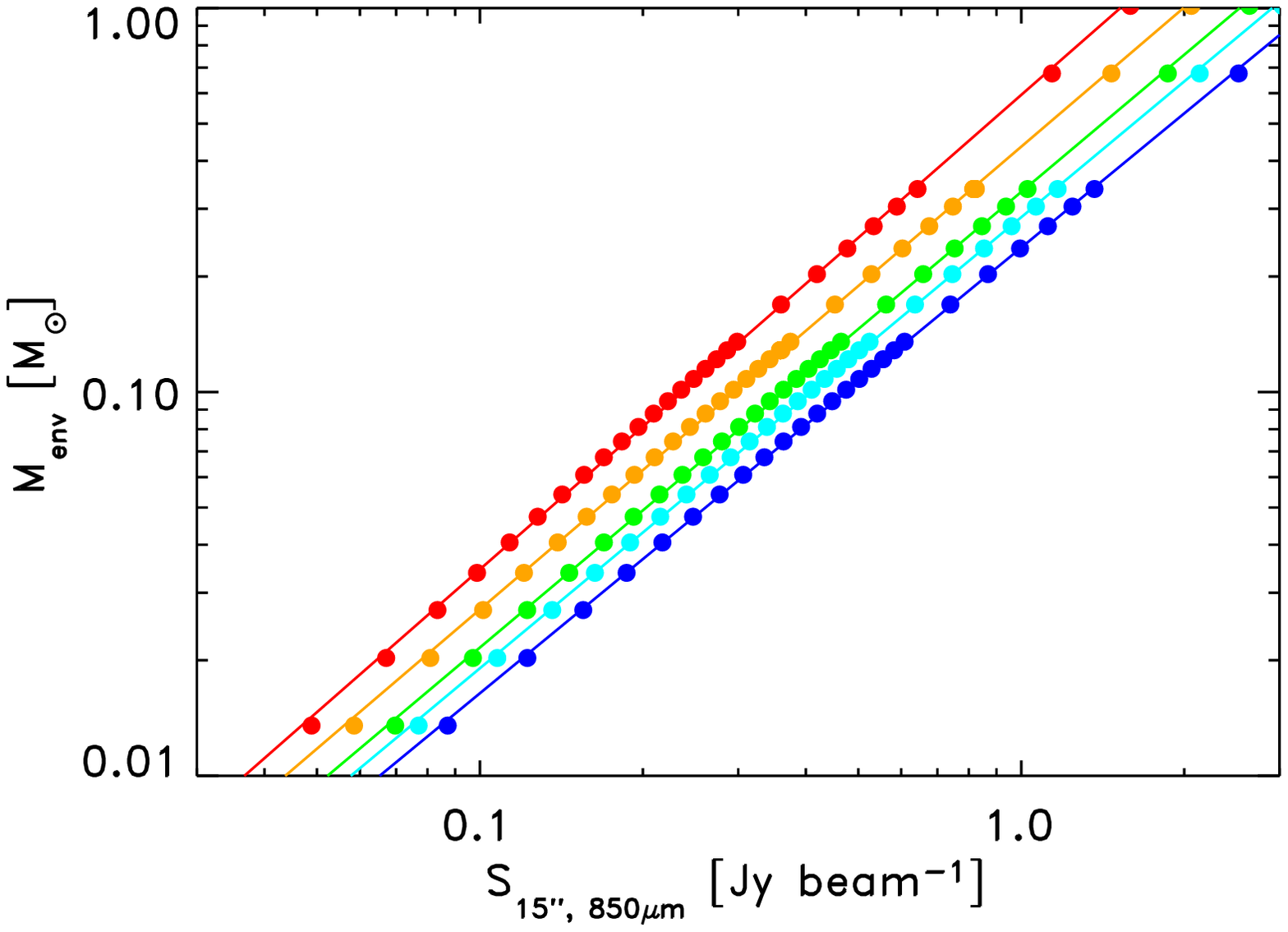}}
\resizebox{\hsize}{!}{\includegraphics{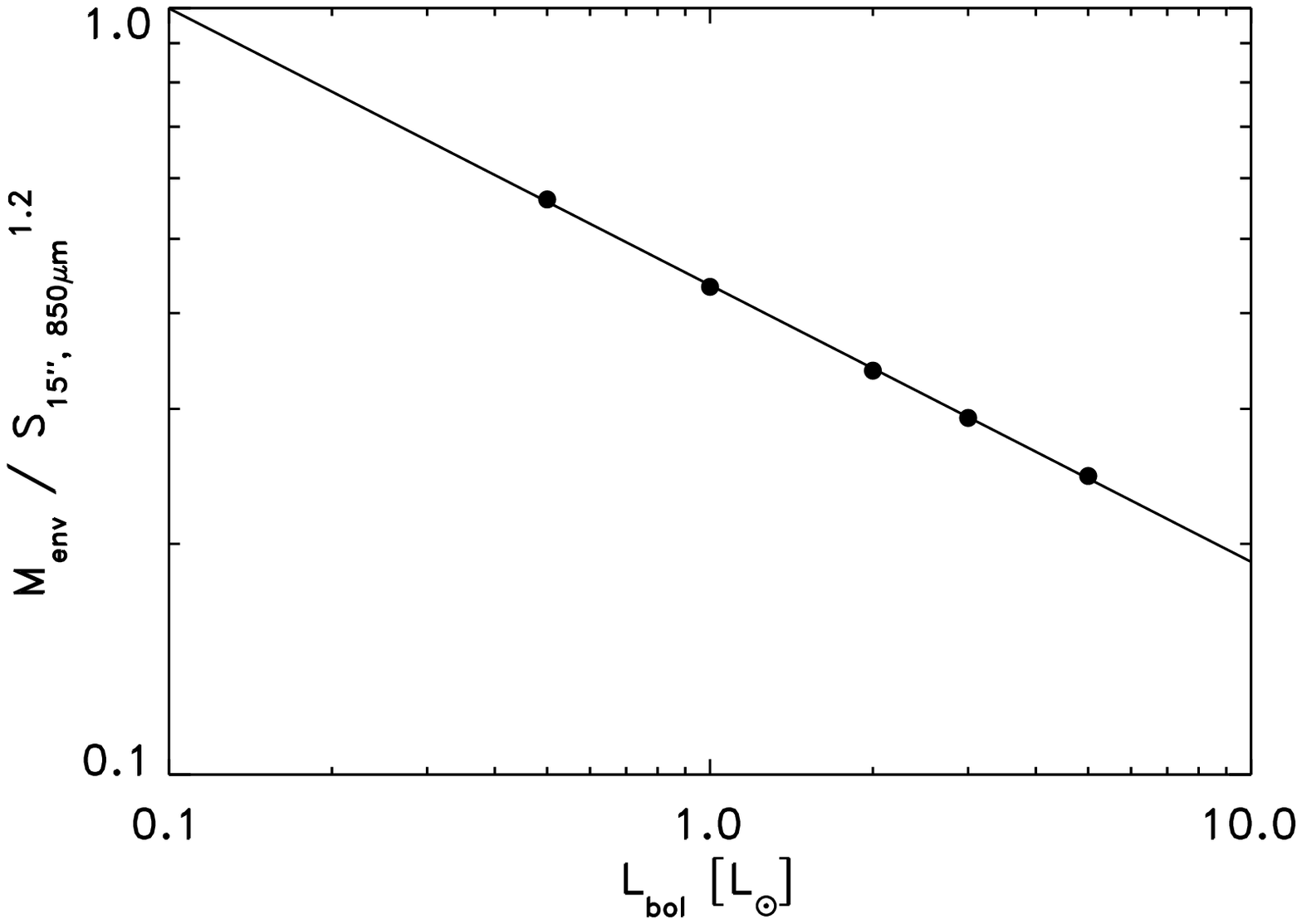}}
\caption{Results of dust radiative transfer calculations for simple
  envelope models at $d=125$~pc. \emph{Upper:} the relation between
  envelope mass and peak flux in the JCMT beam for envelope models
  with 0.5, 1, 2, 3 and 5~$L_\odot$ from left (red) to right (blue):
  each point corresponds to a model in our grid with increasing
  line-of-sight optical depths at 100~$\mu$m ranging from 0.01 to 0.20
  in steps of 0.01, from 0.20 to 0.50 in steps of 0.05 and at
  1.0. Overplotted also are power-law fits with $M_{\rm env} \propto
  S_{\rm 15\arcsec,\, 850 \mu{\rm m}}^{1.2}$. \emph{Lower:} Relation
  between the $M_{\rm env} / S_{\rm 15\arcsec,\, 850 \mu{\rm m}}^{1.2}$
  normalization factor and luminosity. Overplotted is the best fit
  $M_{\rm env}/S_{\rm 15\arcsec,\, 850 \mu{\rm m}}^{1.2}\propto L_{\rm
    bol}^{0.36}$ power-law relation between these
  quantities.}\label{dustcontfits}
\end{figure}

A way to circumvent these problems is again to use the results
  from the detailed dust radiative transfer models:
Fig.~\ref{dustcontfits} compares the relations between the envelope
mass, peak flux and luminosity for the protostellar envelopes
determined from the radiative transfer modeling.  As shown these dust
radiative transfer models give rise to simple power-law relationships
between the envelope mass and observables which can be expressed as:
\begin{equation}
M_{\rm env} = 0.44 M_\odot\, \left(\frac{L_{\rm bol}}{1\, L_\odot}\right)^{-0.36}\,\left(\frac{S_{15\arcsec,\, 850 \mu{\rm m}}}{1\, {\rm Jy}\, {\rm beam}^{-1}}\right)^{1.2}\,\left(\frac{d}{125\, {\rm pc}}\right)^{1.2}\label{envelopemass}
\end{equation}
Together with the luminosity, the above expression makes it possible
to evaluate the mass of each protostellar envelope from its peak flux
only -- once the contribution from the disk emission to the
single-dish flux has been subtracted (see also
  Sect.~\ref{models:separation}). This relation is
particularly useful in crowded regions where it is possible to
estimate the peak flux of a given source, but where it is difficult to
evaluate its outer radius and thus its extended flux due to
confusion. The luminosity can either be derived from the full spectral
energy distribution of each source or by using the empirical
correlation between source internal luminosity and 70~$\mu$m Spitzer
flux by \cite{dunham08}.

These results are generally in good agreement with the models
discussed by \cite{terebey93}: they find a relationship between the
flux and luminosity and distance which for a given envelope mass
scales roughly as $F\propto L^{0.2} d^{-0.7}$ compared to our results
which for a given envelope mass are equivalent to $F\propto L^{0.3}
d^{-1.0}$. The different slope with respect to distance is caused by
the slightly flatter density profile predicted by the \cite{terebey84}
model: for a power-law envelope similar to ours, \cite{terebey93} find
that $F\propto d^{-0.9}$. The remaining difference in the exponent of
this relation as well as the difference in the exponents for the
luminosity is caused by the envelopes being slightly optically thick
to their own radiation which increases the temperature (and thus
emission) on small scales and causes the underlying temperature
profile to depart from the single power-law applicable for an envelope
fully optically thin to its own radiation.

The derived expressions are particularly useful for interpreting large
samples of sources detected, e.g., through the large-scale
submillimeter maps provided by ongoing and upcoming surveys with
bolometer arrays such as LABOCA and SCUBA~2. One of the main goals of
those surveys are naturally to determine core masses and this is often
done using single isothermal cores with two temperatures: one for
cores with and one without embedded protostellar sources
\citep[e.g.,][]{enoch07,hatchell08imf}. However, with
Eq.~\ref{envelopemass} we can take the dependence of the source
submillimeter flux on the luminosity into account: according to
Eq.~\ref{envelopemass} a factor 2.3 decrease in mass is expected due
to the increase of the luminosity by a factor 10 for the same peak
flux -- or put in another way: if the mass for a 1~$L_\odot$ core is
derived using a temperature of 15~K, that for a 10~$L_\odot$
protostellar core should adopt a temperature of 23.5~K. Assuming a
similar temperature for all protostellar cores would therefore lead to
a mass-luminosity relation, $L \sim M^{2.8}$. This is slightly steeper
but close to the relation found by \cite{hatchell07} with $L \sim
M^{1.96}$ with an uncertainty of $\pm 0.36$ in the power-law index of
the observationally derived relation.

\subsection{Determining envelope and disk masses from continuum measurements}\label{models:separation}
\begin{figure*}
\resizebox{\hsize}{!}{\includegraphics{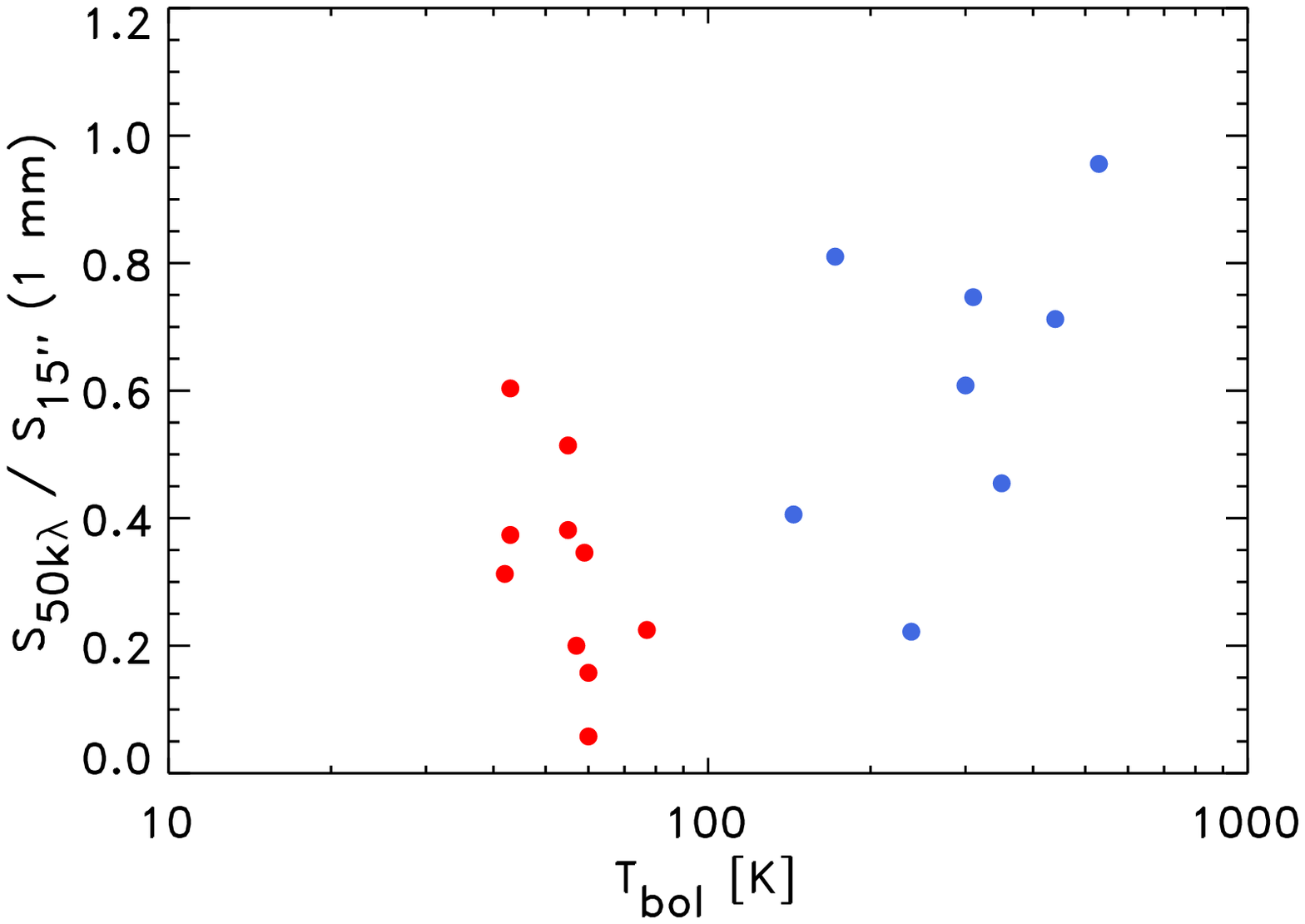}\includegraphics{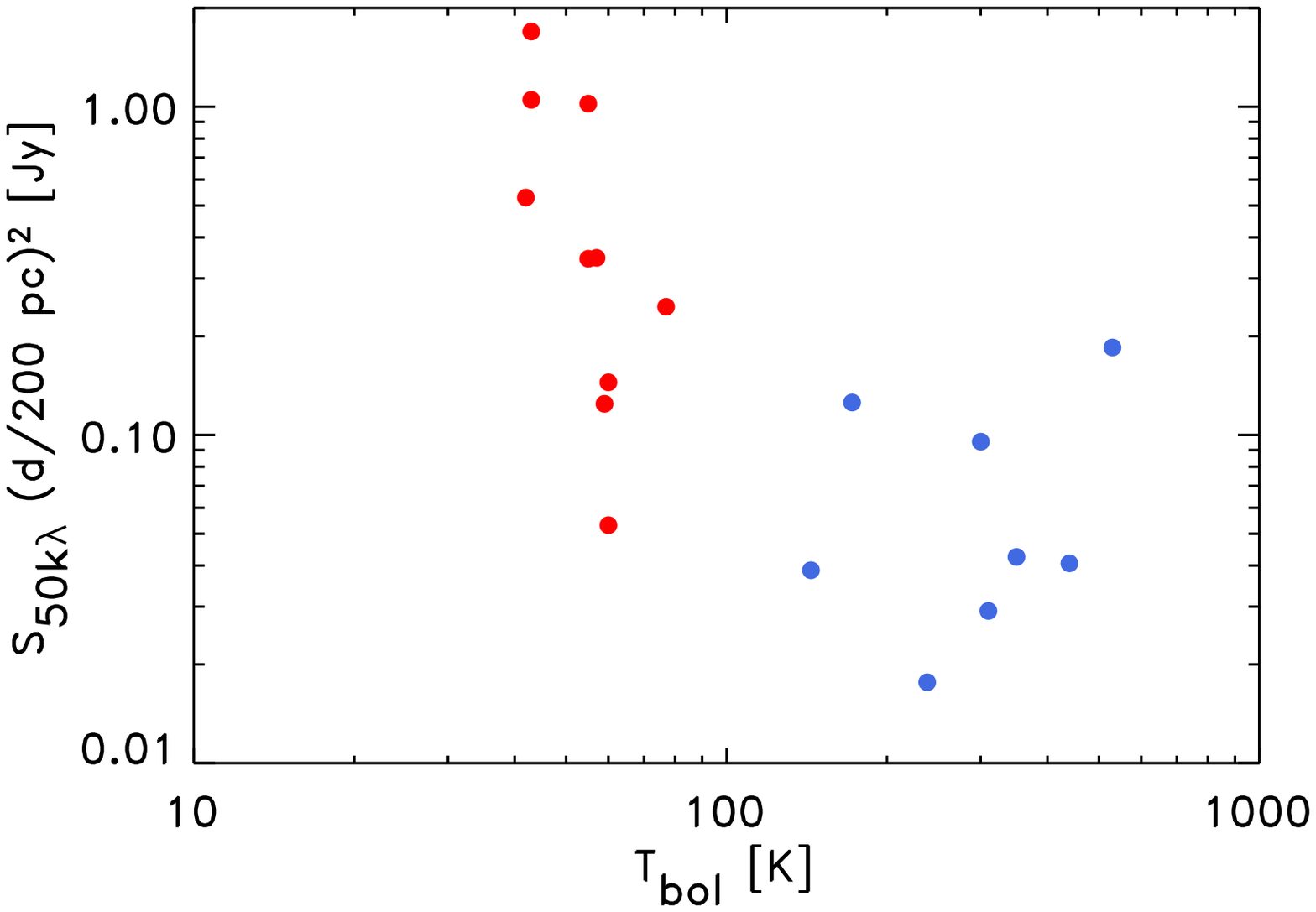}}
\resizebox{\hsize}{!}{\includegraphics{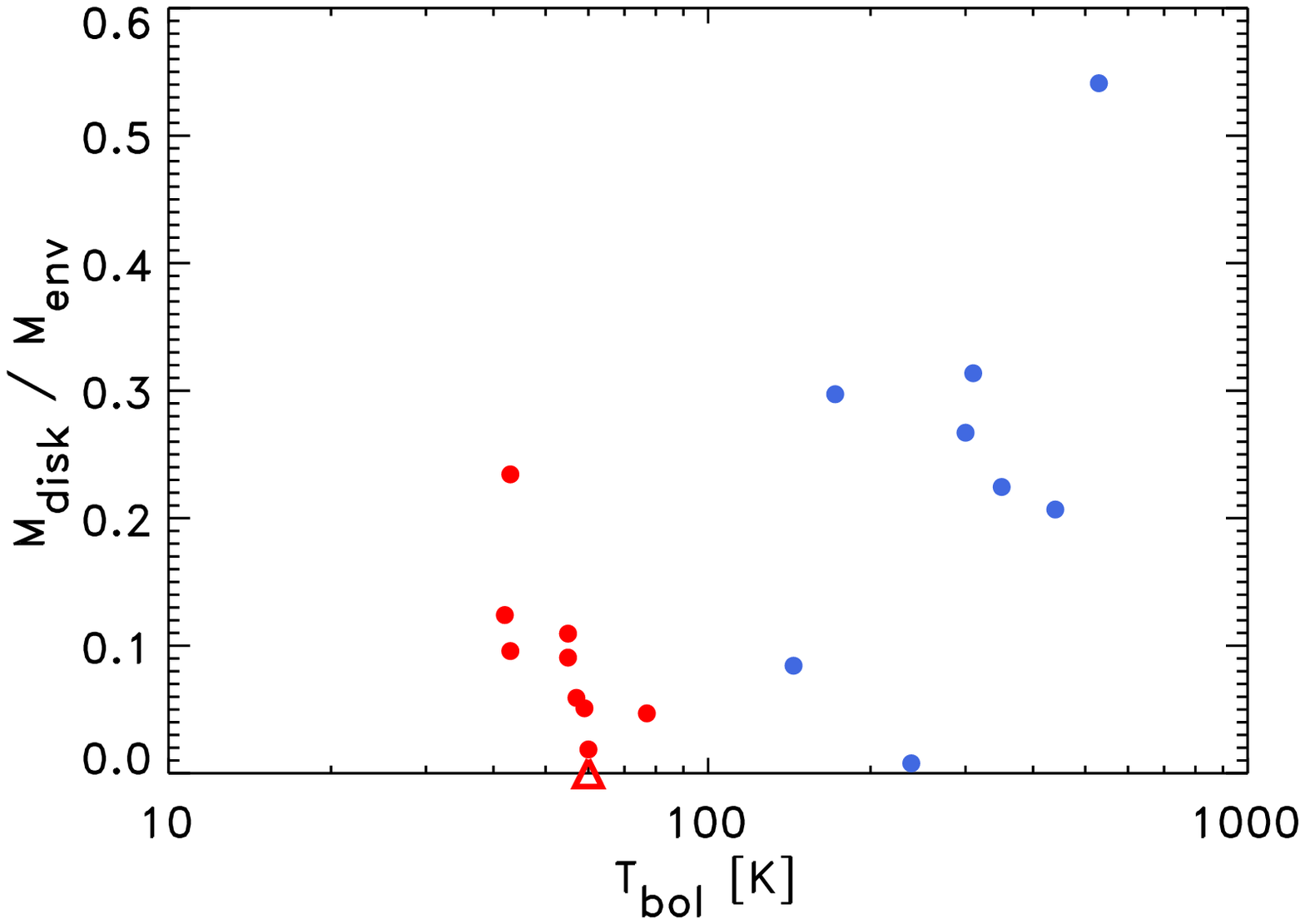}\includegraphics{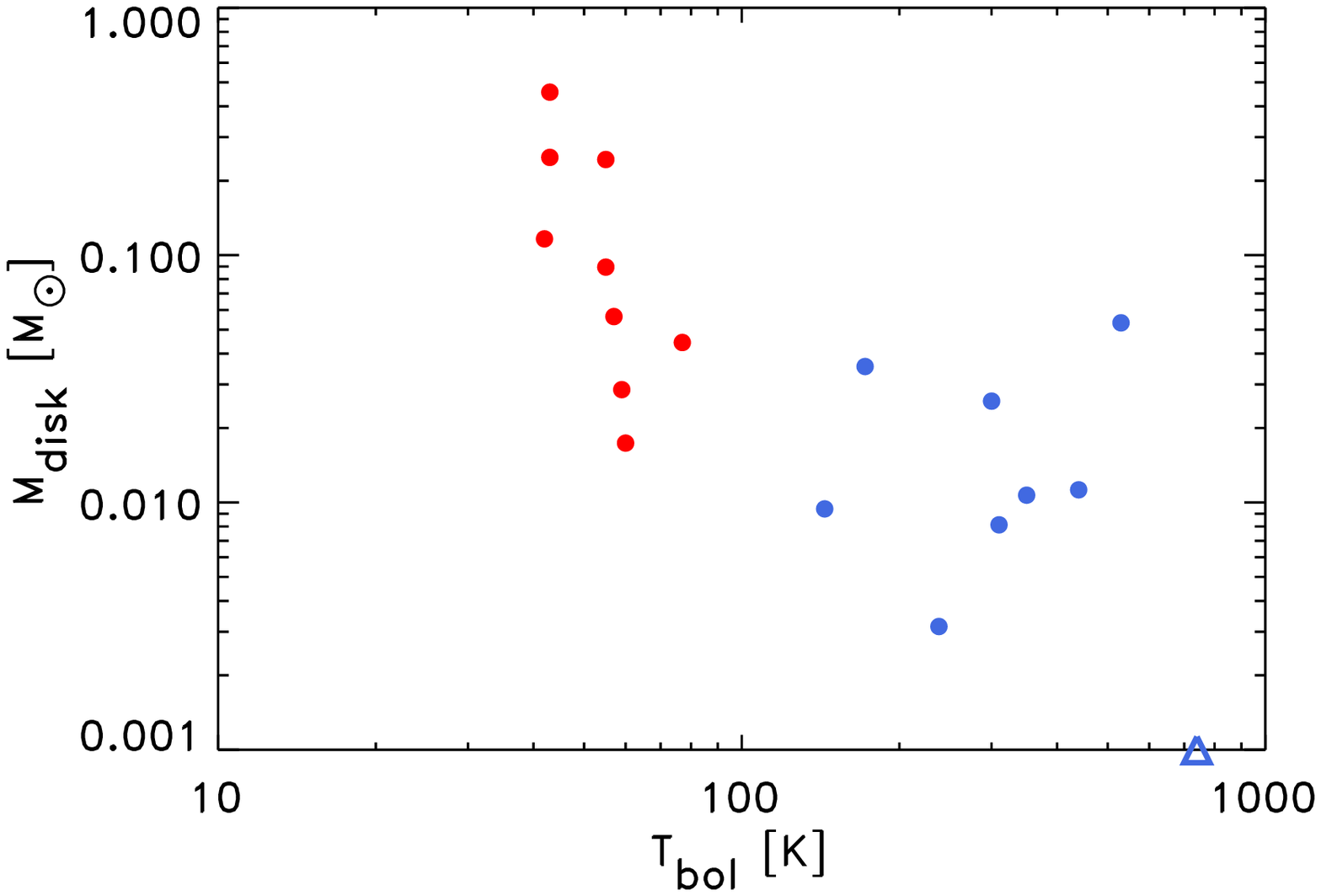}}
\caption{Single-dish and interferometer fluxes as well as disk and
  envelope masses as function of source bolometric temperatures. The upper panels show the ratios of the interferometer
    and single-dish flux \textit{(left)} and the interferometer flux
    \textit{(right)} and the lower panels the ratio of the disk and
    envelope mass \textit{(left)} and disk mass \textit{(right)}.
  Class~0 sources are shown with red symbols and Class~I with blue
  symbols. In the lower left panel, L483 (Class 0 source with
  envelope-only emission; see discussion in text) is shown with the
  red triangle. In the lower right panel, IRS~54 (Class I source with
  no detected interferometric or single dish emission) is shown with a
  3~$\sigma$ upper limit to its disk mass (0.001~$M_\odot$) from the
  SMA observations with the blue
  triangle.}\label{flux_results}\label{mass_results1}
\end{figure*}

With the resuts from the previous section we can now compare the
derived compact and single-dish fluxes across the sample. The Class~0
and I sources were not observed at exactly the same wavelengths at the
SMA. The Class~0 sources were observed at 1.3~mm (230~GHz) and 0.8~mm
(345~GHz) whereas the Class~I sources discussed here were observed at
1.1~mm (270~GHz). Class~0 and I sources have been observed at
850~$\mu$m with SCUBA on the JCMT. To take this difference into
account we interpolate the Class~0 point source fluxes estimated on
baselines longer than 40~k$\lambda$ from \cite{prosacpaper} using the
derived spectral slope of 2.5 from that paper, i.e., less steep than
for the envelope. With this correction the fluxes of each source at
similar angular scales are compared directly
(Fig.~\ref{flux_results}).

To estimate the masses of both components, a few more steps are
required: both the envelope and disk contribute to the observed
emission in both the single-dish and interferometer beams. However,
with the above results we can quantify the maximum contribution from
the envelope on the 1.1~mm interferometric flux at 50~k$\lambda$ as no
more than about 4\% of the envelope single-dish peak flux at
850~$\mu$m. Using this upper limit, we can write the two equations
with the envelope flux at 850~$\mu$m, $S_{\rm env}$, and disk flux at
1.1~mm, $S_{\rm disk}$ as two unknowns that can easily be solved for:
\begin{eqnarray}
S_{{\rm 50 k}\lambda} & = & S_{\rm disk}+c\cdot S_{{\rm env}} \label{twoeqa} \\ 
S_{15\arcsec} & = & S_{\rm disk}\,(1.1/0.85)^\alpha+S_{\rm env} \label{twoeqb}
\end{eqnarray}
In these equations $\alpha$ is the spectral index\footnote{Not to be confused with the
  infrared slope, $\alpha_{\rm IR}$, measured from the Spitzer data at
  2.2 to 24~$\mu$m.} of the dust continuum emission at
millimeter wavelengths in the disks found to be $\approx 2.5$ from the observations of the Class 0
  sources  in our sample \citep{prosacpaper}. $c$ is the fraction of the envelope
850~$\mu$m single-dish peak flux observed at 1.1~mm by the
interferometer at 50~k$\lambda$. For this discussion we adopt the
upper limit to $c=0.04$ from the above discussion for
  $p=1.5$. We note that this contribution generally is small enough
that the direct measurement by the interferometer is a good first
estimate of the actual compact flux not associated with the infalling
($n\propto r^{-1.5}$) envelope.

To derive envelope masses, a few additional effects need to be taken
into account here: as shown above the envelope masses scale with
distance as $d^{1.2}$ for a fixed peak flux, whereas the disk masses
(largely unresolved) are expected to scale as $d^2$. In addition,
since the submillimeter continuum emission measures the dust mass
weighted by temperature, the derived envelope masses are found to
depend on the luminosity (see, Eq.~\ref{envelopemass}). Taking this
into account, however, it is possible to directly estimate the masses
of the envelopes from the observed fluxes once the contribution from
the disk has been estimated through Eqs.~\ref{twoeqa} and \ref{twoeqb}
(Table~\ref{tab:massresults}). For the disk masses we assume that the
dust continuum emission is optically thin and coming from an
isothermal disk with a temperature of 30~K and a dust opacity at
1.1~mm similar to that of the dust in the protostellar envelopes. We
return a discussion of these assumptions in \S\ref{diskassumptions}
\begin{table}
\caption{Envelope and disk masses for the full sample of Class 0 and I sources.}\label{tab:massresults}
\begin{tabular}{llll}\hline\hline
Source      & $M_{\rm env}$$^{a}$ & $M_{\rm disk}$$^{b}$  & $M_{\rm star}$$^{c}$ \\
            & [$M_\odot$]   & [$M_\odot$] & [$M_\odot$] \\ \hline
\multicolumn{4}{c}{\emph{Class 0}} \\[1.0ex]
L1448-mm    & 0.96 & 0.044 & $\ldots$ \\
IRAS2A      & 1.0  & 0.056 & $\ldots$ \\
IRAS4A1     & 4.5$^{d}$  & 0.46  & $\ldots$ \\ 
IRAS4A2     & --$^{d}$   & 0.25  & $\ldots$ \\
IRAS4B      & 2.9$^{d}$  & 0.24  & $\ldots$ \\
IRAS4B$'$   & --$^{d}$   & 0.089 & $\ldots$ \\
L1527       & 0.56 & 0.029 & $\ldots$ \\
L483        & 0.82 & 0.0$^e$   & $\ldots$ \\
B335        & 0.93 & 0.017 & $\ldots$ \\ 
L1157       & 0.94 & 0.12  & $\ldots$ \\[1.0ex]
\multicolumn{4}{c}{\emph{Class I}} \\[1.0ex]
L1489-IRS   & 0.11          & 0.018       & 1.3$^{f}$    \\
TMR1        & 0.11          & 0.0094      & $\ldots$    \\
TMC1A       & 0.12          & 0.035       & 0.35--0.7$^g$ \\
GSS30-IRS3  & 0.096         & 0.026       & $\ldots$    \\
WL~12        & 0.054         & 0.011       & $\ldots$    \\
Elias 29    & 0.047         & 0.011       & 2.5$^{h}$         \\
IRS~43       & 0.026         & 0.0081      & 1.0$^{i}$         \\
IRS~54$^j$   & $\ldots$      & $<$0.001    & $\ldots$    \\
IRS~63       & 0.098         & 0.053       & 0.36$^{h}$        \\ \hline
\end{tabular}

$^{a}$Envelope mass determined from the direct scaling of the JCMT peak flux
(minus the disk contribution). $^{b}$Disk mass assuming a uniform dust temperature of
30~K. $^{c}$Central mass from Keplerian velocity profiles (see specific references for details). $^{d}$IRAS4A1 and IRAS4A2,
respectively IRAS4B and IRAS4B$'$ have larger scale circumbinary
envelopes.  $^e$The compact flux in L483 is consistent with being solely from the
larger scale envelope. $^{f}$Stellar mass from \cite{brinch07b}, envelope and disk masses calculated under same
assumptions as for the remainder of the sources using the continuum data from \citeauthor{brinch07b}. $^g$\cite{brown99chandler}. 
$^{h}$\cite{lommen08}, $^{i}$This work. $^j$IRS~54 is undetected in both interferometric and single-dish maps. A 3$\sigma$ upper limit to its disk mass is estimated from the SMA observations.
\end{table}

\subsection{Resulting disk and envelope masses}\label{models:results}
Fig.~\ref{flux_results} compares the interferometric and single-dish
fluxes scaled to 1.1~mm as well as disk and envelope masses for all
the sources in our sample. As expected the deeply embedded, Class~0,
protostars have larger fluxes in the single-dish relative to the
smaller interferometric beam, reflected in a lower $S_{\rm 50 {\rm
    k}\lambda}/S_{\rm 15''}$ ratio (upper left panel of
Fig.~\ref{flux_results}). A similar effect was noted by
\cite{wardthompson07spie} who compared interferometric and single-dish
fluxes at 850~$\mu$m from literature studies of a sample of 9 YSOs (6
Class~0, 2 Class~I and 1 Class II sources). Taken by itself this can
naturally reflect one or two effects: \emph{(i)} the envelope masses
for these sources are larger also resulting in their redder SEDs or
\emph{(ii)} the disk masses are smaller for the more deeply embedded
protostars (e.g., if significant disk growth occurs from the Class 0
through I stages). The interferometric fluxes do not show such an
increase going from the Class 0 to I stages, however (upper right
  panel of Fig.~\ref{flux_results}). This argues against a
significant increase in disk masses being the explanation.

The lower panels of Fig.~\ref{mass_results1} compare the
envelope and disk masses for all the sources in the Class~0 and I
samples as a function of bolometric temperature. Again, the
disk/envelope mass ratio is seen to be lower for the more deeply
embedded Class~0 protostars reflecting their higher envelope masses at
comparable disk masses as those in the Class~I stages (lower left
  panel of Fig.~\ref{mass_results1}). Typically the Class~0
protostars have disks that contain about 1--10\% of the total
disk+envelope mass whereas the disks around the Class~I sources
contain 20--60\% of the disk+envelope mass. In both stages most of the
derived disk masses are in the range of $\sim$0.01 to 0.1~$M_\odot$
with the median Class~0 disk mass of 0.089~$M_\odot$ and the median
Class~I disk mass of 0.011~$M_\odot$. The drop in envelope masses from
the Class 0 to I stages and the relatively low disk masses, suggests
that a significant fraction of the mass of the central star is
attained already during the most deeply embedded stages.

The results above do not show evidence for an increasing disk mass
going from the Class 0 sources to the Class I sources in the observed
samples. Thus, either the disks build up rapidly during the early,
deeply embedded stages or their physical properties (e.g., grain
properties, temperatures) are significantly different than those in
more evolved stages. \cite{andrews05,andrews07} examined the relation
between the disk masses for a large sample of Class I and II sources
in Ophiuchus and Taurus and found evidence for a decrease of the disk
masses by a factor 3 between these evolutionary stages from a median
disk mass of 0.015~$M_\odot$ to 0.005~$M_\odot$.

Although most sources show compact emission, there are exceptions: the
interferometric flux of the L483 Class~0 source is found to be
consistent with envelope-only emission and does not require a central
compact source. This is similar to the conclusion of \cite{l483art}
who modeled 3~mm continuum emission of the source from OVRO
observations. It is also in agreement with the measured spectral slope
of the SMA continuum flux between 1.3 and 0.8~mm \citep{prosacpaper},
which shows an index of $\approx 4$, consistent with optically thin
dust in a larger scale envelope, contrasting the other Class~0 sources
in the sample which have lower spectral indices in the range of 2--3
(Fig.~3 of \citealt{prosacpaper}). It is puzzling that no direct
evidence of a dust disk is seen for this source, given that it has a
well-established protostellar outflow -- which even seems to be in the
process of actively dispersing the envelope. This may be a case,
however, where we overestimate the contribution by the envelope on the
longest baselines from the radiative transfer models. If the envelope,
for example, has a larger flattened inner region or cavity, the
envelope contribution on the longer baselines would be smaller and a
central circumstellar disk required to explain the submillimeter
emission. The mass of this disk would in that case be
$\lesssim$0.01~$M_\odot$ -- derived from the long baseline
interferometric data without taking the larger scale envelope
contribution into account \citep{prosacpaper}.

In a recent study \cite{girart09} modeled SMA continuum data for the
Class~0 protostellar binary L723 and likewise found that a disk was
not required to explain the compact continuum emission. Scaling their
observed flux at 1.35~mm from the SMA at baselines of about
50~k$\lambda$ to our reference wavelength at 1.1~mm gives a flux of
about 35~mJy -- or about 3--4\% of the observed SCUBA peak flux of
1.1~Jy~beam$^{-1}$ from the JCMT/SCUBA legacy catalog
\citep{difrancesco08}. Within our framework this fraction is exactly
what is expected for envelope only contribution at these baselines,
and in agreement with the interpretation by \cite{girart09} that no
disk is required to explain the SMA continuum observations, but with a
note that if the envelope would have an inner flattened region, an
additional central dust component would be required -- again with a
typical mass $\lesssim$~a few$\times$0.01~$M_\odot$.

\section{Analysis of line data: stellar masses}\label{lineanalysis}
In the previous section masses were determined for two of the three
YSO components -- their disks and envelopes. The resolved line
observations obtained for the Class~I sources provide information
about the dynamical structure of their inner envelopes and disks and
thereby a handle on the mass of the third component, the central star.

\begin{figure}
\centering
\resizebox{0.49\hsize}{!}{\includegraphics{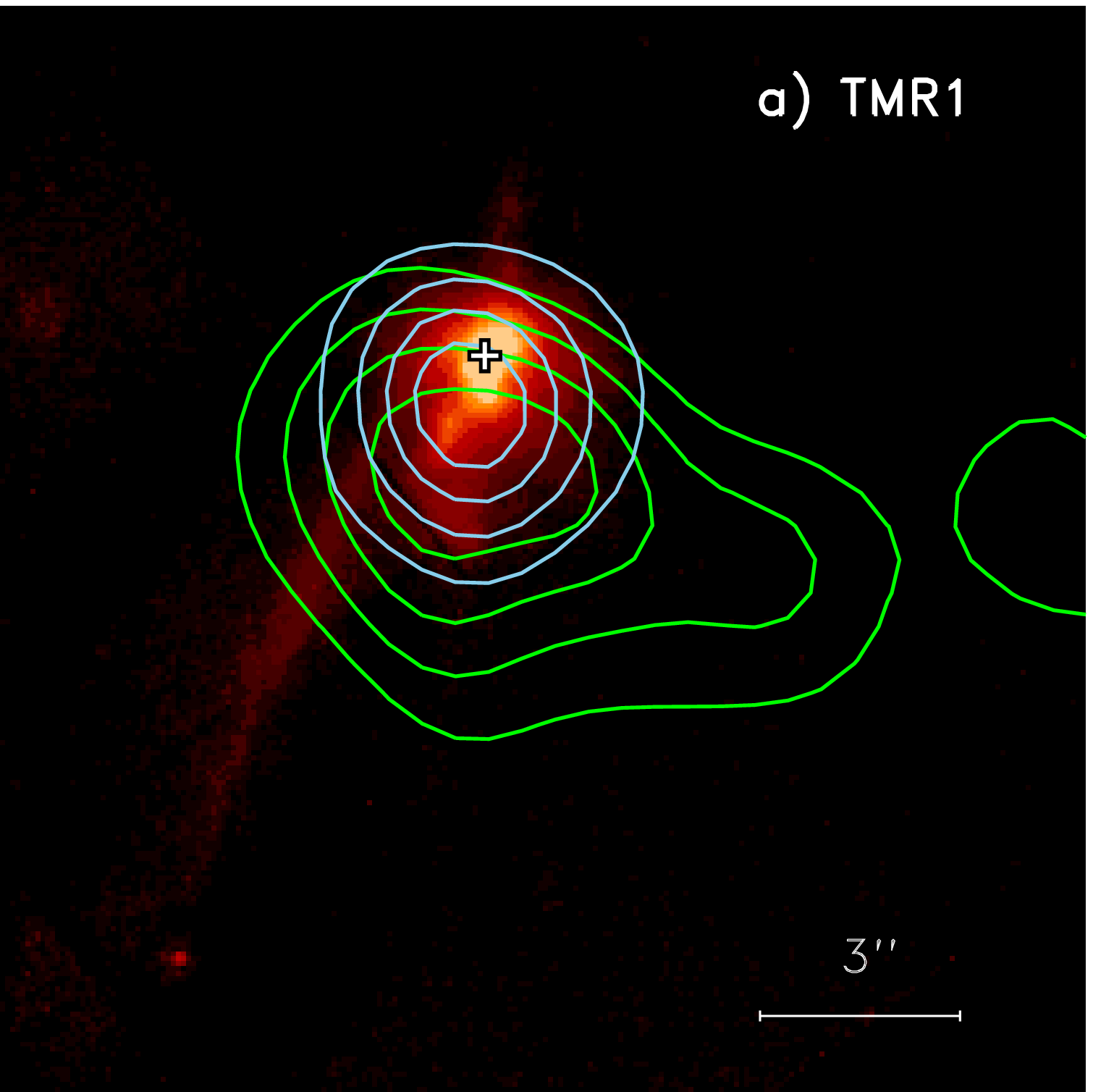}}\hspace{1mm}\resizebox{0.49\hsize}{!}{\includegraphics{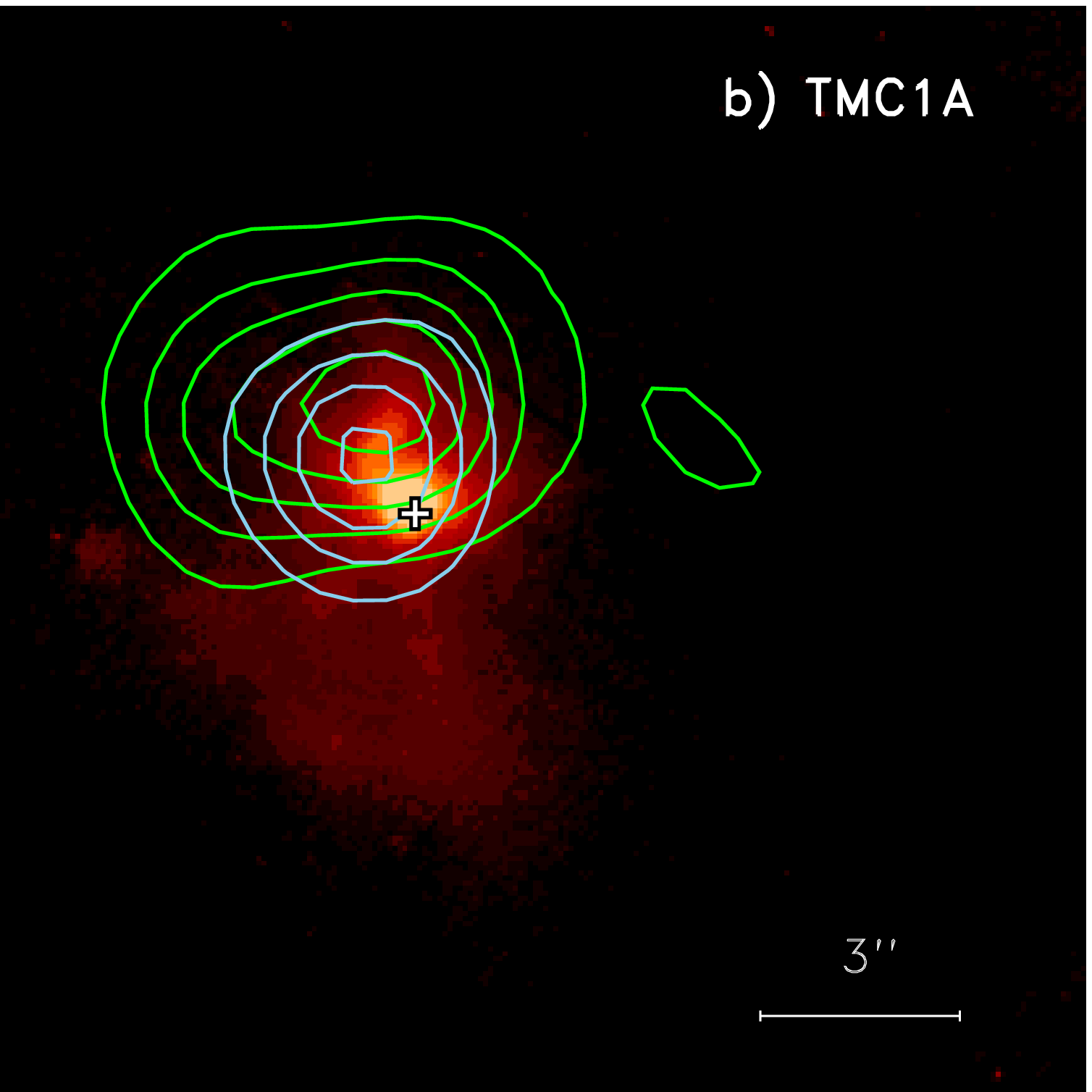}}
\resizebox{0.49\hsize}{!}{\includegraphics{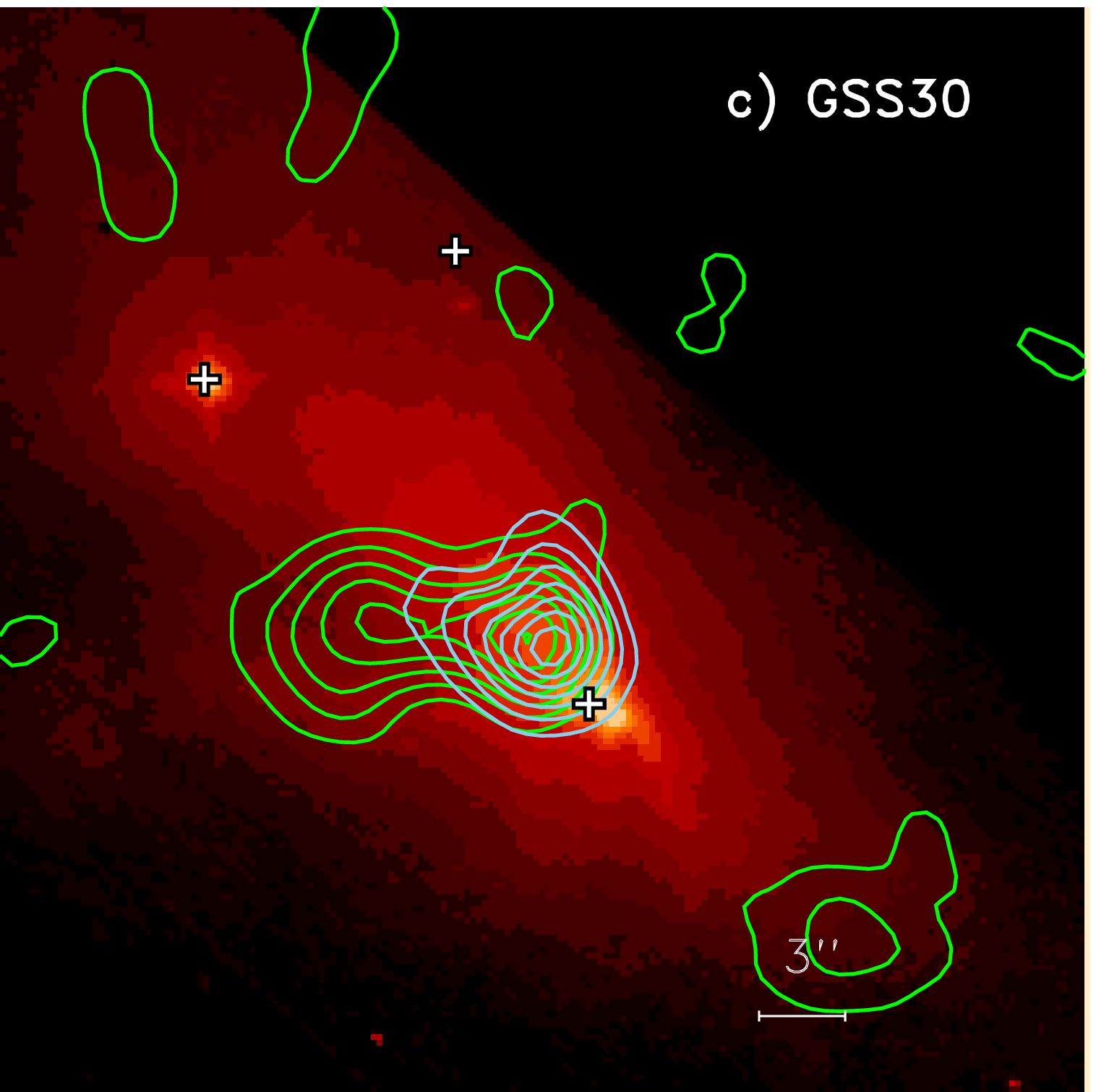}}\hspace{1mm}\resizebox{0.49\hsize}{!}{\includegraphics{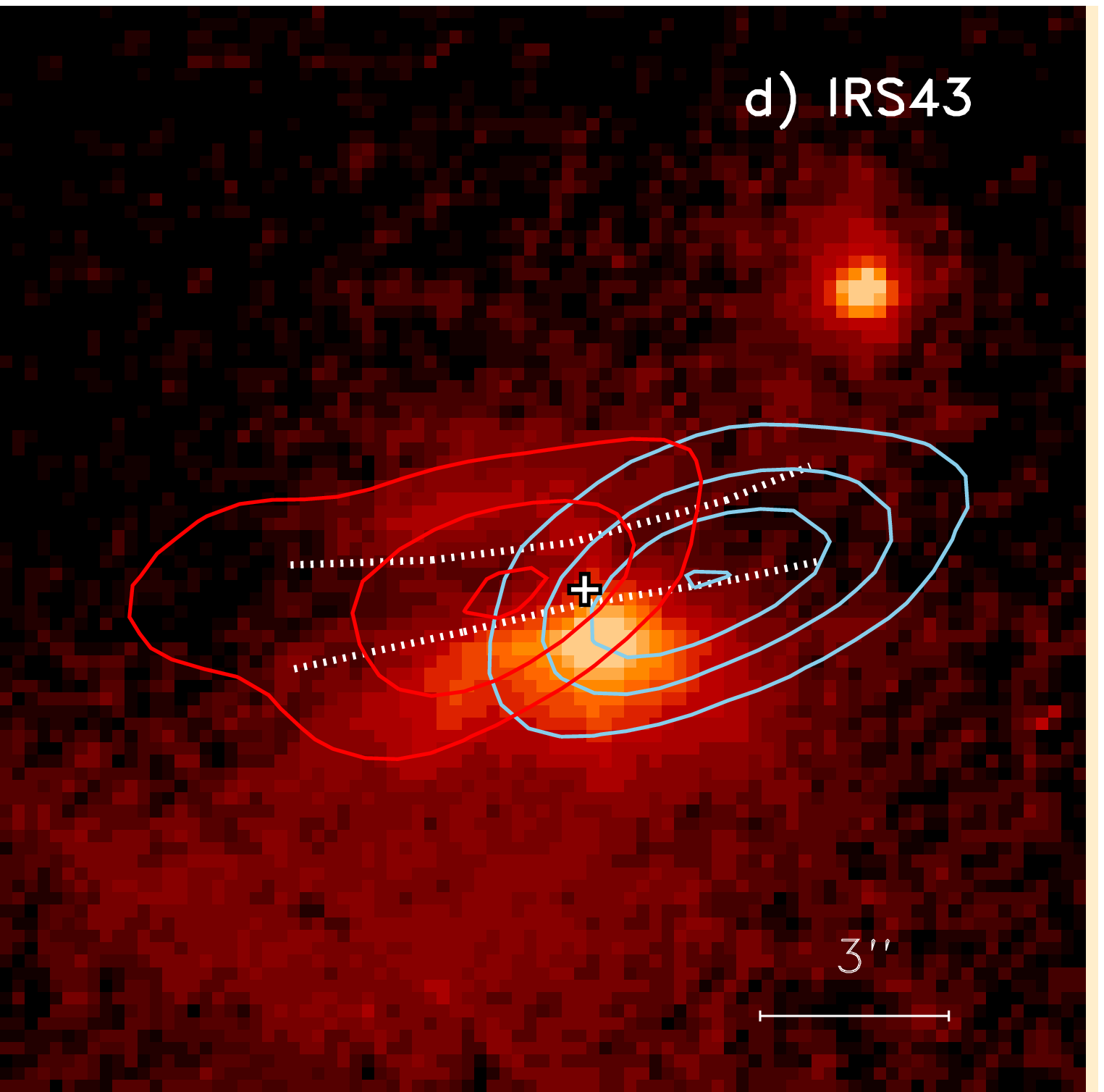}}
\caption{Archive 1.6~$\mu$m HST/NICMOS images of TMR1, TMC1A, GSS30
  and IRS~43 (see \citealt{terebey98} and \citealt{allen02} for details
  about the data) with HCO$^+$ 3--2 emission contours from the SMA
  overlaid. Panel a--c show TMR1, TMC1A and GSS30 with
  blue and green contours corresponding to blue-shifted emission
  integrated from $V_{\rm LSR}-2\Delta V$ to $V_{\rm LSR}-\Delta V$
  and $V_{\rm LSR}-\Delta V$ to $V_{\rm LSR}$ with $\Delta
  V$=1.0~\kms\ (TMR1 and TMC1A) and $\Delta V$=1.5~\kms\ (GSS30). For
  IRS~43 in panel d the blue and red contours correspond to
  blue- and red-shifted emission integrated from $V_{\rm LSR}-2\Delta
  V$ to $V_{\rm LSR}-\Delta V$ and $V_{\rm LSR}+\Delta V$ to $V_{\rm
    LSR}+2\Delta V$ with $\Delta V$=1.5~\kms. In that panel the white
  dotted lines indicate the rough extent of the dark lane in the HST
  image.}\label{hst_hcoP}
\end{figure}

\begin{figure*}\centering
\resizebox{0.3\hsize}{!}{\includegraphics{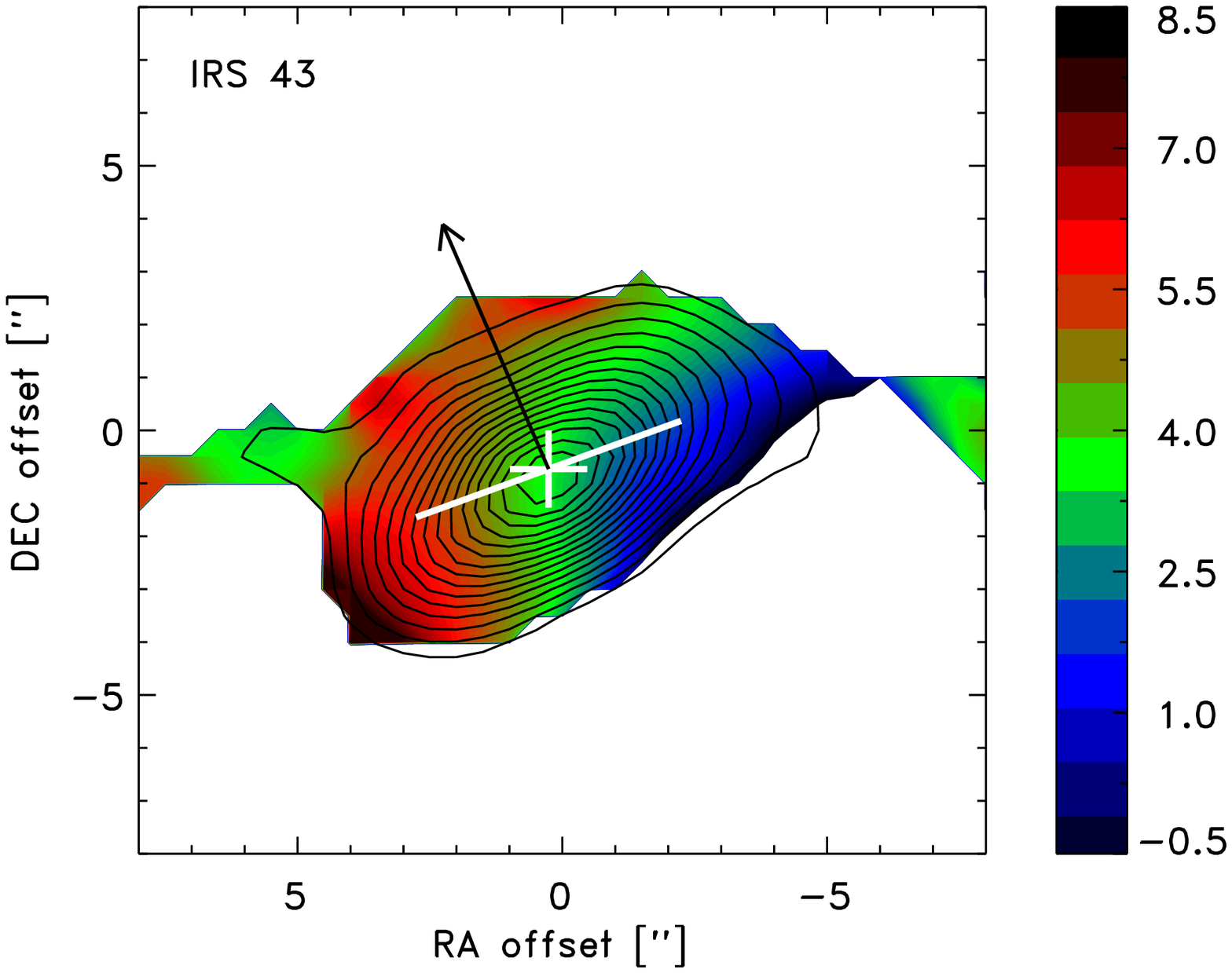}}
\resizebox{0.3\hsize}{!}{\includegraphics{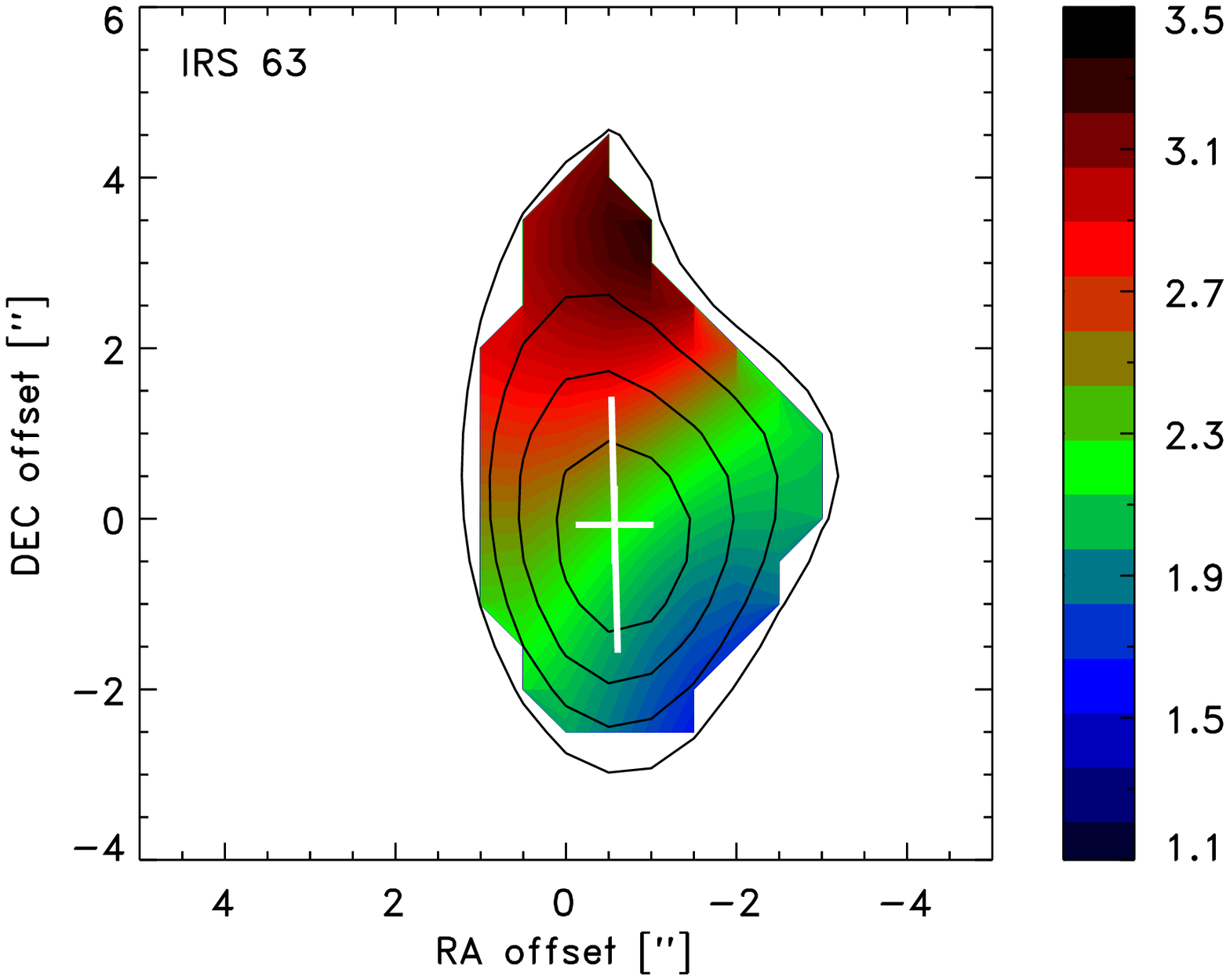}}
\resizebox{0.3\hsize}{!}{\includegraphics{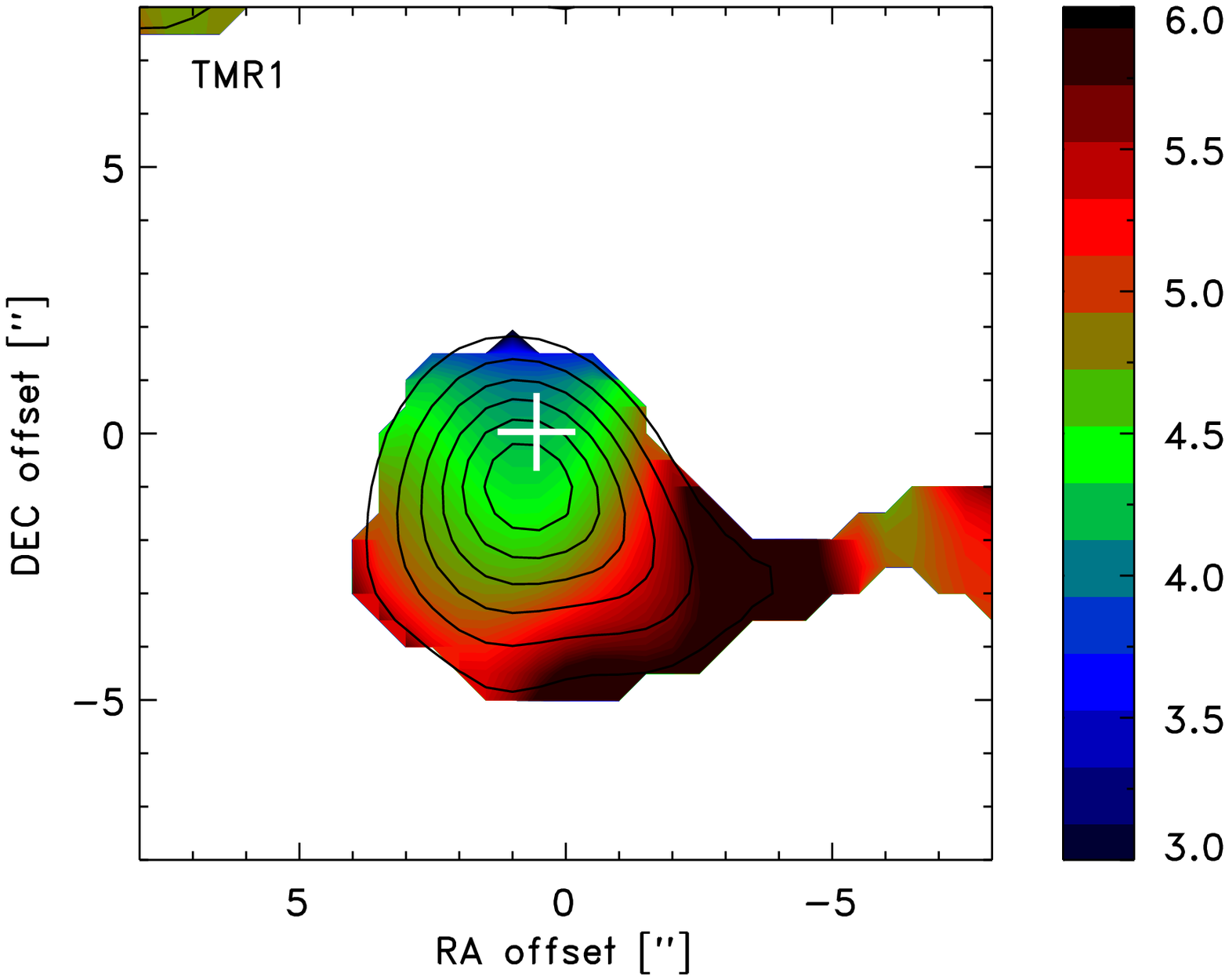}}
\resizebox{0.3\hsize}{!}{\includegraphics{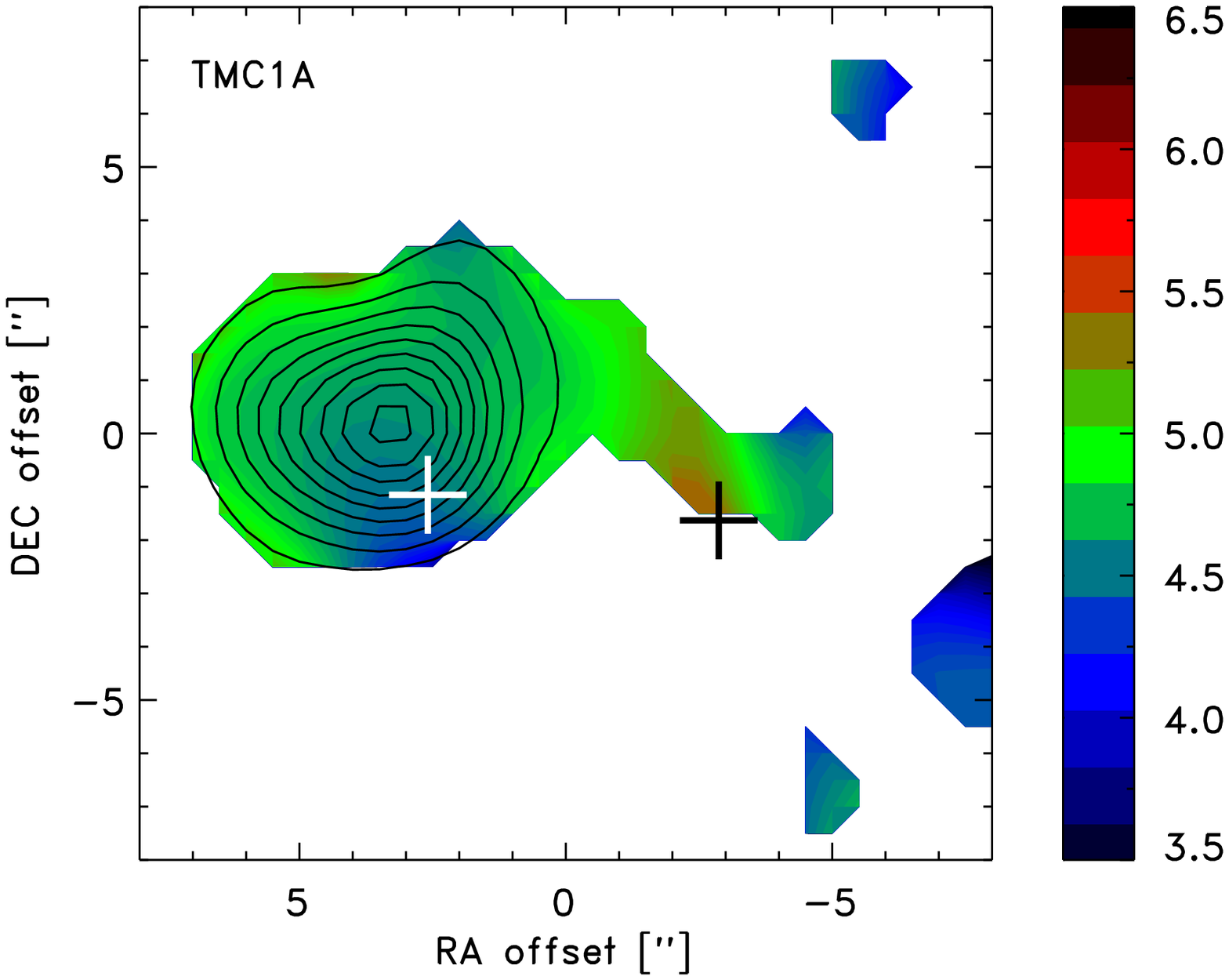}}
\resizebox{0.3\hsize}{!}{\includegraphics{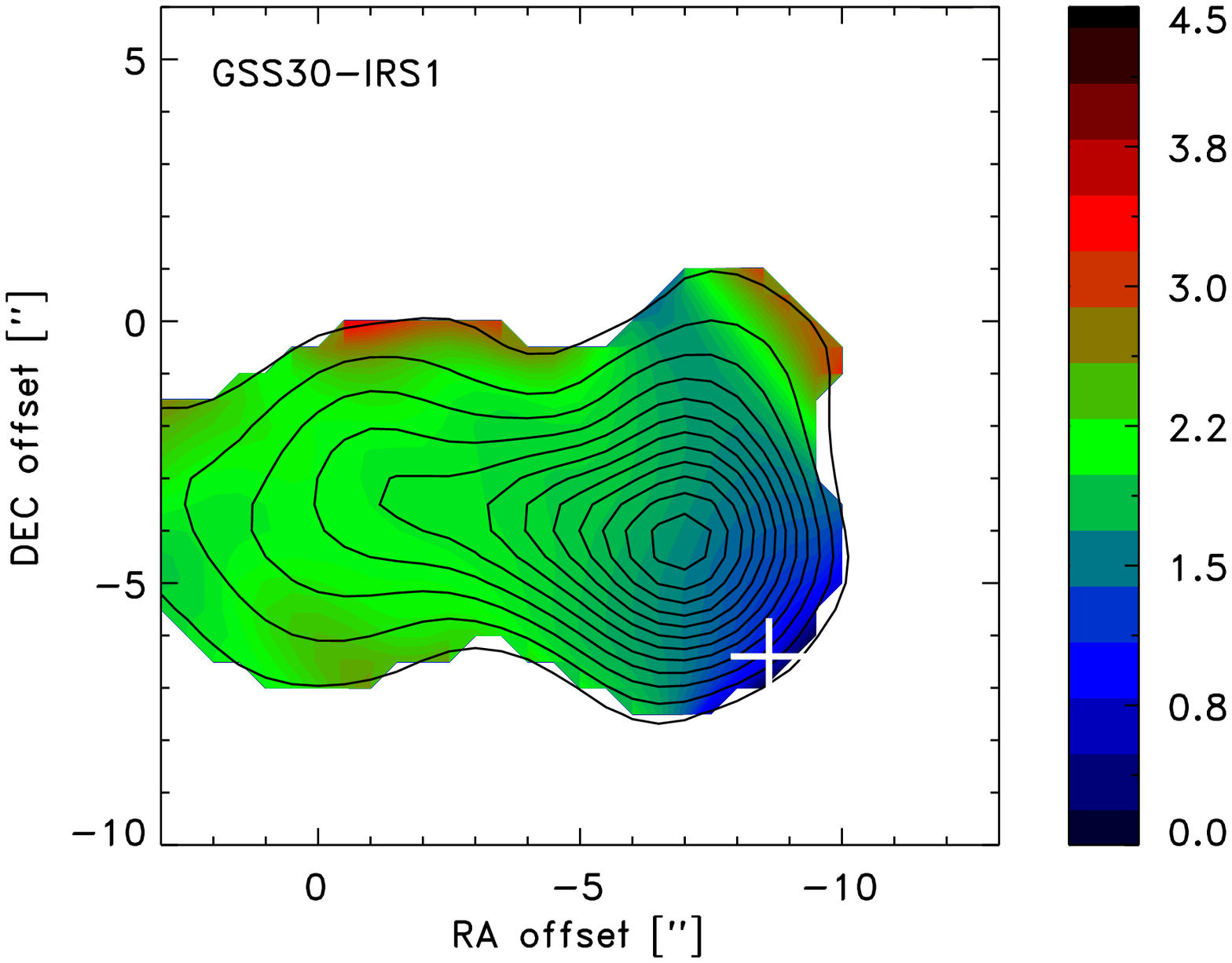}}
\caption{Moment one (velocity) maps of the HCO$^+$ emission
  for IRS~43, IRS~63, TMR1, TMC1A and GSS30-IRS1. In all panels the
  plus-signs indicate the location of the continuum positions (for
  GSS30-IRS1 the near-/mid-infrared source). In the IRS~43 and IRS~63
  panels the white line shows the direction of the continuum
  structure. In the IRS~43 panel, the black arrow furthermore shows
  the direction of the embedded near-IR Herbig-Haro objects
  \citep{grosso01} and thermal jet
  \citep{girart00}.}\label{irs43}\label{moment1}
\end{figure*}

As mentioned above, significant differences exist between the
components traced by the line observations for the different sources
in the sample. Fig.~\ref{hst_hcoP} compares the near-infrared
1.6~$\mu$m images to the HCO$^+$ emission in TMR1, TMC1A, GSS30-IRS1
and IRS~43. The first three of these sources show a clear alignment
with the near-infrared scattering nebulosities whereas the fourth,
IRS~43, shows the HCO$^+$ (and also HCN) emission aligned with the
dark lane across the source. Likewise, in the first order moment
(velocity) maps (Fig.~\ref{moment1}), TMR1, TMC1A and GSS30 show
significantly different morphologies from IRS~43 and IRS~63: the
former sources show evidence of blue-shifted emission over most of
their maps, with velocities decreasing toward the systemic velocities
at the largest distances. In contrast IRS~43, in particular, shows a
clear large-scale velocity gradient relative to the continuum peak
position and systemic velocity, potentially indicative of rotation as
also inferred for IRS~63, Elias 29 \citep{lommen08} and L1489-IRS
\citep{brinch07b}. In the following we focus on the use of
  HCO$^+$ as a tracer of rotation in these sources and include a small
  discussion of the sources showing outflow cavities and the
  limitations of HCO$^+$ as a tracer in Appendix~\ref{app:a}. Some
  details about the individual Class~I sources are given in
  Appendix~\ref{app:b}.

\subsection{Keplerian rotation and stellar mass in IRS~43} Four
  sources in our sample, IRS~43, IRS~63 and Elias~29
  (Fig.~\ref{hcoP_moment_overview}) as well as L1489-IRS
  \citep{brinch07b}, show elongated HCO$^+$ emission stretching
across the location of the YSO. As discussed by \cite{brinch07b} and \cite{lommen08} the
velocity gradients toward L1489-IRS, IRS~63 and Elias~29 can be
interpreted as having their origin in the inner envelope/circumstellar
disk around each source. A number of effects point toward this also
being the case for IRS~43:
\begin{enumerate}
\item Both HCN and HCO$^+$ show an elongated structure with a large
  scale velocity gradient around the continuum peak/systemic velocity.
\item A Gaussian fit to the continuum data in the $(u,v)$ plane shows
  a structure which is elongated in the same direction with a position
  angle of $-$70$^\circ$ (measured from north toward east): its
  deconvolved major axis is about 2$''$ (280~AU) and its minor axis
  about a tenth of an arcsecond, consistent with emission from a
    disk seen at a high inclination angle, close to edge-on.
\item The direction of the Herbig-Haro objects \citep{grosso01} and
  proposed radio thermal jet \citep{girart00} are with a position
  angle of 20--25$^\circ$, perpendicular to this extended structure
  (Figs.~\ref{irs43} and \ref{spitzer_irs43}).
\end{enumerate}

\begin{figure}
\resizebox{\hsize}{!}{\includegraphics{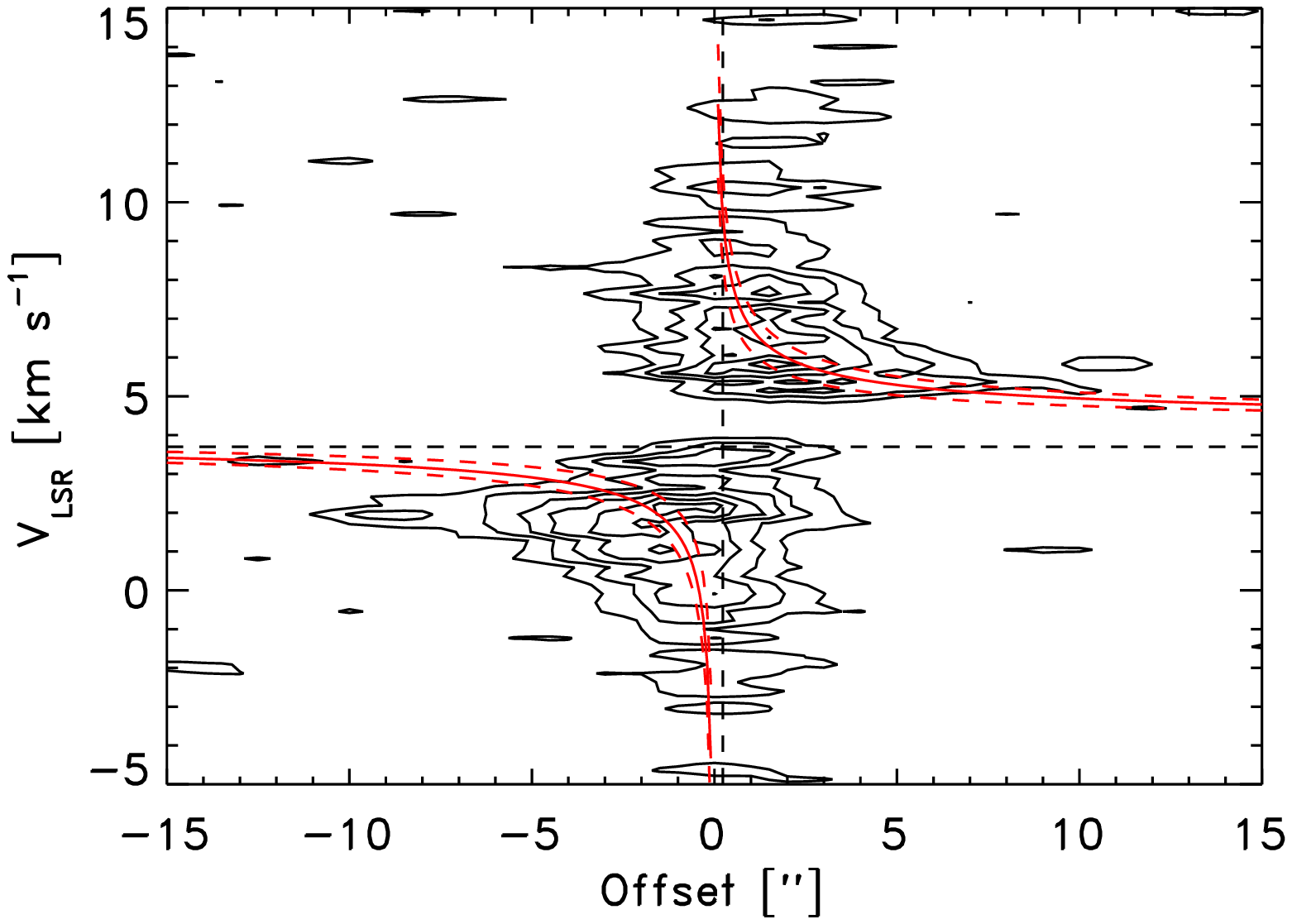}}
\resizebox{\hsize}{!}{\includegraphics{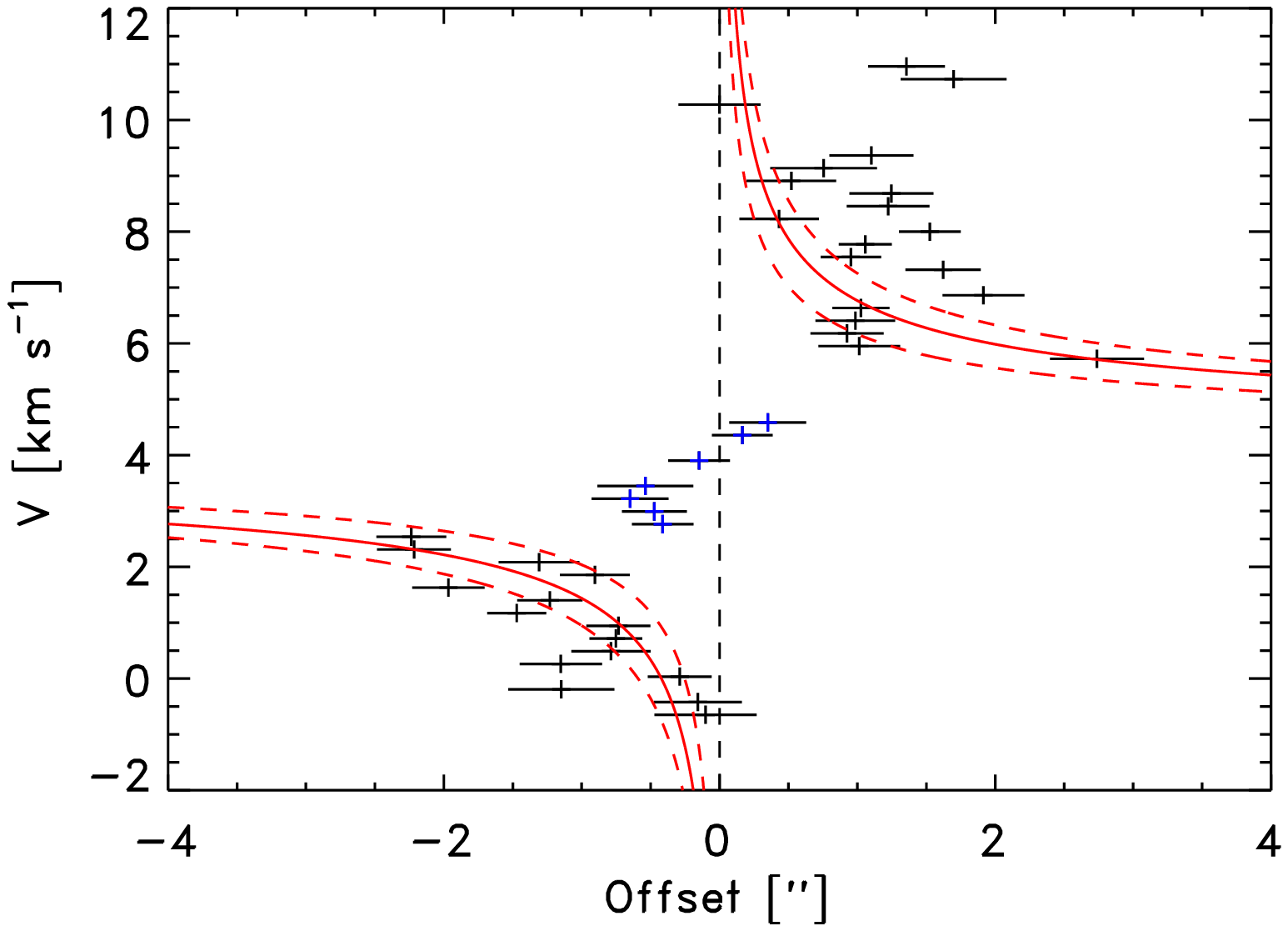}}
\caption{Position-velocity (PV) diagram for the HCO$^+$ emission in
  IRS~43 directly from the image-plane (upper panel) and from fitting
  the individual channels in the $(u,v)$-plane (lower panel). The PV
  diagram has been measured along the major axis of the HCO$^+$
  emission at a position angle of -70$^\circ$ (measured from north
  toward east). Overplotted are the predicted curves for Keplerian
  rotation around a 1.0~$M_\odot$ central object (solid) line (best
  fit to $(u,v)$-data), which can be compared to the dashed lines
  showing the same predictions for 0.6 and 1.4~$M_\odot$ central
  objects.}\label{irs43_pv}
\end{figure}

The narrow continuum structure and the alignment with the HCO$^+$ 3--2
emission is also seen in the IRS~63 data from \cite{lommen08}. For
that source, the velocity field could be well-fit with a Keplerian
profile around a 0.37~$M_\odot$ central object for an inclination of
30$^\circ$ (as hinted by modeling of the spectral energy distribution
of the source). A position-velocity plot of the HCO$^+$ 3--2 emission
toward IRS~43 is shown in Fig.~\ref{irs43_pv}. The position velocity
diagram shown here has been extracted along the major axis of the
HCO$^+$ emission at a position angle of $-70^\circ$. The velocity
profile is consistent with Keplerian rotation.

\begin{figure}
\resizebox{\hsize}{!}{\includegraphics{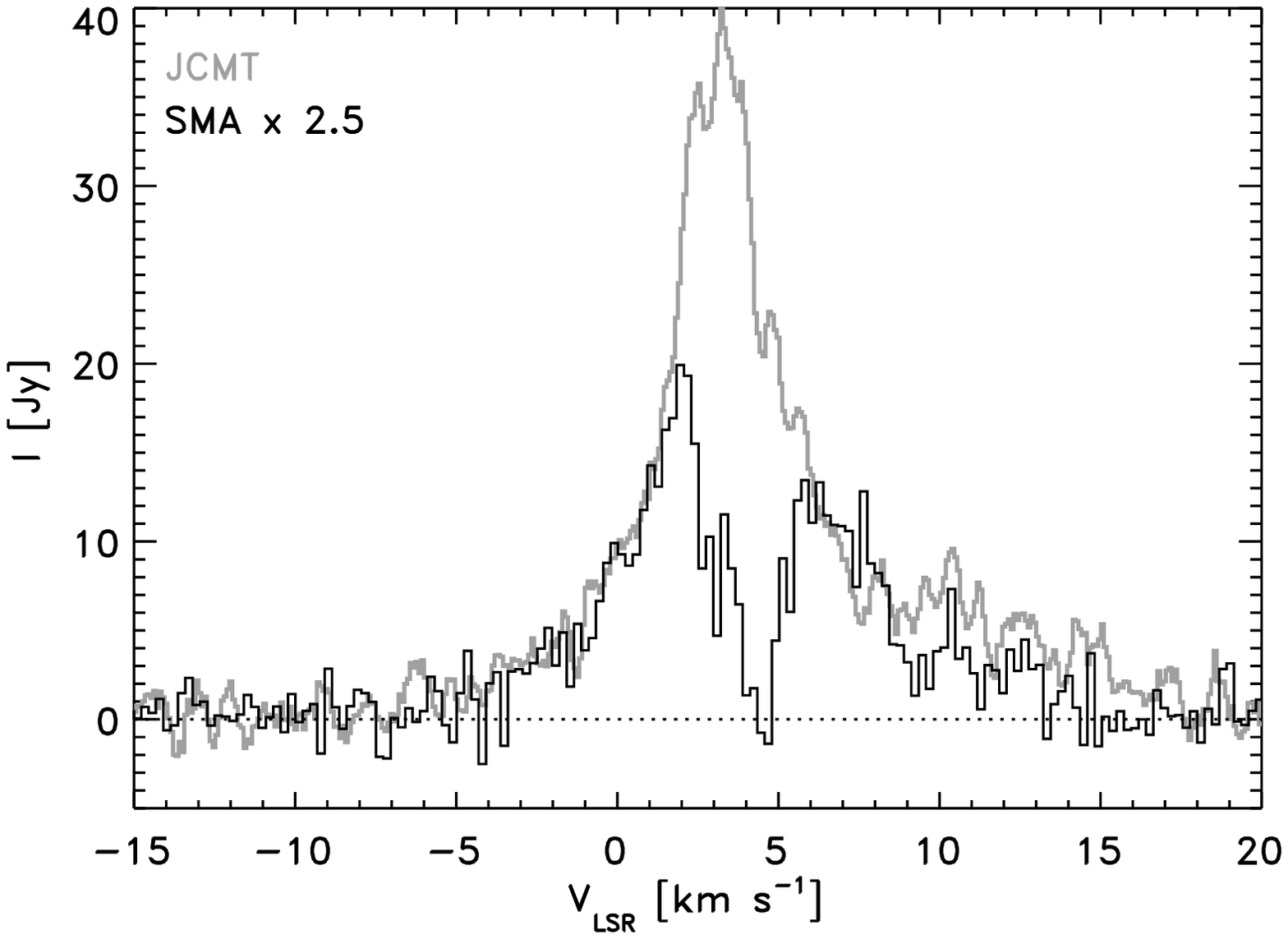}}
\caption{IRS~43: comparison between HCO$^+$ 3--2 spectrum from the SMA
  data integrated over the single-dish beam and JCMT observations. The
  interferometric spectrum has been scaled by a factor
  2.5.}\label{irs43_sd_int}
\end{figure}

The extent of the HCO$^+$ emission ($\approx 15''$) suggests that its
origin is not purely in the central disk, but also the inner regions
of the rotating envelope ($r \lesssim$~1000~AU). This is further
supported by the comparison between the single-dish spectrum and a
spectrum extracted from the interferometric data integrated over the
extent of the JCMT single-dish beam (Fig.~\ref{irs43_sd_int}). About
40\% of the single-dish flux is recovered by the interferometer,
suggesting the presence of some extended emission resolved out by the
interferometer -- although this fraction is significantly less than
typically seen, e.g., in the more deeply embedded Class~0 protostars
\citep[e.g.,][]{prosacpaper}.

To derive the mass of the central object a position-velocity curve is
extracted by fitting the position of the peak emission for each
channel in the $(u,v)$-plane at baselines larger than 20~k$\lambda$
and projecting this on the major axis of the HCO$^+$ emission (right
panel, Fig.~\ref{irs43_pv}). To this curve, a $\chi^2$-fit is
performed with a straightforward Keplerian rotation curve with two
free parameters, the systemic velocity and central mass. A best fit is
obtained for a central mass of 1.0~$M_\odot$ (2$\sigma$ confidence
levels of $\pm 0.2$~$M_\odot$) and a systemic velocity of
4.1~\kms. This mass estimate is of the ``enclosed'' mass within the
resolution of the interferometer, but since it is significantly higher
than the best estimate of the envelope and disk masses of 0.026 and
0.0081~$M_\odot$ respectively (see \S\ref{continuumanalysis}), it is
likely close to the total mass of the central star. This is also a
lower limit since it is not corrected for inclination, but given the
small minor axis relative to the major axis from the Gaussian fit to
continuum emission, it is not unreasonable to assign a high
inclination angle corresponding to an edge-on disk. 

In summary, we detect signatures of Keplerian rotation in the
  HCO$^+$ 3--2 emission for 4 Class~I sources allowing direct
  determinations of their central stellar masses, which range from
  about 0.3 to 2.5~$M_\odot$. As discussed in Appendix~\ref{app:a2}, the
  100~AU scales of embedded protostars can only be traced with HCO$^+$
  3--2 in sources with envelope masses less than about 0.1~$M_\odot$,
  however. Future high sensitivity ALMA observations will be required
  to observe more optically thin isotopes, to determine the dynamical
  structure of the more deeply protostars.

\section{Discussion}\label{discussion}
\subsection{Effects of assumed dust properties and temperatures on
  disk masses}\label{diskassumptions} 
\begin{figure}
\resizebox{\hsize}{!}{\includegraphics{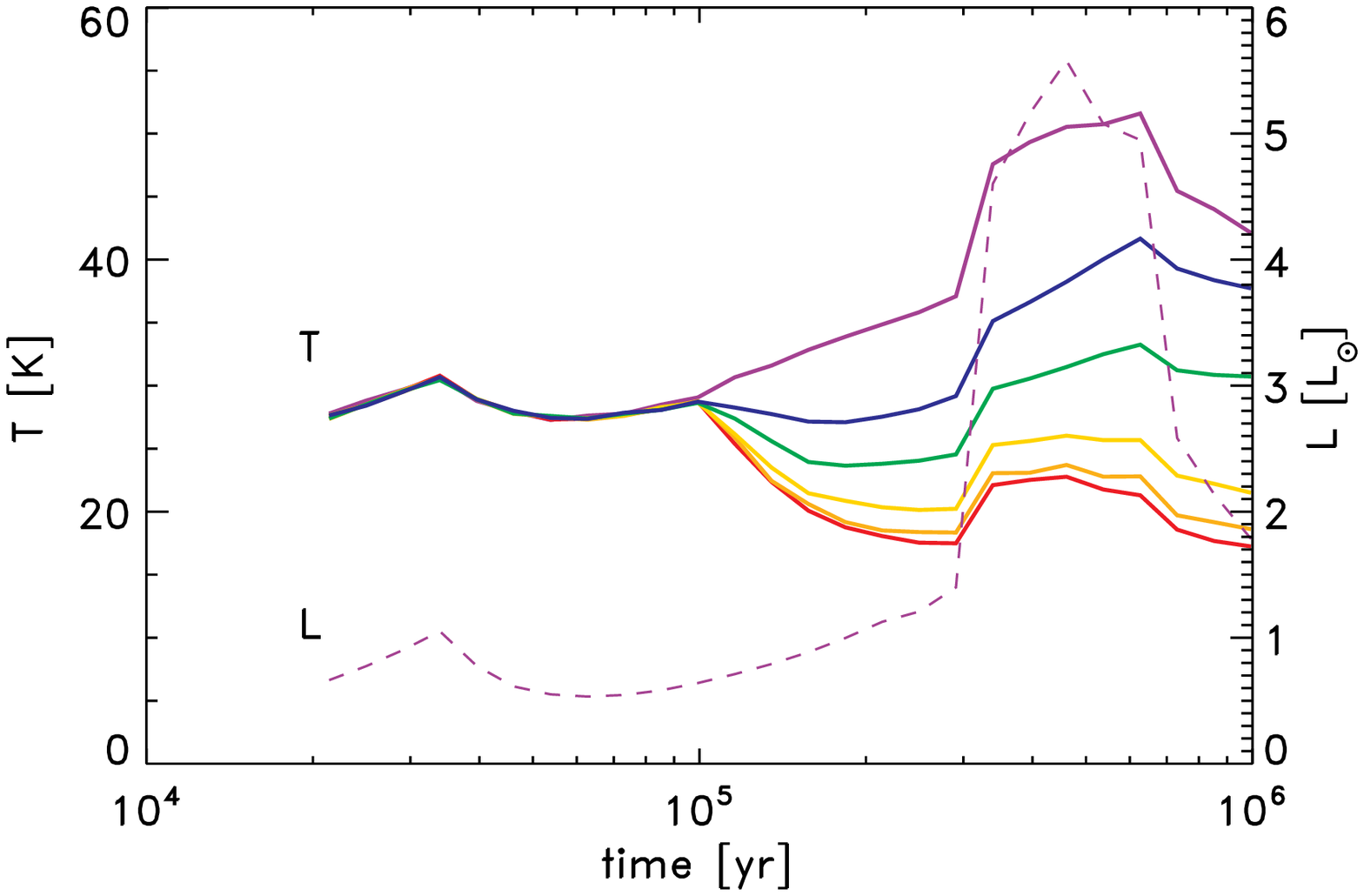}}
\caption{Temporal evolution of the temperature and luminosity in the
  simulation of \cite{visser09}. The solid lines shows the temperature at 200
  AU along rays inclined with 0, 5, 10, 15, 20 and 90$^{\circ}$ relative to
  the disk plane (from bottom to top; i.e., the line with the highest
  temperature corresponds to an inclination of 90$^\circ$ -- perpendicular to
  the disk plane). The dashed line shows the corresponding evolution of the
  luminosity.}\label{temp_profile}
\end{figure}

As in most other
  studies in the literature, two assumptions in our study are that the disk can be characterized by a
uniform temperature and by a single unchanging dust opacity law
throughout its extent and evolution. Both these assumptions may
  affect the systematic trends observed in the data: for the
  temperature an increase in luminosity increases the mass-weighted
temperature of the disk, while the shielding by the disk itself will
lead to a lower mass weighted temperature with increasing mass. As an
example, Fig.~\ref{temp_profile} compares the temporal evolution of
the temperature at a radius of 200~AU between rays through the model
with different inclinations with respect to the disk plane in the
simulation of \cite{visser09}. Also shown is the evolution of the
luminosity in the simulations. Early in the evolution, before the disk
is formed, the temperature tracks the luminosity changes closely, but
as the disk forms (at about $10^5$~years) the temperature at
inclinations directed through the disk drops sharply, even though the
luminosity continues to increase. The strong increase in total
luminosity at about 3$\times 10^5$~years is directly reflected in the
temperature profiles.

\begin{figure}
\resizebox{\hsize}{!}{\includegraphics{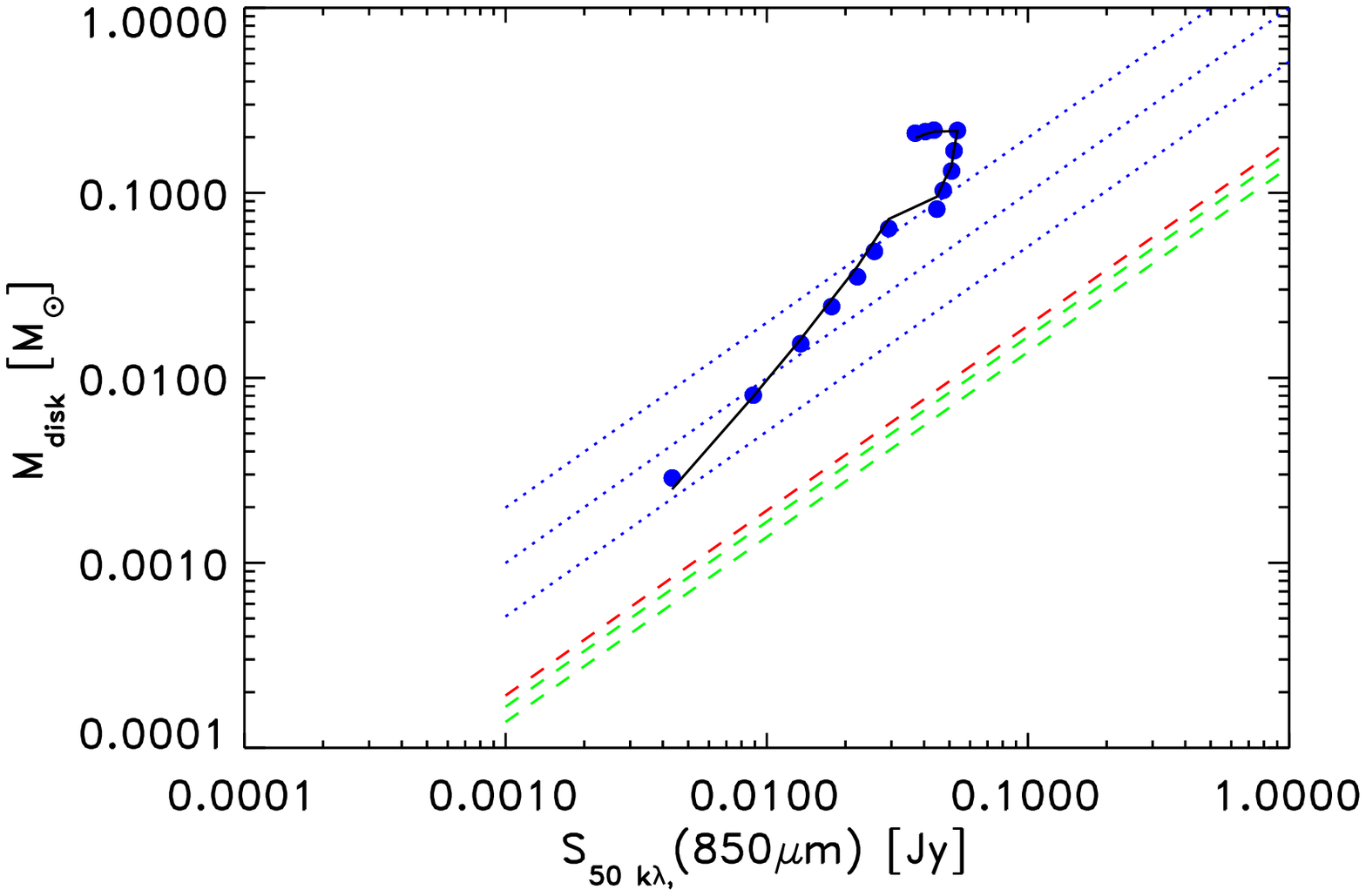}}
\caption{Relationship between disk mass and 850~$\mu$m    interferometer flux assuming a distance of 250~pc: the three
  lower straight lines indicate the linear relationships from
  \cite{terebey93} (lowest), \cite{andrews07} (middle) and
  \cite{prosacpaper} (top) scaled to the same distance and
  wavelength. The filled symbols indicate the relationship found for
  the radiative transfer modeling of the \cite{visser09} collapse
  simulation discussed in the text. The solid line through these
  points indicates a relationship where the temperature used to relate
  the disk mass and flux follows Eq.~\ref{fluxmass_temp}. Finally, the
  three dotted lines indicate predicted linear relationships using the
  dust properties from the simulations of \cite{visser09} with
  temperatures of 8, 11 and 16~K (from top to
  bottom).}\label{fm_model}
\end{figure}

Fig.~\ref{fm_model} compares the relationships between (sub)millimeter
flux and disk mass adopted by \cite{terebey93}, \cite{andrews07} and
\cite{prosacpaper} to the relationship predicted by the radiative
transfer modeling of the collapse simulation from \cite{visser09}: it
is clearly seen that the models require a steeper relationship between
the submillimeter flux and disk mass than what is obtained with a
single constant temperature: to calculate the disk mass from the
submillimeter flux in the model, one would need a temperature scaling
as:
\begin{equation}
T = 11.3\,\, {\textrm K} \left(\frac{M_{\rm disk}}{0.01 M_\odot}\right)^{-0.25}\left(\frac{L_\star}{L_\odot}\right)^{0.1}\label{fluxmass_temp}
\end{equation}
as shown with the solid line in Fig.~\ref{fm_model} -- i.e., an
increasing temperature with increasing luminosity (and thus heating)
and a decreasing temperature with increasing disk mass (and thus
shielding at a given radius). Although the variations in the
temperature with disk mass and stellar luminosity are slow, they still
introduce a systematic gradient in the derived disk masses by up to a
factor of 4 through the evolution of the embedded stages of the
YSOs. It thus appears that the disk masses in the later stages
relative to the early stages can be underestimated by about this
factor -- solely due to the evolution of the protostellar
system. This may explain the slightly higher masses for the Class 0
sources in Fig.~\ref{mass_results1}.

\begin{figure}\centering
\resizebox{\hsize}{!}{\includegraphics{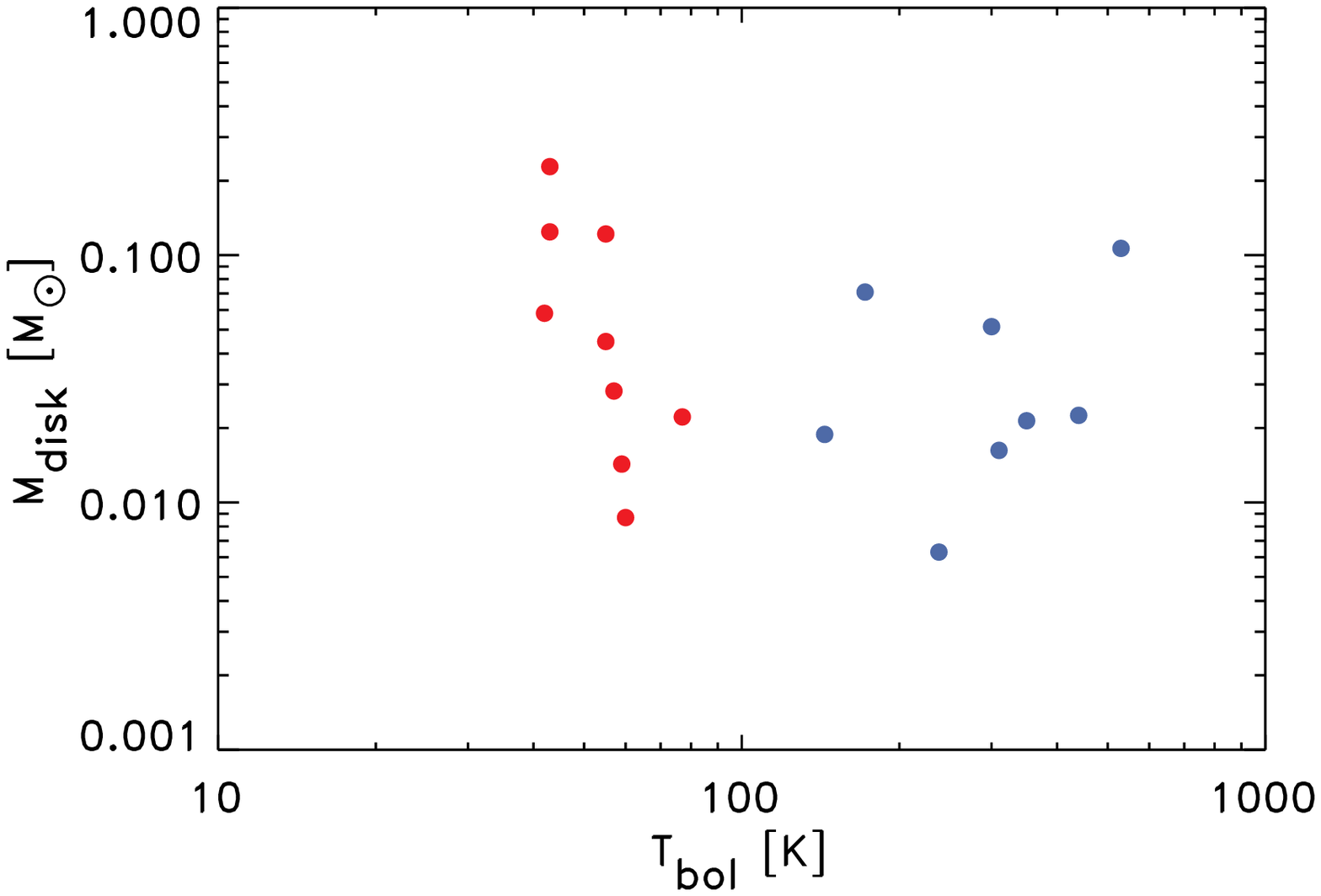}}
\resizebox{\hsize}{!}{\includegraphics{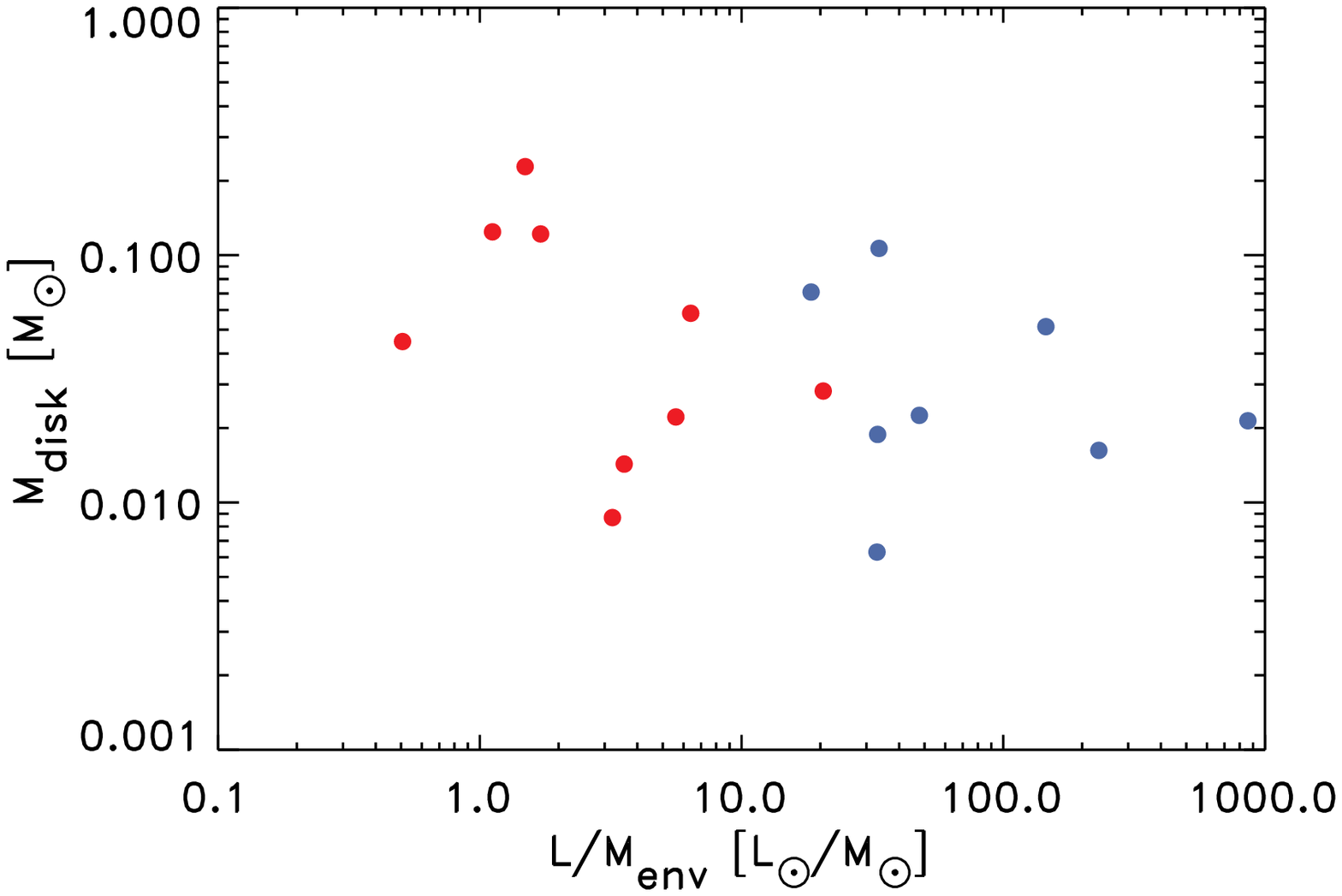}}
\caption{Disk masses as function of bolometric temperature
  \emph{(upper)} and source luminosity/envelope mass ratio
  \emph{(lower)}. In comparison to Fig.~\ref{mass_results1}, the Class
  0 disk masses have been decreased by a factor 2 and the Class I disk
  masses increased by a similar amount -- simulating the evolutionary
  effect in the relationship between flux and disk mass suggested by
  Fig.~\ref{fm_model}. }\label{mass_results2}
\end{figure}

Fig.~\ref{mass_results2} compares the derived disk masses to
bolometric temperature as well as the ratio of the source luminosity
over the envelope mass, $L_\star / M_{\rm env}$. The latter ratio is
an alternative evolutionary tracer, increasing with time as long as
the YSO luminosity is dominated by accretion and the stellar mass (and
luminosity) increases, whereas the envelope dissipates and its mass
decreases. In this figure the Class 0 disk masses are decreased by a
factor of 2 and the Class I disk masses increased by a factor of 2 to
simulate the temperature effect discussed above. With this correction,
the Class 0 and I disk masses are roughly constant with a median mass
of about 0.04~$M_\odot$ for the full sample.

The assumption of a constant unchanging dust opacity law similar to
the large-scale envelope dust may also be wrong: in particular, grain
growth may increase the dust opacity at submillimeter wavelengths
leading to a lower derived disk mass for a given interferometric
flux. That something like this may be going on is hinted by the
observation that the compact components around the Class 0 sources
show spectral slopes, $\alpha$, of the emission $F_\nu \propto
\nu^\alpha$, at 230 and 345~GHz in the range 2--3
\citep{prosacpaper}. This suggests that the disks are either optically
thick or have significantly different dust opacities than standard ISM
dust. As discussed in \citeauthor{prosacpaper}, the disk masses
implied for the Class 0 objects suggest that only the most massive
ones may be marginally optically thick. Resolved observations at
multiple wavelengths and measurements of their sizes are required to
address this. On the other hand, the low spectral indices confirm that
the compact components around the Class~0 sources are not associated
with the larger scale envelopes for which an $\alpha$ of 3.5--4 is
expected.

Other dust opacity laws or temperatures have been used in literature:
\cite{andrews07} adopted an empirical scaling based on modeling of the
disk SEDs, corresponding to an equivalent dust temperature of 20~K and
a dust opacity-law following \cite{beckwith90} normalized to the dust
opacity at 1000~GHz of 0.1~cm$^2$~g$^{-1}$ (for a dust-to-gas ratio of
1:100). The lower temperature leads to higher disk masses for a given
1.1~mm flux by a factor of 2.3 whereas their adopted dust opacity-law
corresponds to lower disk masses by a factor 2.0 compared to
ours. Likewise, \cite{terebey93} adopted a dust temperature of 40~K
and adopted a dust opacity law with a mass opacity at 250~$\mu$m of
0.1~cm$^2$~g$^{-1}$. If compared at a common wavelength of 850~$\mu$m,
these three choices of parameters result in disk masses that agree
within $\pm$15\% from the middle value from \cite{andrews07}.

However, systematic changes in the dust properties occurring during
the evolution of the young protostars could have an important effect
on our derived disk masses and observed trends. For example, the dust
opacities at 500~$\mu$m--1~mm from \cite{ossenkopf94} increase by as
much as an order of magnitude depending on the density at which the
grains have coagulated, although the effect is significantly smaller
for grains with ice mantles. As shown by \cite{dalessio01} grain
growth increases the dust opacity at a specific wavelength, $\lambda$,
as long as the largest grains are smaller than $\lambda$. In a
scenario where grains grow toward sizes inferred for T~Tauri disks
($\sim 1$mm; e.g., \citealt{dalessio01}) as the protostellar system
evolve through its embedded stages the derived disk masses for the
more evolved sources would be overestimated under our assumptions of
an unchanging dust opacity law. This would work in the opposite
direction of the temperature effect -- toward even more massive disks
in the Class 0 compared to Class I stages. It cannot be ruled out that
grain growth would appear in a fashion not straightforwardly
increasing with ``age'' of the studied young stellar objects, e.g., if
the observed dust in the Class 0 sources is located in regions with
relatively higher densities for a significant fraction of time. Based
on the current observational data, it is not possible to separate
these scenarios, however: further resolved multi-wavelength images of
the dust continuum emission from the envelopes and disks around
protostars or of well-calibrated gas tracers (taking their chemistry
into account) are required.

\subsection{Collapse and disk formation models}\label{models:complex}
Power-law density profiles are simplified parameterizations of the
underlying envelope density profile and the envelopes are not
spherical due to, e.g., rotation and the action of protostellar
outflows.

In the framework of the \cite{shu77} and \cite{terebey84} models for
inside-out collapsing envelopes, the density profile flattens from a
stationary $r^{-2}$ density profile through a free-falling $r^{-1.5}$
profile to a flattened inner envelope where the angular momentum
roughly balances gravity. \cite{terebey93} examined the millimeter
emission from such envelope models using analytical descriptions for
the envelope density and temperature profiles and used those to
compare IRAM~30 m single-dish continuum observations (11$''$; at
1.3~mm) to OVRO interferometric observations (4-10$''$; at
2.7~mm). Consistent with Fig.~\ref{comp_ext} they found that the
emission in the interferometric observations is completely dominated
by a central compact component whereas most of the observed
single-dish flux in fact has its origin in the collapsing
envelope. The dominance of the disk to the interferometric flux was
found by \citeauthor{terebey93} to be due to \emph{(i)} the smaller
beam of the interferometric data and negligible envelope mass on those
scales and \emph{(ii)} the steeper wavelength dependence found for the
dust emission in the optically thin envelope.

\begin{figure}
\resizebox{0.9\hsize}{!}{\includegraphics{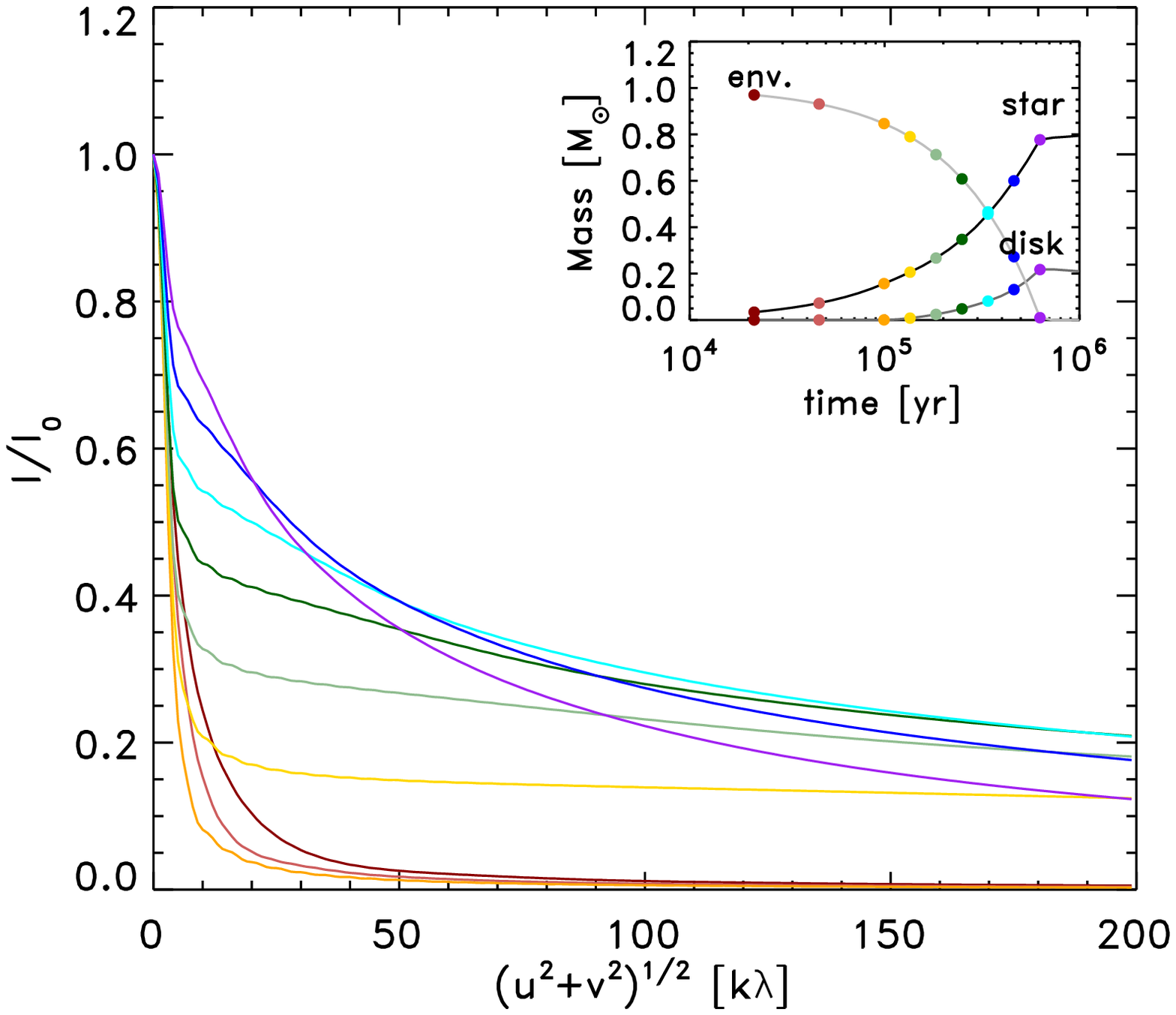}}
\resizebox{0.9\hsize}{!}{\includegraphics{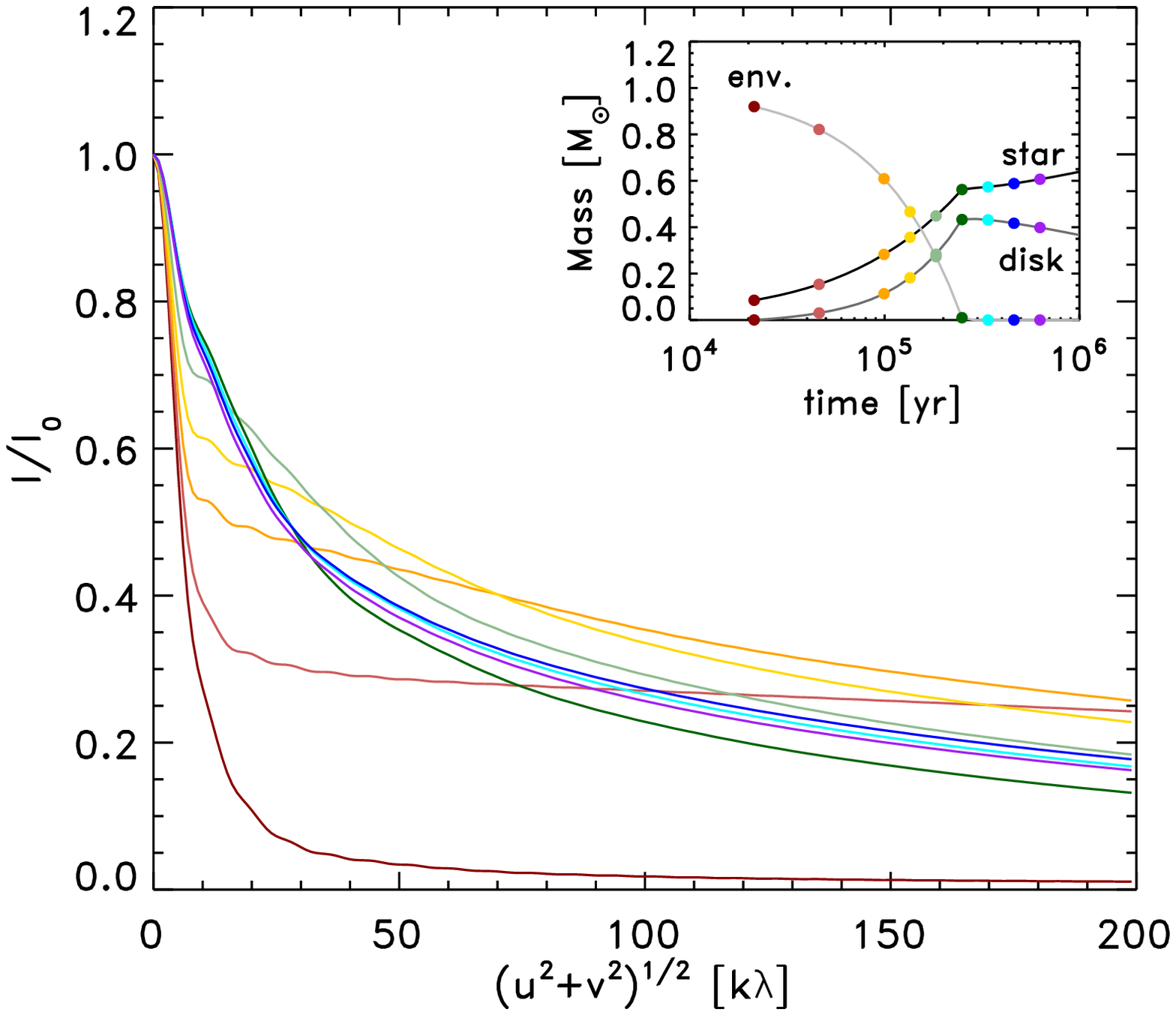}}
\caption{Evolution of the predicted visibility amplitude curves (normalized by
  the zero-spacing flux) as function of projected baseline lengths for two
  models for envelope collapse and disk formation from \cite{visser09}
  characterized by the solid-body rotation rate, $\Omega_0$, and the effective
  sound speed $c_s$. \emph{Upper:} $\Omega_0 = 10^{-14}$~s$^{-1}$ and
  $c_s=0.19$~km~s$^{-1}$. \emph{Lower:} $\Omega_0 = 10^{-13}$~s$^{-1}$ and
  $c_s=0.26$~km~s$^{-1}$. The different curves correspond to snapshots at
  different times indicated in the upper right insert in each panel, also
  showing the evolution of the envelope, disk and stellar
  masses.}\label{ruud_circ}
\end{figure}

We expand this analysis by introducing the formation of a
circumstellar viscous disk in the center using the models of
\cite{visser09} and following the evolution of the submillimeter
brightness profiles observed by an interferometer:
\citeauthor{visser09} adopted the framework of \cite{cassen81} and
\cite{terebey84} to describe a collapsing envelope and introduced the
formation and growth of a 2-dimensional central disk and star similar
to, e.g., \cite{yorke99}. The models of \citeauthor{visser09} also
include the presence of an outflow cavity with a growing opening angle
through the evolution consistent with observations
\citep[e.g.,][]{arce06}. Using the axisymmetric radiative transfer
code ``RADMC'' \citep{dullemond04}, \citeauthor{visser09} calculated
the temperature profile for time-steps in the above model description
and furthermore followed the chemical history of particles during the
collapse and disk formation. For this purpose, we adopt the
temperature and density profiles of the envelope (with outflow cavity)
and disk from \citeauthor{visser09}, as well as their prescription for
the evolution of the stellar luminosity and effective temperature and
dust parameters, and ray-trace these. Fig.~\ref{ruud_circ} shows the
evolution of the predicted visibility curves at different time-steps
for two representative models from the grid of \citeauthor{visser09}

Both panels clearly illustrate the effect discussed above: until the
formation of the circumstellar disk more than 95\% of the total flux
at a given wavelength is emitted on baselines shorter than 50
k$\lambda$ -- corresponding to a ratio between the 1.1~mm
interferometric flux and the peak 850~$\mu$m single-dish flux of less
than a few \% (depending on the exact spectral slope between the
observations at these differing wavelengths) in accordance with the
results above. Once the disk is formed, the longer baseline flux
rapidly increases -- first simultaneously on all baselines and then
more rapidly on the shorter baselines as the disk grows to sizes large
enough to be resolved by the interferometric observations. It should
be noted that even though the outflow cavity is opened up early in the
evolution of the collapsing protostar, it does not provide significant
compact emission, e.g., due to heated material in the outflow cavity
walls, to affect the evolution of the observed interferometric flux.

An alternative explanation to the disk hypothesis favored in this work
is that there is an enhancement of the density in the envelope profile
at small scales. As an example, \cite{chiang08} adopted the
theoretical collapse models of \cite{tassis05} and compared those to
2.7~mm interferometric data from \cite{looney00}. The models of
\cite{tassis05} simulate the formation of a protostellar object from a
cloud core in the presence of magnetic fields. In their simulation,
ambipolar diffusion leads to the formation of a thermally and
magnetically super-critical core which contracts
dynamically. Initially this core has a density profile $n\propto
r^{-1.7}$, but as mass and magnetic flux is accreted from the envelope
onto the central protostar, a region of enhanced magnetic fields is
formed driving an outward-propagating shock and causing a density
enhancement growing quasi-periodically from scales of
$\sim$100~AU. This density enhancement on small scales will result in
a compact component observable by the interferometer which is slightly
resolved on long baselines and thereby in some cases alleviates the
need for a circumstellar disk on those scales as shown by
\cite{chiang08}.

Still, the conclusions of both studies are similar: a density
enhancement is required in the collapsing envelope on small scales -
above what would be expected from, e.g., a free-falling envelope. The
question is whether this component is inherent to the collapse profile
or the first sign of the formation of a circumstellar disk. A
straightforward, although currently observationally challenging, way
to address this issue is through resolved line observations of the
innermost envelope: the resolved line observations of the Class I
sources in this sample \citep{brinch07b,lommen08} show clear
signatures of Keplerian rotation, which argues in favor of the
  disk scenario for the sources in those evolutionary stages at
  least. For the more deeply embedded protostars future ALMA
observations are needed to confirm the dynamical nature of the central
compact dust component.

\subsection{Constraints on evolutionary models}
\begin{figure}
\resizebox{\hsize}{!}{\includegraphics{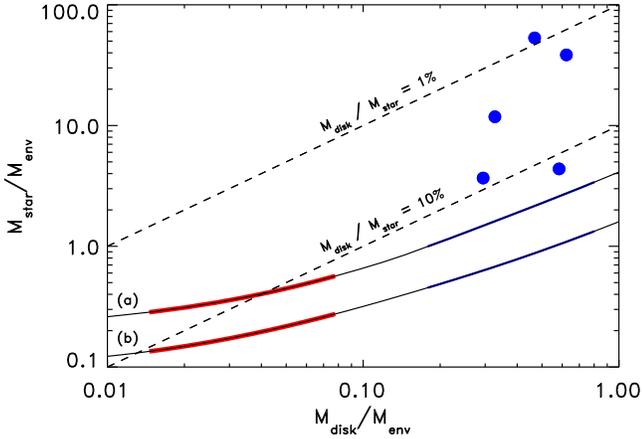}}
\caption{Predicted stellar mass, $M_{\rm star}$, vs. disk mass,
  $M_{\rm disk}$, both measured relative to the envelope mass $M_{\rm
    env}$ in the models of \cite{visser09} with $\Omega_0 =
  10^{-14}$~s$^{-1}$ and $c_s$ of (a) 0.19~km~s$^{-1}$ and (b)
  0.26~km~s$^{-1}$ (solid lines; see also Fig.~\ref{ruud_circ}). On
  this model track, the thickened lines indicate where the Class 0 and
  I objects would be located given their distribution of measured
  $M_{\rm disk}$/$M_{\rm env}$ values (red and blue,
  respectively). The sources for which stellar, disk and envelope
  masses are measured are shown with symbols. Finally, the dashed
  lines indicate $M_{\rm disk}/M_{\rm star}$ ratios of 1\% (upper) and
  10\% (lower).}\label{ruud_model}
\end{figure}

The presented observations and methodology make it possible to
directly constrain models for the evolution of young stellar objects -
in particular, concerning the transport of matter from the larger
scale envelope to the central star. This is in principle also the most
direct way for estimating the evolutionary status of individual young
stellar objects -- rather than indirect indicators such as their
spectral energy distributions or similar. Fig.~\ref{ruud_model}
compares the predicted stellar and disk masses relative to the
envelope masses (i.e., $M_{\rm star}/M_{\rm env}$ vs. $M_{\rm
  disk}/M_{\rm env}$) from the models of \cite{visser09} to our
observed values. The thickened part of the lines in this plot
indicates where the Class 0 and I sources would fall on the
evolutionary track given their measured $M_{\rm disk}/M_{\rm env}$
ratios.

For the observed ratios of $M_{\rm disk}/M_{\rm env}$ for the Class 0
sources, the models predict ratios of stellar to envelope masses
$M_{\rm star}/M_{\rm env}$ in the range $\approx 0.2-0.6$, i.e.,
systems that have accreted $\approx$~15--40\% of their final mass --
in accordance with the phenomenological definition of objects in the
Class~0 stage. Likewise the Class~I sources are predicted to have more
massive stars than envelopes -- again, in accordance with the
phenomenological distinction between Class~0 and I objects. For the
Class I sources with measured stellar masses, the observed
disk-to-stellar mass ratios are in the 1--10\% range, whereas the
models predict a higher ratio of 20--25\% in the late Class~I stage
(i.e., overestimate the disk mass relative to the stellar mass).  It
is seen that the predicted stellar masses for the Class 0 sources with
this model are a few$\times 0.1~M_\odot$.

The low disk masses relative to stellar masses suggest that
material is quickly and efficiently accreted from the envelope through
the disk onto the star. This is also directly illustrated by the
comparison between the observed and modeled stellar-to-disk mass
ratios for the five Class~I sources for which both have been
determined. For the Class I sources the observed disk masses are
1--10\% of the corresponding stellar masses whereas the models predict
ratios of 20--25\%. This suggests that the material is processed more
quickly through the circumstellar disks than what is given by the
models. \cite{looney03} suggested that this was the case based on
their conclusion that Class 0 protostars have disks that are not more
massive than their Class~I counterparts. Our results are in agreement
with this general conclusion.

It is worth emphasizing though, that our analysis does find evidence
for disks around the Class~0 sources. If the Class~0 sources indeed
are precursors to the Class~I sources in this sample, this in turn
would suggest that circumstellar disks are formed early in the
evolution of low-mass protostars. Without resolved line observations
it is not possible to address whether these disks in fact are
rotationally supported Keplerian disks. The absence of a clear
rotation signature in the disk around NGC~1333-IRAS2A \citep{brinch09}
combined with its large apparent size from the dust continuum
measurements \citep{iras2sma} could perhaps be consistent with it
being an unstable ``pseudo-disk'' as in the framework of
\cite{galli93a,galli93b} -- although it may also simply be that the
evolutionary time-scale for IRAS2A still has been too short for the
disk to establish Keplerian rotation \citep{brinch09}.

An open question in the comparison to the theoretical models is
  how important outflows are in carrying away envelope material. The
  observed increasing opening angles of outflows with increasing age
  of the protostar has been interpreted as evidence for the erosion of
  the protostellar envelope by the outflow
  \citep[][]{velusamy98,arce06} in part explaining the observed shift
  in the peak between the core and initial mass function
  \citep[e.g.,][]{myers08}. In the models of \cite{visser09} and shown
  in Fig.~\ref{ruud_model} the amount of material carried away by the
  outflow only amounts to a few percent. Generally, a more efficient
  outflow dispersal of the core would decrease the disk and stellar
  masses relative to the envelope masses - corresponding to a diagonal
  shift of a given modeled object in Fig.~\ref{ruud_model} - roughly
  along the evolutionary tracks. Therefore, unless the outflow
  preferentially alter the accretion onto the disk relative to the
  star, the ratio between the masses of the two will not be changed
  and thus also not the implications discussed here.

A final interesting possibility is that the flat distribution of disk
masses from the Class~0 to I stages reflects the stability of the
disks, given the suggestion that the luminosity problem of low-mass
protostars can be explained if mass is accreted on large scales in the
disks and from there onto the central star in highly episodic events:
in their models for collapse of a dense core and formation of a
protostar and disk, \cite{vorobyov06} find strong episodic accretion
once the disk has formed which keeps its mass lower than the stellar
mass at all points during the protostellar evolution. Given that the
recent Spitzer/c2d results suggest that the luminosity problem extends
to the earliest Class 0 stages \citep{evans09}, it is tempting to
suggest that material already at these stages indeed ends up in
circumstellar disks and from there is accreted toward the central star
in bursts when a local instability criterion is exceeded -- e.g.,
related to comparison between the disk size/mass and stellar
mass. This could in turn flatten the distribution of disk masses as
function of bolometric temperatures, with an upper envelope of disk
masses of 0.05--0.1$M_\odot$ corresponding to the maximum values
observed for our sample of Class 0 and I protostars. It should be
emphasized that since our sample is rather small, biases may have been
introduced in its selection. Future studies of larger samples of
protostars with current and upcoming (sub)millimeter interferometers
will be needed to address this potential bias.

\section{Summary and outlook}\label{summary}
We have presented a comprehensive analysis of the dust continuum
envelope emission of a sample of 20 embedded low-mass YSOs (Class 0
and I) based on submillimeter interferometric observations from the
Submillimeter Array probing scales of a few hundred AU and single-dish
continuum observations from the JCMT (scales larger than
2000~AU). In addition we discuss HCO$^+$ 3--2 line emission for
  the Class I sources in the sample. With these data we have
characterized the evolution of the three main components YSOs -- the
envelope, disk and stellar masses -- and the relation between these
compared to theoretical models. The main conclusions are as follows:
\begin{enumerate}
\item The millimeter dust continuum emission from both the
  interferometric observations (on scales of 250--500~AU) and
  single-dish observations (scales larger than 2000~AU) decreases in
  strength from the more deeply embedded Class~0 to the more evolved
  Class~I sources. Relative to the single-dish data, the
  interferometric continuum flux at millimeter wavelengths increases
  toward the more evolved sources.
\item Through comparison to predictions from dust radiative transfer
  models, an upper limit on the contribution from protostellar
  envelope to the interferometric continuum flux is estimated -
  relative to the continuum flux measured on larger scales with
  single-dish telescopes. For the Class I sources about 87\% of the
  observed SMA continuum flux (median value for the sample of Class I
  sources) has its origin in the circumstellar disks with the
  remainder coming from the large-scale envelope. For the Class 0
  sources, 68\% of the interferometric emission is related to the
  circumstellar disks. 
\item Using empirical relationships between the envelope and disk
    fluxes and their masses from the radiative transfer models, the
    observations show that the disk-to-envelope mass ratio increases
    from about 1--10\% in the Class 0 stage to about 20--60\% in the
    Class I stage, reflecting the dissipation of the envelope in the
    more evolved stages.
\item The compact interferometric fluxes due to disks are
    observed to decrease from the Class 0 to the Class I
    objects. However, this decrease is largely due to the changing
    physical properties, in particular disk temperature, during this
    phase. If these effects are taken into account, the disk mass
    shows no significant trends from the Class 0 to I stages and is
    typically 0.05 M$_\odot$, albeit with significant scatter. Thus,
    no evidence is seen for significant disk mass growth from the
    Class~0 to the Class~I stages.
\item Four Class~I sources in the sample show evidence of Keplerian
  rotation in HCO$^+$ 3--2, and for IRS~43 also in HCN 3--2. IRS~63
  and IRS~43 show evidence for elongated ``continuum disks'' at
  1.1~mm. For IRS~43 both the ``continuum disk'' and the HCO$^+$ 3--2
  emission coincides with a dark lane in the HST 1.6~$\mu$m images --
  and orientated perpendicular to the direction of the protostellar
  outflow. For the four Class I sources we derive central masses of
  0.3--2.5~$M_\odot$ from the Keplerian rotation curve.
\item The high excitation HCO$^+$ and HCN lines appear to be good
  tracers of the motions in the inner envelopes and disks, although in
  the most massive envelopes their use for this purpose is limited by
  the lines of the main isotopes being optical thick. Future
  observations with ALMA will allow similar studies using the less
  abundant isotopologues.
\item We compare the derived envelope and stellar masses to models for
  the mass evolution in a collapsing protostellar system including the
  formation and growth of a central circumstellar disk
  \citep{visser09}. The data are in agreement with phenomenological
  definition of Class 0 sources having accreted less, and Class I
  sources more, than half of their final mass. The models cannot
  explain the observed stellar masses for the handful of Class~I
  sources for which these have been measured from the disk dynamical
  structure: generally the models produce too massive disks relative
  to their stellar masses.
\item Taken together these results suggest that circumstellar disks
  are established early in the evolution of low-mass protostars, but
  at the same time transport material rapidly from the larger scale
  envelopes onto the central stars. This may occur in episodic events,
  which may explain why the ``luminosity problem'' (that embedded YSOs
  are under-luminous compared to the theoretical predictions from pure
  accretion models) appears already during the Class 0 protostellar
  stages \citep[e.g.,][]{evans09}.
\end{enumerate}

An important test of the theoretical models for the evolution of
embedded young stellar objects will be to measure the disk dynamical
structure even in the most deeply embedded stages. In this way it will
be possible to address whether these disks have in fact settled into a
rotationally stable pattern -- and if so what central stellar masses
are for the central newly formed star -- or if the disk dynamics
perhaps suggest some other configuration, for example, unstable
``pseudo-disks''. In this context, it remains a fundamental question
whether the central stellar masses of the Class 0 objects indeed are
significantly lower than those of the Class~I sources, which would
confirm their earlier evolutionary status.

Future ALMA observations will provide the sensitivity to molecules
that are less abundant, and thus not severely optically thick in the
outer envelopes, and thus a tracer of this structure. With ALMA it
will also be possible to perform larger surveys of a statistically
more unbiased sample of YSOs to constrain the evolution more
directly. This will for example address whether traditional
classification schemes for young stellar objects need to be
recalibrated, whether the episodic accretion scenarios for YSOs are
tied to the properties of their circumstellar disk, whether there are
significant variations between accretion histories of protostars in
different star forming environments, and what fraction of the envelope
material is accreted onto the star and what fraction is carried away
by the outflows.

\begin{acknowledgements}
  We thank the referee for good suggestions about the presentation of
  these results. It is also a pleasure to thank the Submillimeter
  Array staff for help in carrying out observations and technical
  assistance with the resulting data. The Submillimeter Array is a
  joint project between the Smithsonian Astrophysical Observatory and
  the Academia Sinica Institute of Astronomy and Astrophysics, and is
  funded by the Smithsonian Institution and the Academia Sinica. This
  research was based in part on observations from the James Maxwell
  Telescope archive at the Canadian Astronomy Data Centre operated by
  the National Research Council of Canada with the support of the
  Canadian Space Agency. It is also used observations made with the
  NASA/ESA Hubble Space Telescope, obtained from the data archive at
  the Space Telescope Institute. STScI is operated by the association
  of Universities for Research in Astronomy, Inc. under the NASA
  contract NAS 5-26555. Astrochemistry in Leiden is supported by a
  Spinoza Grant from the Netherlands Organization for Scientific
  Research (NWO) and a NOVA grant. The research of TLB was supported
  by NASA Origins grant NNX09AB89G.
\end{acknowledgements}

\Online
\appendix

\section{The use of HCO$^+$ as a dense gas tracer}\label{app:a}
\subsection{Outflow cavities in TMR1, TMC1A and GSS30-IRS1}\label{app:a1}
 The
  HCO$^+$ 3--2 alignment with the near-infrared scattering
nebulosities in the HST images for TMR1, TMC1A and GSS30-IRS1
(Fig.~\ref{hst_hcoP}) is similar to what has been found in high
angular resolution observations of the lower excitation 1--0
transition in both Class~I \citep{hogerheijde97} and Class~0
\citep{l483art} sources.

\begin{figure}
\resizebox{\hsize}{!}{\includegraphics{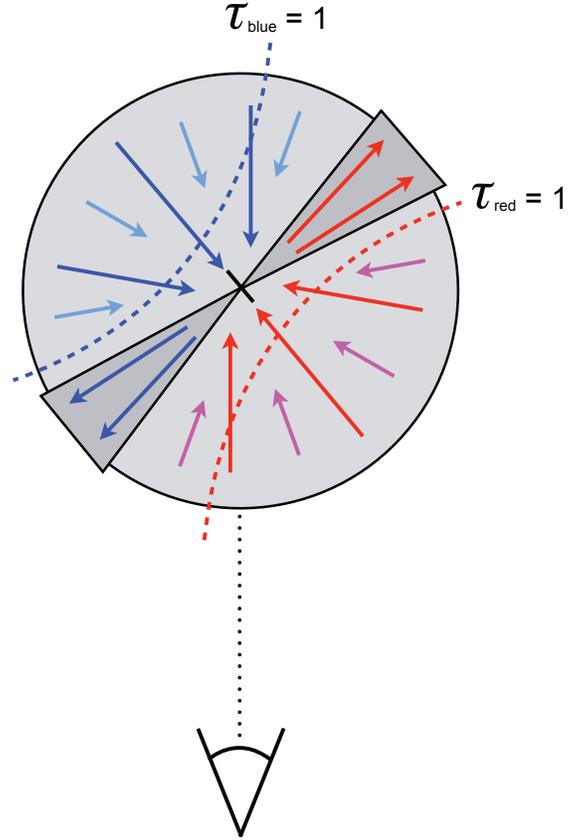}}
\caption{Schematic figure of a collapsing envelope with an outflow
  sweeping up material. The dashed lines indicate the $\tau = 1$
  surfaces for a red- and blue-shifted velocity ($\tau_{\rm red}$ and
  $\tau_{\rm blue}$, respectively) for a line which becomes optically
  thick at low column densities through the envelope. As illustrated
  in the figure, the blue-shifted part of the outflow cone may appear
  in front of the $\tau_{\rm blue} = 1$ surface, while the red-shifted
  part is behind the $\tau_{\rm red} = 1$ surface, obscured by
  material in the large scale collapsing envelope, which in turn may
  be resolved out by the interferometer.}\label{cartoon_fig}
\end{figure}

The patterns in the velocity maps (Fig.~\ref{moment1}) support the
interpretation that the HCO$^+$ 3--2 emission has its origin in
one-side of the outflow cone: in all maps the HCO$^+$ emission is
blue-shifted with respect to the systemic velocity with the most
extreme velocities (most blue-shifted) narrowly confined around the
continuum position and material closer to the systemic velocity
extending around this. This signature can best be understood in a
simple geometry where HCO$^+$ probes material in a large infalling
protostellar envelope being swept up by the outflow, as illustrated in
Fig.~\ref{cartoon_fig}. If the velocities in the swept-up material are
relatively small (comparable to the infalling velocities in the
envelope) and HCO$^+$ 3--2 optically thick, the front-side of the
envelope (toward us) will be slightly red-shifted due to the infall
and thus obscure the outflowing material in the red-shifted cone on
the far-side of the envelope (Fig.~\ref{cartoon_fig}), which in turn
is resolved-out. In contrast, the outflow cone pointed toward us will
be unobscured by the envelope. Likewise, the near-infrared scattered
light from the red-lobe would largely be blocked -- in contrast to the
scattered emission from the outflow cavity on the front-side. An
alternative explanation is that all the sources are at the far-side of
their respective clouds, the red-shifted part of the outflow cone
breaking out into a low-density medium not observable in the line
tracers. This would require a rather sharp density gradient, however.

\subsection{Limitations of HCO$^+$ 3--2 as a tracer of disks}\label{app:a2}
The results above suggest that HCO$^+$ 3--2 is a good tracer of
motions in disks for envelopes that have already been largely
dispersed. For the more embedded YSOs, HCO$^+$ 3--2 emission may be of
limited use as a tracer of rotation in the disk due to its optical
thickness. A simple estimate can be made of the typical envelope mass
below which HCO$^+$ 3--2 line emission is optically thin in the
envelope and can be used as a tracer of the kinematics in the
innermost regions of the protostar. First, for typical envelope
parameters we expect temperatures in the range of 15--40~K and
densities of $10^4$--$10^7$~cm$^{-3}$. Using the Radex escape
probability code \citep{vandertak07radex}\footnote{\tt
  http://www.sron.rug.nl/$\sim$vdtak/radex/radex.php} for these
parameters, the HCO$^+$ 3--2 line has an optical thickness $\approx 1$
for an HCO$^+$ column density of $5\times 10^{12}$~cm$^{-2}$ and a
line width of 1~km~s$^{-1}$. Secondly, we can estimate what density,
$n_{\rm 100 AU}$ is required to reach that optical thickness in a
free-falling envelope with a power-law density profile, $n = (r/{100
  \rm AU})^{-1.5}$ at a size scale of 100~AU corresponding to the
region where the disk potentially forms, i.e., by simply estimating
$n_{\rm 100 AU}$ from
\begin{equation}
N_{\rm HCO+} = \int_{10^2 {\rm AU}}^{\rm 10^4 AU}n_{\rm 100 AU} (r/100 {\rm AU})^{-1.5}\, {\rm d}r
\end{equation}
with $N_{\rm HCO+}=5\times 10^{12}$~cm$^{-2}$. Assuming a typical
abundance of HCO$^+$ of $1\times 10^{-9}$ with respect to H$_2$
\citep[e.g.,][]{paperii}, we find $n_{\rm 100 AU}=2\times
10^6$~cm$^{-3}$, which in turn corresponds to an envelope mass of
0.1~$M_\odot$. This is the maximum envelope mass for which we can use
the HCO$^+$ 3--2 to trace scales smaller than about 100~AU without
worrying about opacity effects in the larger scale envelope. It indeed
corresponds to the estimate of the masses of the most deeply embedded
Class~I sources in our sample for which the outflow emission is
seen. For these sources and the even more deeply embedded Class~0
sources a less abundant or more optically thin tracer is therefore
required to trace the dynamical structure on less than 100~AU
scales. Future observations with ALMA will provide the sensitivity to
image, e.g., the optically thin 3--2 and 4--3 submillimeter
transitions of H$^{13}$CO$^{+}$ in the disks around even more deeply
embedded protostars and thus provide stellar masses for protostars in
the deeply embedded stages.

\section{Summary of observed Class I sources}\label{app:b}
\subsection{IRS~43}
IRS~43 \citep[YLW15;][]{young86} has been the target of numerous
studies across the full wavelength range. At radio wavelengths, IRS~43
is resolved into two separate components, YLW15-VLA1 and -VLA2 with a
separation of 0.6\arcsec\ \citep{girart00}. From multi-epoch high
angular resolution radio data, these binary components were found to
show relative proper motions indicative of orbital motions in a
1.7~$M_\odot$ total mass binary
\citep{curiel03}. \citeauthor{girart00} found that one of the two
components, VLA1, were resolved in observations at 3.6 and 6~cm with
the VLA with a deconvolved size of about 0.4\arcsec\ along its major
axis and suggested that it was associated with a thermal radio jet. In
addition, IRS~43 is a peculiar X-ray emitter showing quasi-periodic
X-ray flares \citep{tsuboi00} consistent with magnetic shearing and
reconnection between the central star and an accretion disk, with the
period suggesting a mass of the central star in the range
1.8--2.2~$M_\odot$ \citep{montmerle00}.

\begin{figure*}\centering
\resizebox{0.8\hsize}{!}{\includegraphics{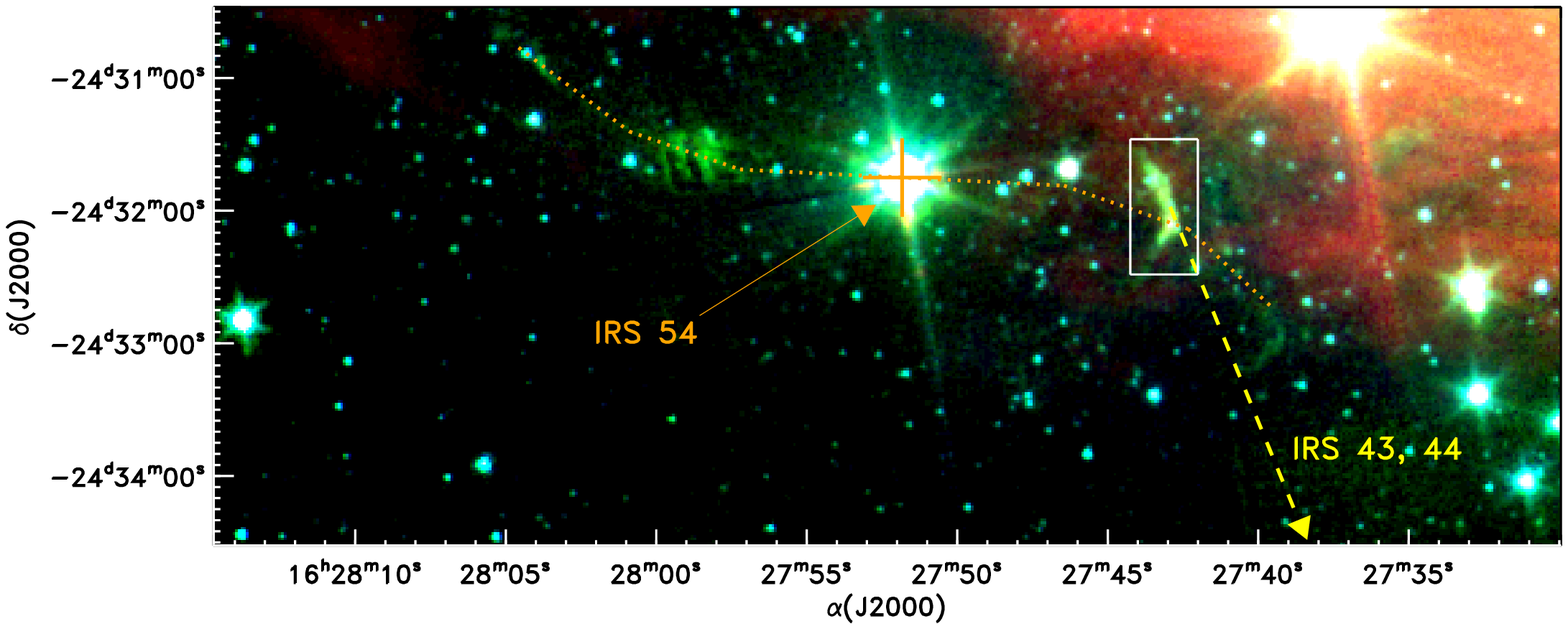}}
\resizebox{0.8\hsize}{!}{\includegraphics{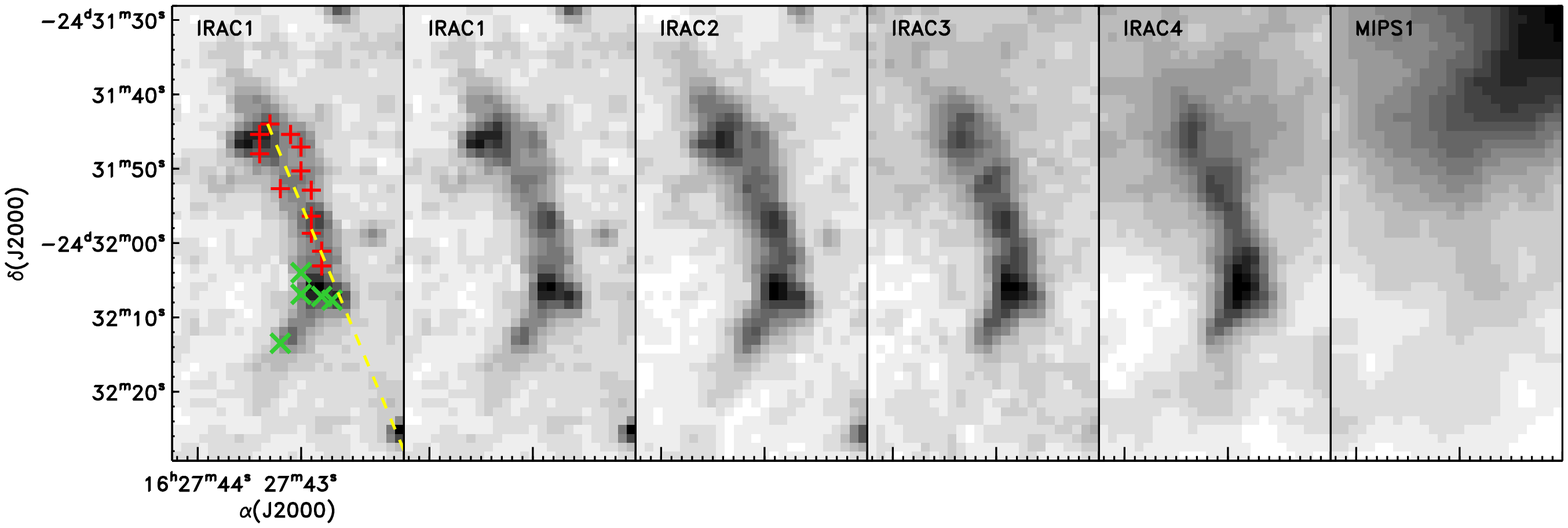}}
\caption{\emph{Upper panel:} The region of the IRS~54 outflow as well
  as the near-infrared Herbig-Haro objects from Spitzer observations
  with 3.6~$\mu$m in blue, 4.5~$\mu$m in green and 8.0$~\mu$m in
  red. In addition we show the suggested direction of the precessing
  IRS~54 outflow (dotted line) and the direction to IRS~43 and IRS~44
  (dashed arrow). \emph{Lower panels:} images of the near-infrared
  Herbig-Haro knots from 3.6-8.0~$\mu$m (IRAC1-4) and 24~$\mu$m
  (MIPS1). The location of the Herbig-Haro knots suggested by
  \cite{grosso01} to be associated with the IRS~54 outflow are shown
  with the crosses, whereas the knots proposed to be associated with
  the IRS~43 outflow are shown with plus-signs in the first
  panel.}\label{spitzer_irs43}
\end{figure*}

The region was imaged in the near-infrared by \cite{grosso01} who
discovered a set of ``embedded'' Herbig-Haro objects, well aligned
with the thermal jet candidate from the centimeter observations of
\cite{girart00}. Spitzer Space Telescope images of the embedded
Herbig-Haro objects (Fig.~\ref{spitzer_irs43}) show the knots clearly
along a line pointing directly back toward the group of objects where
IRS~43 and another embedded protostar, IRS~44, is located. As also
pointed out by \citeauthor{grosso01}, the association of the
Herbig-Haro knots to IRS~43 is not unique: the nearby protostar,
IRS~54 (see also \S\ref{irs54}), also drives a large-scale, precessing
outflow clearly seen in the Spitzer images in
Fig.~\ref{spitzer_irs43}.

The dotted lines in Fig.~\ref{spitzer_irs43} show the propagation of
the IRS~54 outflow, following the H$_2$ emission in the eastern lobe
and assuming that the precession of the IRS~54 outflow is symmetric
around the protostar -- thereby giving the expected location of the
western lobe.  The western lobe in this case passes through the
location of the southern part of the near-infrared Herbig-Haro
objects: based on their morphologies in the higher resolution
near-infrared images, \cite{grosso01} suggested that these knots
(crosses in the lower left panel of Fig.~\ref{spitzer_irs43}) are
indeed caused by the IRS~54 outflow with the northern knots (plus
signs in the lower left panel of Fig.~\ref{spitzer_irs43} related to
the IRS~43 outflow -- the lower spatial resolution Spitzer images do
not reveal this distinction clearly, though. Some support for the
interpretation of \citeauthor{grosso01} can be taken from the
morphology of another bow-shock seen in the Spitzer images at
$(\alpha,\delta)_{\rm J2000} = (16^{\rm h}27^{\rm m}39.5^{\rm s},
-24^\circ33'02'')$ to the west of the near-infrared Herbig-Haro
knots. Together with the near-infrared Herbig Haro knots detected by
\citeauthor{grosso01}, this bow-shock is on the southern side of the
dashed-line suggesting that the dashed line indeed is delineating the
northern edge of the western lobe (and the southern edge of the
eastern lobe) of the IRS~54 outflow. It is also possible that another
source in the group encompassing IRS~43 could be responsible for the
Herbig-Haro knots: the main argument for the association to IRS~43 is
the coincidence with the radio jet and the identification of a
bow-shock south of IRS~43 in the near-infrared, possibly representing
the counter-jet \citep{grosso01,girart00,girart04}.

We detect IRS~43 in both HCO$^+$ 3--2, HCN 3--2 and continuum. The HCN
and HCO$^+$ 3--2 emission is elongated in the W/NW-E/SE direction with
a clear velocity gradient along its major axis. The major axis of the
HCO$^+$ and HCN 3--2 emission is close to the axis of the continuum
emission -- and perpendicular to the direction toward the Herbig-Haro
objects (Fig.~\ref{spitzer_irs43}; \citealt{grosso01}) and the thermal
continuum jet \citep{girart00}.

\subsection{TMR1 and TMC1A}
Images of TMR1 using the HST/NICMOS at 1.6-2.1~$\mu$m by
\cite{terebey98} show s triple system consisting of a close binary
(TMR-1A and TMR-1B) with a projected separation of 0.3$''$ and a third
source TMR-1C offset from the binary by about 10$''$ but connected
with this through an extended filament of near-infrared
nebulosity. \cite{terebey98,terebey00} discuss the option that TMR-1C
is a runaway giant planet, but conclude based on near-infrared
spectroscopy that it is most likely is a background star. TMC1A is an
embedded protostar with an associated scattering nebulosity
\citep[e.g.,][]{chandler96,hogerheijde98} aligned with blue-shifted
outflow emission.

Both TMR1 and TMC1A are detected in continuum, together with HCO$^+$
and HCN 3--2. Both show clear evidence of the line emission being
offset from the continuum peaks toward the outflow nebulosity observed
in the near-infrared HST/NICMOS images -- similar to the case for
GSS30-IRS1. The offsets in the HCO$^+$ 3--2 emission toward the two
sources are similar to those observed in HCO$^+$ 1--0 by
\cite{hogerheijde98}.

\subsection{GSS30}
The GSS30 region is the most complex of the studied fields associated
with a clear near-infrared nebulosity. At least three young stellar
objects (GSS30-IRS1, -IRS2 and -IRS3) are seen at infrared
wavelengths, located within the SMA primary beam field of view,
\citep{weintraub93} and observed by Spitzer
\citep[e.g.,][]{scubaspitz2}. GSS30-IRS1 is located at the center of
this outflow cone and has a mid-infrared SED slope consistent with a
Class I protostar. GSS~30-IRS3 (LFAM1) was detected in high-resolution
images at 2.7~mm \citep{zhang97} and 6~cm \citep{leous91}. GSS30-IRS2
is likely a more evolved (T~Tauri) young stellar object also detected
at 6~cm.

The SMA continuum observations identify one protostar GSS30-IRS3
(LFAM1) whereas the two other protostars remain undetected down to the
noise level of $\approx$~3~mJy~beam$^{-1}$. The same was the case for
the similar in the 2.7~mm observations by \cite{zhang97}. The
single-dish continuum emission toward GSS30 is often attributed solely
to GSS30-IRS1 \citep[e.g.,][]{zhang97}, but appears to be peaking at
GSS30-IRS3 consistent with the suggestion that this source is more
deeply embedded.

The HCO$^+$ line emission shows a distinct peak at the location of
GSS30-IRS1 with a narrow line detected toward GSS30-IRS2
(Fig.~\ref{hcoP_overview}). HCN is also clearly detected toward GSS30,
again toward GSS30-IRS1, but offset from the HCO$^+$ 3--2
emission. Toward IRS1, both HCO$^+$ and HCN appear to trace one side
of the outflow cone directed toward us.

\subsection{WL~12}
WL~12 is a ``standard'' Class~I YSO: it is neither a binary, nor does
it have a spectacular outflow. It is clearly detected in the
interferometric continuum data, but not in the HCO$^+$ or HCN 3--2
maps. It is identified as a core in SCUBA maps and included in the
list of embedded YSOs by \cite{scubaspitz2} with red mid-infrared
colors characteristic of those sources, as well as bright HCO$^+$ 4--3
emission \citep{vankempen09}.

\subsection{IRS~54}\label{irs54}
IRS~54 is found to be a binary with a separation of about 7\arcsec\ in
deep near-infrared wavelength observations \citep{haisch04,duchene04},
possibly also responsible for the precession of the outflow observed
in the mid-infrared Spitzer images (Fig.~\ref{spitzer_irs43}). IRS~54
is the only source in our sample, which is not detected in either
continuum and only shows a tentative detection in HCO$^+$ 3--2. It is
associated with a faint SCUBA core (peak of 0.1~Jy~beam$^{-1}$; total
integrated flux in a 40$''$ radius aperture of 0.4~Jy) below the
threshold for core detection in \cite{scubaspitz2}. It is also not
identified as a separate millimeter core in the Bolocam maps of
\cite{young06}. Based on its HCO$^+$ 4--3 line emission,
\cite{vankempen09} classify it as a ``late Stage 1'' source in
transition to the T~Tauri stage.

\subsection{IRS~63 and Elias 29} 
The results for IRS~63 and Elias 29 are discussed in more detail in
\cite{lommen08}. In summary, the two sources are detected in both
continuum and HCO$^+$ 3--2. IRS~63 shows the strongest continuum
emission by about a factor 8 whereas the Elias 29 is stronger in
HCO$^+$ 3--2 also by about a factor 8. Where Elias 29 clearly is
associated with strong \emph{extended} continuum emission, witnessed
both by its brightness profile in the SMA data and by comparison to
single-dish observations, IRS~63 is dominated by emission from a
compact, unresolved component, consistent with the suggestion that it
is a source in transition between the embedded and T~Tauri stages with
$M_{\rm env} \sim M_{\rm disk}$.

\end{document}